# Neutrino mass limit from tritium β-decay

E.W. Otten

*Johannes Gutenberg-Universität Mainz, Institut für Physik, D-55099 Mainz, Germany*
*Email: ernst.otten@uni-mainz.de*

*and* C. Weinheimer

*Westfälische Wilhelms-Universität Münster, Institut für Kernphysik, Wilhelm-Klemm-Str. 9, D-48149 Münster, Germany*
*Email: weinheimer@uni-muenster.de*

**Abstract**

The paper reviews recent experiments on tritium β-spectroscopy searching for the absolute value of the electron neutrino mass $m(\nu_e)$. By use of dedicated electrostatic filters with high acceptance and resolution, the uncertainty on the observable $m^2(\nu_e)$ has been pushed down to about 3 eV$^2$. The new upper limit of the mass is $m(\nu_e) < 2$ eV at 95% C.L. In view of erroneous and unphysical mass results obtained by some earlier experiments in β-decay, particular attention is paid to systematic effects. The mass limit is discussed in the context of current neutrino research in particle- and astrophysics. A preview is given of the next generation of β-spectroscopy experiments currently under development and construction; they aim at lowering the $m^2(\nu_e)$-uncertainty by another factor of 100, reaching a sensitivity limit $m(\nu_e) < 0.2$ eV.

## 1. Introduction

The existence of the neutrino was postulated by Wolfgang Pauli [1] in 1930, in order to explain why the electron spectrum of β-decay is continuous. This hypothetical neutrino would be emitted in the process of β-decay, together with the electron, and would be neutral and massless, or at least much lighter than the electron. It should have spin ½ and should interact with matter much more weakly than any other particle known at the time. These hypotheses led Fermi [2] to his well-known theory of weak interaction, which explained the pricipal features of β-decay quantitatively. With a modification accounting for maximum parity violation, Fermi`s theory is still valid at low energies. It took about a quarter of a century to prove the existence of the neutrino directly when, in 1956, reactor electron antineutrinos $\bar{\nu}_e$ were detected by Cowans and Reines [3]. Soon thereafter, Goldhaber showed that neutrinos participate in the weak interaction as left-handed particles only [4]. With the discovery of all the 12 fundamental fermions of the Standard Model of particle physics, three different neutrino flavours $\nu_e$, $\nu_\mu$, $\nu_\tau$ were found [5]; these being the partners of the three charged leptons e$^-$, μ$^-$, τ$^-$ in weak charged current reactions, which convert a neutrino into its corresponding charged lepton or vice versa, respectively. Consequently, neutrinos are incorporated into the Standard Model of particle physics as massless, left-handed and neutral spin ½ particles.

Although the Standard Model could describe the properties of the neutrino very well, it was discussed quite early on, whether neutrinos were indeed massless, or whether they carried some small mass to explain, to a significant fraction at least, the missing dark matter in the universe (e.g. [6]); The velocity distribution in spiral galaxies and other observations show that there is much more gravitating matter in the universe than can be seen in the form of luminous stellar objects. Comparison of the abundance of light elements in the universe, and the theory of the synthesis of light elements within the first three minutes after the Big Bang, excluded baryonic matter as making up most of this dark matter. At this point, neutrinos with masses of a few eV (using the convention $\hbar = c = 1$ we present masses in units of eV) were considered to be the ideal candidates for the missing non-baryonic dark matter, as the universe should be full of neutrinos: According to the Big Bang theory, a relic (still not detected) neutrino density of 336 neutrinos per cm$^3$ should exist throughout the universe, similar to cosmic microwave background radiation, as an imprint of the early universe, when the neutrinos decoupled from the hot particle plasma. In recent years, observations of the cosmic microwave background radiation and of the distribution of matter in the universe at different scales, suggest that most of the missing non-baryonic dark matter is so-called 'cold dark matter': massive particles, which were already non-relativistic during structure formation [7]. These particles have to differ from the neutrinos $\nu_e$, $\nu_\mu$, $\nu_\tau$, although massive neutrinos might still have played an important role in the evolution of the universe.

Contemporaneously, the question of neutrino masses developed as an emerging field of inquiry in particle physics: Deficits in the $\nu_e$-flux from the sun, already established in 1968 [8] as well as deficits in the $\nu_\mu$-flux from the atmosphere [9,10] observed in deep underground laboratories, gave early hints of the possible existence of neutrino oscillations. In 1998, the observation of atmospheric neutrinos by the Super-Kamiokande experiment showed a clear deficit of up-going muon neutrinos[11]. This and many more underground experiments (Super-Kamiokande [12], Kamiokande [13] Gallex [14], SAGE [15], SNO [16], Borexino [17], KamLAND [18], K2K [19], MINOS [20]) with atmospheric, solar, accelerator and reactor neutrinos proved, in the last decade, that a neutrino flavour state (e.g. a $\nu_\mu$ in the case of the previously mentioned atmospheric neutrinos) can oscillate into another neutrino flavour state (e.g. into a $\nu_\tau$,) during flight. Thus, a flight path dependent variation of the neutrino reaction rate will be observed if the detector is not equally sensitive to the different oscillating neutrino flavour states. Neutrino oscillation could explain all these experimental findings and thus consistently solve the long-standing solar and atmospheric neutrino puzzles.

Neutrino oscillation can only occur if the neutrino flavour states $\nu_e$, $\nu_\mu$, $\nu_\tau$, are non-trivial mixings of three neutrino mass states $\nu_1$, $\nu_2$, $\nu_3$, whose masses $m_1$, $m_2$, $m_3$,, differ from each other. This is not entirely surprising, as we have been accustomed to the fact, for some decades now, that the weak quark eigenstates are superpositions of the mass or strong quark eigenstates connected by the Cabbibo-Kobayashi-Maskawa matrix.

This discovery of neutrino oscillation clearly points to physics beyond the Standard Model of particle physics, although neutrino masses could be incorporated into the Standard Model without major changes. But we will discuss in section 2, that such a "minimal extended Standard Model" remains unsatisfactory and that more changes are only to be expected. This circumstance strongly correlates questions concerning neutrino masses to the type of theories beyond the Standard Model. Knowledge of neutrino masses is not only of great importance, therefore, for cosmology and astrophysics, but also very significant for particle physics.

Unfortunately, neutrino oscillation experiments are only sensitive to the neutrino mixing angles and to differences between squared neutrino masses $\Delta m_{ik}^2 = |m_i^2 - m_k^2|$ but not to the neutrino mass values $m_i$ themselves.

Information on neutrino masses can be obtained by three different methods:
1) cosmological observations
2) search for neutrino-less double beta decay
3) direct determination of the neutrino mass by kinematics

Although Methods 1 and 2 are very sensitive to neutrino masses, their results are model-dependent as we will point out in section 2. On the other hand, direct neutrino mass determination from the kinematics of weak decays is essentially based on energy and momentum conservation only, and thus model-independent. As we will discuss in section 2, each of these three methods has its advantages and disadvantages, and each of them gives complementary information on the neutrino masses.

The most sensitive direct neutrino mass search is based on the very precise investigation of a β-spectrum near its endpoint, as was pointed out by Fermi when he developed the theory of β-decay in 1934 [2]. At given energy resolving power of the β-spectrometer $E/\Delta E$ a low β-endpoint energy is favoured, as is found in the case of tritium decay with $E_0 \approx 18.6 \, \text{keV}, T_{1/2} = 12.3 \, \text{a}$. The long history of neutrino mass searches in tritium ß-decay numbers about a dozen experiments, commenced by Curran et al. in the late forties, which yielded an upper limit of $m(v_e) < 1 \, \text{keV}$ [21] (We do not differentiate between the masses of particles and antiparticles assuming CPT-symmetry. A definition of $m(v_e)$ in terms of neutrino mass eigenvalues is given in section 2). The most recent limit of

$$m(v_e) < 2 \, \text{eV} \; (95\% \, \text{C.L.}) \qquad (1)$$

Which has been published by the particle data group [5] is based on the results of the latest generation of experiments performed at the University of Mainz [22] and at the Institute of Nuclear Research at Troitzk near Moscow [23]. Considering that the observable in such a mass search is $m^2$ rather than $m$, the limit has been improved, in the meantime, by 5 orders of magnitude.

The first breakthrough toward improved mass limits was achieved by Bergkvist in the early seventies [24] by building a dedicated "$\sqrt{2}\pi$"-magnetic spectrometer with both high luminosity and high resolution resulting in $m(v_e) < 55\,\text{eV}$. For the first time, the influence of the excitation of the electron shell of the daughter molecule, which extends well above this limit, was taken into account in the analysis. The next major event, a decade later, was the claim of the discovery by the ITEP group around Ljubimov in Moscow of a non-zero neutrino mass of $30\,\text{eV}$ [25]. The group used for the first time, the new, powerful Tretyakov spectrometer [26] with its superior luminosity and resolution when compared to previous spectrometers. This instrument was based on a long, toroidal magnetic field which accepted the full azimuth of emitted β-particles, like the well-known orange spectrometer, but enhanced the momentum resolution by aligning four 180° orange spectrometers in series. A thin film of tritiated valine served as the source. The ITEP-result was received with some scepticism; this criticism was focussed on the analysis, but could not offer clear-cut grounds to dismiss the findings. The ITEP group responded, in a later publication, with a slightly revised, but still finite mass value of $17 < m(v_e)/\text{eV} < 40$, based on a result for the squared mass of

$$m^2(v_e) = (970 \pm 60_{\text{stat}} \pm 160_{\text{syst}})\,\text{eV}^2$$ [27]. In retrospect one may identify two sources of error which could have falsified the ITEP result: (i) the inelastic scattering correction was probably overestimated which shifted the endpoint as well as $m^2(v_e)$ upwards through correlations similar to those discussed in section 3.5. (ii) The confidence of the authors in their apparently too high T/$^3$He-mass difference of $\Delta M(\text{T},^3\text{He}) = 18600(4)\,\text{eV}$ which resulted from their spectrum was backed, unfortunately, by another, at the time, new result from direct mass spectroscopy yielding $\Delta M(\text{T},^3\text{He}) = 18599(2)\,\text{eV}$ [28]. Both turned out later to be significantly wrong, when van Dyck *et al*. published a doublet splitting of 18590.1(1.7) eV obtained by cyclotron resonance in a Penning trap [29]. The latter value has been confirmed by a recent result $\Delta M(\text{T},^3\text{He}) = 18589.8(1,2)\,\text{eV}$ using the same technique [30].

Three groups decided to check the ITEP result by setting up dedicated experiments based on advanced Tretyakov spectrometers and new source concepts. The first to challenge the ITEP claim was the group around W. Kündig at the University of Zürich with a result $m^2(v_e) < 310\,\text{eV}^2$ [31]. Their wide experience in the field of condensed matter led them to build a sophisticated source which consisted of a very homogeneous Langmuir-Blodget film of $CT_2$-chains fixed by a single bond at one end to a silicon substrate for the second phase of the experiment. The final state spectrum of daughter excitations was computed by quantum chemistry methods [32] and considered in the analysis. The final result of this experiment was $m(v_e) < 11.7\,\text{eV}$ derived from the measured observable

$$m^2(v_e) = (-24 \pm 48_{\text{stat}} \pm 61_{\text{syst}})\,\text{eV}^2 \text{ at 95\% C. L. [33]}.$$

A group at Los Alamos National Laboratory took the next big step forward in source development by choosing a windowless gaseous tritium source (WGTS) [34]. They fed molecular tritium ($T_2$) into a long tube with open ends. β-particles were guided along a strong axial magnetic field out of the tube into a Tretyakov spectrometer. It was even considered converting this source into an ideal, atomic one by dissociating the $T_2$ within a plasma beforehand. But this procedure has to be quantitative because an uncertain ratio of $T/T_2$ would be highly disturbing in the analysis as the two endpoints differ by 10.0 eV (see figure 5). The final state spectrum of $T_2$ has been calculated by various authors with increasing precision over the years, to the extent that its uncertainty has never dominated experimental error bars to the present. A WGTS has also been chosen for the follow-up experiment at the Lawrence Livermore National Laboratory [35], as well as for the Troitzk experiment [36,37]. At present, an advanced molecular WGTS is under construction for the forthcoming KATRIN experiment at the Forschungszentrum Karlsruhe [38,39] (see section 5.1.). The Los Alamos group published a final result of $m(v_e) < 9.3\,\text{eV}$ at $95\%\,\text{C.L.}$ derived from the measured observable

$$m^2(v_e) = (-147 \pm 68_{stat} \pm 41_{syst})\,\text{eV}^2$$ [40]. An upgraded and, in certain respects, modified version of the Los Alamos experiment has been set up at Livermore, as mentioned above. Their full data set fitted to a squared mass of $m^2(v_e) = (-120 \pm 20)\,\text{eV}^2$; from a restricted set, the authors derived an upper mass limit of $m(v_e) < 8\,\text{eV}$ [35].

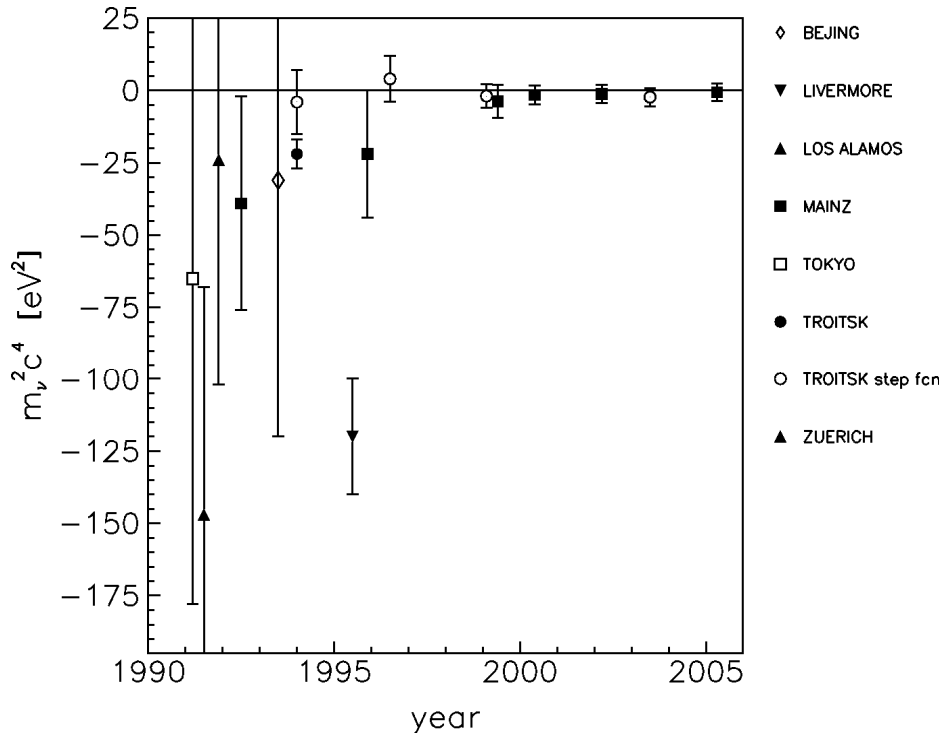

Figure 1: Squared neutrino mass values obtained from tritium β-decay in the decisive period 1990-2005 plotted against the year of publication (see text for the references). The results from the more recent experiments in Mainz and Troitzk, presented in section 4, are already included.

The $m^2(v_e)$-values obtained by the three checking experiments did agree among themselves within their combined error bars (compare figure 1); the error bars also excluded the ITEP-result. One could, however, recognize another problem which subsequently troubled the community for a long time: The mean values now fell into the unphysical negative region of the $m^2$-plot! Somehow the new experiments seemed to have overshot the mark. This feature was not significant for the Zürich-result where the error bar still extended into the positive sector. This also holds true for the results from the Tokyo- and Beijing-experiments, which were obtained from conventional spectrometers with modest luminosity [41,42].

But in view of the smaller errors, the results from Los Alamos, and particularly from Livermore, have a problem, caused, in all likelihood, by some unrecognised systematic error source. How should such a result be interpreted? Before 1998 one followed the so-called Bayesian approach, which was recommended by the Particle Data Group; it gave the following guidance : (i) The respective Gaussian error curve is centred at the place of the mean experimental value in the unphysical region and the fraction of its area which extends into the physically allowed region is determined; this fraction is considered the chance of the unphysical value found to be just a statistical fluctuation instead of being caused by some unrecognised systematic error. (ii) The residual area in the physically allowed sector is split into parts 95% to 5% (90% to10%) and the position of the split is considered the upper limit of the quantity in question with 95% (90%) confidence level (C. L.). Since 1998 the Particle Data Group favours the so-called frequentist approach [43], which gives similar results close to the physically allowed region.

Of course, such limits are subject to the disadvantage of being increasingly 'contaminated', the further the result is located in the unphysical region in terms of units of its uncertainty $\sigma$. Moreover, this ought not to be the only criterion of evaluation: looking again at the $m^2(v_e)$-plot of figure 1, one can see that later, much more obviously, precise results also populate the negative sector. Furthermore, their distance from the physical limit has been reduced in proportion to the error, and the latest result is fully compatible with 0 [5]! This is a typical situation in precision experiments: Residual contaminations of a result are uncovered only at higher levels of experimental and statistical accuracy and are eventually eliminated only then. These very precise results have been obtained at Mainz and Troitzk using a novel type of electrostatic filter, which features higher luminosity and resolution than magnetic spectrometers. A huge instrument of this kind will also serve the forthcoming KATRIN-experiment [39]. These experiments form the centrepiece of this review. The subject has already been covered in part in the context of a recent, more general review on neutrino masses by Weinheimer [44] or by Robertson and Wilkerson [45]. The earlier experiments, which we have recalled here briefly, have been discussed in reviews by Holzschuh [46] and by Robertson and Knapp [47]

Our review is structured as follows: In section 2 of this paper, we present briefly the phenomena of neutrino oscillation and models to describe neutrino masses. In this section, we also present a brief synopsis of the three above mentioned methods of obtaining information on neutrino masses. In section 3, we discuss the β-spectrum with respect to its sensitivity to the neutrino mass and correlated parameters, as well as systematic effects. Section 4 deals with the latest generation of experiments on neutrino mass limits from tritium-β-decay, including side experiments for the study of systematic effects. The paper finishes with a preview on forthcoming experiments, in particular the KATRIN experiment at Karlsruhe.

## 2. Neutrino masses in particle physics, astrophysics and cosmology

*2.1. Neutrino oscillations*

Proceeding under the assumption that the three neutrino generations have finite masses $m_i = m_1, m_2, m_3$, it is natural to assume that (in analogy to the quark sector) the weak interaction is not diagonal in the neutrino *mass* eigenstates $\nu_i = \nu_1, \nu_2, \nu_3$ but produces *flavour* eigenstates $\nu_\alpha = \nu_e, \nu_\mu, \nu_\tau$, as superposition of mass eigenstates connected by an unitary 3×3 mixing matrix $U$:

$$\nu_\alpha = \sum_i U_{\alpha i} \nu_i . \qquad (2)$$

Then, after generating a weak eigenstate $\nu_\alpha$ in a weak reaction, its mass components $\nu_i$ propagate each with different phase velocity leading to observable flavour oscillations along its flight path; the oscillation length will scale with the total neutrino energy $E_{\text{tot }\nu}$ and the reciprocal difference of squared masses $|m_i^2 - m_k^2| = \Delta m_{ik}^2$ [48]:

$$L_{ik} = 4\pi E_{\text{tot }\nu} / \Delta m_{ik}^2 . \qquad (3)$$

The oscillation amplitude is determined by the coefficients $U_{ik}$ or likewise by mixing angles $\Theta_{ik}$. In the latter representation, the unitary neutrino mixing matrix $U$ is decomposed into three rotation matrices

$$U = \begin{pmatrix} 1 & 0 & 0 \\ 0 & \cos\Theta_{23} & \sin\Theta_{23} \\ 0 & -\sin\Theta_{23} & \cos\Theta_{23} \end{pmatrix} \begin{pmatrix} \cos\Theta_{13} & 0 & \sin\Theta_{13}e^{i\varphi} \\ 0 & 1 & 0 \\ -\sin\Theta_{13}e^{-i\varphi} & 0 & \cos\Theta_{13} \end{pmatrix} \begin{pmatrix} \cos\Theta_{12} & \sin\Theta_{12} & 0 \\ -\sin\Theta_{12} & \cos\Theta_{12} & 0 \\ 0 & 0 & 1 \end{pmatrix} . \quad (4)$$

In case neutrinos are identical to their antiparticles, we call them Majorana particles, in contrast to Dirac particles, which differ from their antiparticles. For Majorana neutrinos the matrix $U$ has to be expanded by 2 additional complex Majorana phases $\alpha_2$ and $\alpha_3$:

$$U = \begin{pmatrix} 1 & 0 & 0 \\ 0 & \cos\Theta_{23} & \sin\Theta_{23} \\ 0 & -\sin\Theta_{23} & \cos\Theta_{23} \end{pmatrix} \begin{pmatrix} \cos\Theta_{13} & 0 & \sin\Theta_{13}e^{i\varphi} \\ 0 & 1 & 0 \\ -\sin\Theta_{13}e^{-i\varphi} & 0 & \cos\Theta_{13} \end{pmatrix} \begin{pmatrix} \cos\Theta_{12} & \sin\Theta_{12} & 0 \\ -\sin\Theta_{12} & \cos\Theta_{12} & 0 \\ 0 & 0 & 1 \end{pmatrix} \times$$

$$\times \begin{pmatrix} 1 & 0 & 0 \\ 0 & \exp(i\alpha_2/2) & 0 \\ 0 & 0 & \exp(i\alpha_3/2) \end{pmatrix}.$$

(5)

The phase $\phi$ accounts for a possible CP-violation of neutrino mixing like in the quark sector; $\phi$ but has not been observed in present oscillation experiments as yet. The Majorana phases $\alpha_2$ and $\alpha_3$ are not accessible by neutrino oscillation experiments.

The many neutrino oscillation experiments with evidence for neutrino mixing of type $\Theta_{12}$ and $\Theta_{23}$ together with those experiments observing upper limits for $\Theta_{13}$ [49,50] have yielded a consistent picture of neutrino mixing with the following parameters (as evaluated from the latest results) [5]:

$$\Delta m_{12}^2 = 8.0^{+0.4}_{-0.3} \times 10^{-5}\,\text{eV}^2 \qquad 1.9 < \Delta m_{23}^2/10^{-3}\,\text{eV}^2 < 3.0$$
$$\sin^2 2\Theta_{12} = 0.86^{+0.03}_{-0.04} \qquad 0.92 < \sin^2 2\Theta_{23} \leq 1 \qquad \sin^2 2\Theta_{13} < 0.19.$$

(6)

These results tell that
- Neutrinos have a rest mass in contrast to their description in the Standard Model.
- Their mass differences are small and separate $m_3$ distinctly from a more closely spaced doublet $(m_1, m_2)$.
- The lepton family number $L_i$ is violated, which is not in contradiction to the Standard Model, as there is no underlying symmetry requiring lepton family number conservation.
- The mass eigenstates are strongly mixed in weak interactions; $\Theta_{12} \approx 34^o$ dominates the observed $(\nu_e, \nu_\mu)$ - oscillation in solar and reactor neutrinos; $\Theta_{23} \approx 45^o$ dominates the observed $(\nu_\mu, \nu_\tau)$ -oscillation in atmospheric neutrinos; $(\nu_e, \nu_\tau)$-oscillation is yet to be discovered.

The two differences of squared neutrino masses observed so far in neutrino oscillation experiments provide neither the absolute mass values nor the ordering of $m_3$ with regard to the other two neutrino masses $m_1$ and $m_2$. The ordering of $m_1$ and $m_2$ is known to be $m_2>m_1$ from additional, dispersion-like matter effects in solar neutrino oscillation [51]. Assuming that the lightest neutrino has a mass much smaller than $\Delta m_{23}^2$ (so called hierarchic case), the mass of the heaviest one would be

$$m_{\text{hierarchic, max}} \approx \sqrt{\Delta m_{23}^2} \approx 0.05 \text{ eV} . \qquad (7)$$

The opposite case $\sqrt{\Delta m_{ik}^2} << m_i \approx m(\nu_\alpha) \approx m_\nu$ is called the degenerate case, where all neutrinos have about the same mass $m_\nu$ and their absolute splitting $|m_i - m_k| \approx \Delta m_{ik}^2 / 2m_\nu$ shrink reciprocal to this common mass. Figure 2 shows the possible neutrino mass schemes versus the unknown mass of the lightest neutrino mass for the case $m(\nu_3) > m(\nu_2) > m(\nu_1)$ ("normal hierarchy") and for $m(\nu_2) > m(\nu_1) > m(\nu_3)$ ("inverted hierarchy").

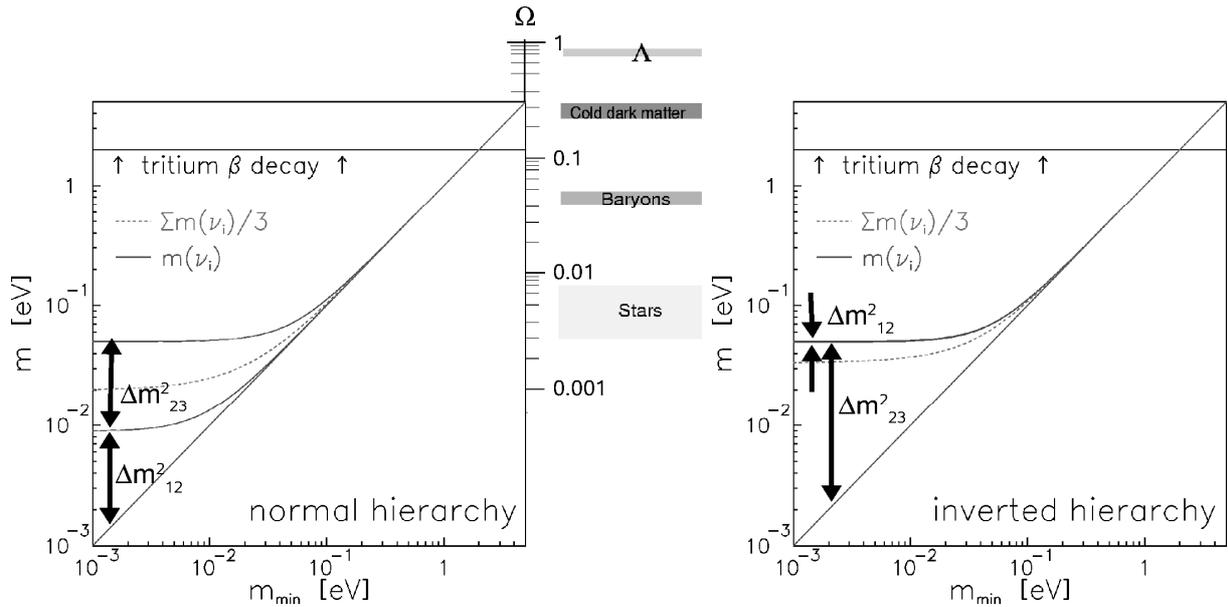

**Figure 2**: Neutrino mass eigenvalues $m(\nu_i)$ (solid lines) and one third of the cosmological relevant sum of the three neutrino mass eigenvalues $\Sigma m(\nu_i)/3$ (dashed line) as a function of the smallest neutrino mass eigenvalue $m_{\text{min}}$ for normal hierarchy $m(\nu_3) > m(\nu_2) > m(\nu_1)$ (left) and inverted hierarchy $m(\nu_2) > m(\nu_1) > m(\nu_3)$ (right). The upper limit from tritium β-decay on $m(\nu_e)$ (solid line), which holds in the degenerate neutrino mass region for each $m(\nu_i)$, and for $\Sigma m(\nu_i)/3$ (dashed line) is also marked. The hot dark matter contribution $\Omega_\nu$ to the universe relating to the average neutrino mass $\Sigma m(\nu_i)/3$ is indicated by the right scale in the normal hierarchy plot and compared to all other known matter/energy contributions in the universe (middle). With the relic neutrino density of 336/cm³ the laboratory neutrino mass limit from tritium β-decay $m(\nu_e) < 2$ eV corresponds to a maximum allowed neutrino matter contribution in the universe of $\Omega_\nu < 0.12$.

*2.2. Models of generating neutrino masses*

Since we still lack a general, valid theory of masses of elementary particles, these fundamental quantities are empirical input parameters to the theoretical model. The experimental situation is furthermore quite complex. In the case of quarks, access to the mass of the free particles is impossible, in the strict sense, as the strong interaction (QCD) confines them to the complex structure of hadrons; asymptotic freedom, on the other hand, is reached only at fully relativistic energy where the rest mass is hardly recognised in the kinetics any more. Hence we have to accept relatively large uncertainties; a factor of 2 even in the case of the first, light generation u, d [5]. Regarding leptons, masses are known to many digits in the charged sector e, μ, τ. But the many attempts in the search for a missing neutrino mass in the kinetics of any kind of reaction or decay involving neutrinos, have yielded only upper limits so far; here, the result for the electron neutrino from tritium decay, $m(\nu_e) < 2$ eV, is by far the most precise [5].

Intuitively, one might take the large mixing angles as indicating degenerate neutrinos. Speculation apart, we possess a first experimental indication for the degenerate case from neutrinoless double β-decay (see next section). In any case, the oscillation parameters (6) definitely imply that the present upper mass limit of 2eV from tritium β-decay applies to any mass- or flavour eigenstate. Experiments, therefore, to improve further on the present limits of direct $m(\nu_\mu)$- or $m(\nu_\tau)$-measurements (see sections 2.5.1. and 2.5.2.) would have limited interest for on-going study.

Present theoretical models cannot give clear preference to either case, but the various neutrino mass scenarios may correspond to different extensions of the Standard Model: usually the charged fermions acquire their mass by the coupling of the left-handed weak isospin fermion-doublet (e.g. $\begin{pmatrix} \nu_L \\ e_L \end{pmatrix}$) to the Higgs-doublet $\begin{pmatrix} \Phi^+ \\ \Phi^0 \end{pmatrix}$ and a right-handed fermion singlet (e.g. $(e_R)$). Spontaneous symmetry breaking $\left( \begin{pmatrix} \Phi^+ \\ \Phi^0 \end{pmatrix} \rightarrow \begin{pmatrix} 0 \\ v/\sqrt{2} \end{pmatrix} \right)$ generates a fermion mass term defined by the Yukawa-coupling (e.g. $f_e$) and the Higgs vacuum expectation value $v$ within the Lagrangian;

$$L_e = -f_e (\bar{\nu}_L \ \bar{e}_L) \begin{pmatrix} \Phi^+ \\ \Phi^0 \end{pmatrix} (e_R) + h.c. = -f_e (\bar{\nu}_L \ \bar{e}_L) \begin{pmatrix} 0 \\ v/\sqrt{2} \end{pmatrix} (e_R) + h.c.$$
$$= -\frac{f_e v}{\sqrt{2}} \bar{e}_L e_R + h.c. := -m_e \bar{e}_L e_R + h.c. \tag{8}$$

The corresponding Yukawa-couplings are rather similar within a fermion family (e.g. $f_e \approx f_u \approx f_d$) yielding similar masses. Within the Standard Model there is no right-handed neutrino singlet, resulting in non-existing neutrino masses. Of course, in order to obtain a non-zero neutrino mass, we may introduce a right handed neutrino singlet $(v_R)$ into an extended Standard Model yielding (via the coupling to the charge conjugate of the Higgs-doublet) a mass term with neutrino mass $m_D$, which we call the Dirac-mass term (see illustration in figure 3a):

$$L_v = -m_D(\overline{v_L}v_R) + h.c. \tag{9}$$

It remains very unsatisfactory, however, that the Yukawa-coupling for neutrinos have to be at least 6 orders of magnitude smaller than the ones for all other charged fermions without any reason. Therefore, this mechanism for generating neutrino masses is generally considered to be rather unlikely.

We will get much more room to manoeuvre, if we allow neutrinos to equal their antiparticles. It is, of course, appreciated that this violates total lepton number conservation. Using the charge and parity conjugated neutrino state $v^c$ in addition two new neutrino mass terms, the so-called Majorana mass terms, are allowed to appear in the Lagrangian $L$ of an extended Standard Model (We denote it for one neutrino family for simplicity. The factor ½ appears because of the double counting of neutrino states and their conjugates):

$$L_v = -\frac{1}{2}\left(m_D(\overline{v_L}v_R + \overline{(v_R)^c}(v_L)^c) + m_{LL}\overline{v_L}(v_L)^c + m_{RR}\overline{(v_R)^c}v_R\right) + h.c. = -\frac{1}{2}(\overline{v_L}, \overline{(v_R)^c}) M \begin{pmatrix}(v_L)^c \\ v_R\end{pmatrix} + h.c. \tag{10}$$

The $m_{LL}$-term transforms a left-handed neutrino into its charge and parity conjugated state, whereas the $m_{RR}$-term couples a right-handed neutrino with its charge and parity conjugate. It is obvious, that such couplings between particles and their charge and parity conjugates could only exist for neutral particles like the neutrinos. The $m_{LL}$-term is not allowed in an extension of the Standard Model with only the Higgs doublet [52], because $v_L$ lives in a weak isospin doublet and the term $\overline{v_L}(v_L)^c$ violates weak isospin. If we then calculate for $m_{LL} = 0$ the mass eigenvalues of the mass matrix $M$, we will obtain the following neutrino masses (the resulting negative mass value $m_{v1}$ can be mirrored into the positive sector by a phase rotation) for $m_{RR} \gg m_D$:

$$m_{v1} \approx \frac{m_D^2}{m_{RR}} \qquad m_{v2} \approx m_{RR}. \tag{11}$$

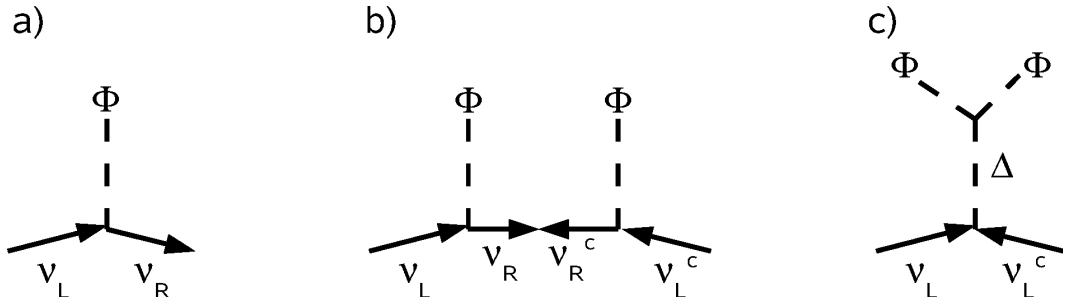

**Figure 3**: Schematic view on different ways to generate light neutrino masses (charge conjugate neutrino states $\nu^c$ are plotted moving backward in time): a) Coupling a left-handed neutrino $\nu_L$ to a right-handed light neutrino $\nu_R$ via the Higgs $\Phi$ (Dirac mass term). b) Coupling a left-handed neutrino $\nu_L$ to a right-handed heavy neutrino $\nu_R$ (via Higgs), transforming to its charge and parity conjugated mass state and back to the conjugate of the left-handed neutrino (via Higgs) within the time allowed by the uncertainty principle (the heavy neutrino is integrated out and gives rise to the suppression of the light neutrino mass $m_{\nu 1}$). This mechanism is termed the Type 1 Seesaw-mechanism. c) Neutrino mass term by coupling of a left-handed neutrino to its charge and parity conjugated state via a Higgs-triplet $\Delta$, which couples twice to the Standard Model Higgs. This mechanism is called Type 2 Seesaw-mechanism.

In this limit the parameter $m_{RR}$ corresponds to the mass of a right-handed heavy neutrino (see figure 3b). If $m_{RR}$ is of the order of the GUT scale and $m_D$ is as large as typical charge lepton masses, then $m_{\nu 1}$ will become very light and just be of the right order. This mechanism is called the Type 1 Seesaw-mechanism [55,56], and describes the smallness of the neutrino mass in a natural way and favours hierarchical neutrino mass scenarios by the $m_D^2$ dependence of $m_{\nu 1}$. Although this mechanism looks quite appealing, it offers no natural explanation for the neutrino mixing to be so large.

As an alternative to the introduction of the right-handed heavy neutrino, we could rather naturally obtain small neutrino masses if we allow a $m_{LL}$-term by introducing additionally to the Standard Model Higgs, a new Higgs-triplet $\Delta$, which couples to the left-handed light neutrino and its conjugate (see figure 3c) [57]. Such a Higgs-triplet exists in most supersymmetric extensions of the Standard Model. This mechanism favours degenerate neutrino mass scenarios and is termed the Type 2 Seesaw-mechanism.

When both extensions (right-handed heavy neutrinos and Higgs-triplet) are present, we can explain any kind of neutrino mass hierarchy.

The diagrams in figures 3b) and 3c) are both implementations of an effective dimension 5 operator with two left-handed fermion doublets, two Higgs fields and an effective heavy scale $\Lambda$ in the denominator, which generates the smallness of the neutrino mass [58]. Therefore, from the particle physics point of view, the determination of neutrino masses and their hierarchy leads necessarily into terra incognita up to very large scales Λ beyond the Standard Model. Another corner stone is the knowledge of the history of the Big Bang, which has run through the relevant high energy phases. For a comprehensive review on the present status of neutrino theory see, for example, ref. [59].

*2.3. Neutrino masses from cosmology*

The Big Bang has left over a dense, blackbody radiation of so-called relic neutrinos, similar to the cosmic microwave background (CMB) (the substance of this section is reviewed, for example, in ref [60]). These have cooled down by adiabatic expansion over time to a temperature of about 2 K, and populate the universe with an average density of 336/cm$^3$, summed over the 3 eigenstates and both helicities with equal abundance. Their number surpasses that of baryons by 9 orders of magnitude. Let us assume, for instance, that the 3 mass eigenvalues sum up to $\sum m_i = 1\,\text{eV}$; then neutrinos would already account for some 2% of the total mass/energy budget of the universe – more than that of the stars. In the current cosmological model, relic neutrinos represent so-called hot dark matter which (in contrast to cold dark matter) was not bound to local gravitational fields during structure formation because of the sufficiently small rest masses of the particles. Therefore, hot dark matter was capable of relaxing local fluctuations in the mass/energy density distribution by a kind of diffusion-like mass/energy transport in-between. The transport would scale with the mass density of hot dark matter and it would relax density fluctuations over short distances more strongly then over large ones, which is typical for diffusion processes.

Non-statistical fluctuations of the mass/energy distribution have been discovered and mapped with great accuracy for CMB [7], as well as for the distribution of a huge sample of observed galaxies [61]. The former represents the early universe some 380 000 years after the Big Bang, the latter the large scale structure (LSS) of the present and recent universe. In case of CMB, the observed angular pattern of the radiation temperature is expanded into spherical harmonics up to angular momentum $l \approx 1000$. In case of LSS, one determines the spatial correlation of the distribution of galaxies and Fourier-transforms it onto *k*-space. The resulting very characteristic spectra are basic input parameters for fitting the parameters of the present model of cosmic evolution. The most spectacular outcome of such fits was the following: About 95% of the mass/energy budget of the universe has to be attributed to unknown sources of which 70% are ascribed to Einstein's cosmological constant Λ or, more generally, to so-called dark energy, which is responsible for the accelerated expansion of the universe. A share of 25% is ascribed to cold dark matter which leads to the clustering of galaxies.

For hot dark matter, the fits have, so far, found only upper, but quite sensitive limits : In 2008, the WMAP-Collaboration published a value [7]

$$\sum m_i \leq 0.61\,\text{eV}.$$

(12)

In the meantime more data on CMB, as well as LSS, has become available, and a number of new upper limits have been published, which cover a range of about (0.2 – 1) eV [5,58]. The sharper limits rely on additional constraints, e. g. from weak gravitational lensing and from the so-called Lyman-α forest; the weaker ones are more conservative and bear lesser risk of model dependence. Recently, analysis of so-called baryon acoustic oscillations in the LSS [59] allows for a less model-dependent analysis of the structure at small scales than using the Lyman-α forest [62].

In the near future the SSDS-survey will have mapped some 1 million galaxies, and the Planck satellite will have refined the CMB map. The fits are predicted to achieve a sensitivity on neutrino masses well below $\sum m_i = 0.1\,\text{eV}$ and even to approach the mass scale of the normal hierarchy . However, this will not be a measurement of the neutrino mass but a result from fitting a model with many parameters, of which some have not yet even found a physical explanation . Consequently, astro- and particle physicists are equally eager to see the model-dependent fitting of the neutrino mass being paralleled by another direct mass measurement with increased sensitivity in the laboratory. In a number of papers, the interplay between absolute neutrino mass values from single- and double β-decay, as well as from cosmology, has been discussed in the context of present results and future targeted experimental outcomes. For three recent examples see [63,64,65].

*2.4. Neutrinoless double β-decay and neutrino mass*

The search for neutrinoless double β-decay (0νββ) is a classical approach to physics beyond the Standard Model, as it would violate,in the first instance, the conservation of lepton number *L*. The mechanism usually considered is as follows: The emission of the first β is accompanied by a virtual ν; the latter is reabsorbed by the intermediate daughter nucleus (A, Z-1) via inverse β-decay, leading to the final state (A, Z-2) + 2β + 0ν (here denoted for the case β⁻β⁻). The neutrino has to be of Majorana type which violates *L* by definition as particle-antiparticle superposition. A finite neutrino mass is required in order to produce in the chirality-selective $(V-A)$-interaction a neutrino with a small component of opposite handedness on which this neutrino exchange subsists. The decay rate would scale with the absolute square of the so called effective neutrino mass $m_{ee}$ :

$$\Gamma_{0\nu\beta\beta} \propto \left|\sum_i U_{ei}^2 m_i\right|^2 = \left|(m_1 \cos^2 \Theta_{12} + m_2 e^{i\alpha_2} \sin^2 \Theta_{12})\cos^2 \Theta_{13} + m_3 e^{i(\alpha_3 + 2\varphi)} \sin^2 \Theta_{13}\right|^2 := m_{ee}^2.$$

(13)

Here $m_{ee}$ represents the coherent sum of the $m_i$-components of the 0νββ-decay amplitudes and hence carries their relative phases. With the given oscillation parameters and the knowledge of the type of neutrino mass hierarchy, the number of unknown parameters of $m_{ee}$ reduces to 4, one mass $m_i$, two phases $\alpha_2, (\alpha_3 + 2\varphi)$, and the mixing angle $\Theta_{13}$. In case of inverted mass hierarchy (i. e. $m_3 \ll m_1, m_2$) or degenerate neutrinos, we may disregard the $m_3$-term in (13) because of the small $\Theta_{13}$. In both cases $m_1$ and $m_2$ are about equal; hence the phase difference $\alpha_2$ plays a decisive role because $\alpha_2 \approx \pi$ leads to a strong cancellation of $m_{ee}$ as $\Theta_{12}$ is quite large (see equation. (6)).

Besides neutrino masses and phases, the nuclear matrix element is a major player in 0νββ-decay, because it has to be summed up coherently over the whole spectrum of states in the intermediate nucleus, to which the parent and the final nucleus connect by β-decay. Great efforts have been made, at many times, to tackle this difficult task, applying different theoretical approaches, and with different results (for a recent article see [66]). To accept these differences as current theoretical uncertainties is a pragmatic, albeit not convincing, approach. If one assigns an accuracy within a factor of order 2 to the most advanced theoretical results, one has probably arrived at a fair compromise between the views of the optimists and those of the pessimists in the field.

Models beyond our Standard Model extended to Majorana neutrinos, offer a range of mechanisms which could contribute at some level to 0νββ-decay with a structure different from (13) (*e.g.* by the exchange of a supersymmetric particle, instead of a massive neutrino) and with an open parameter space. ( So far the situation resembles, somewhat, the well-known case of the muon g-factor anomaly, which integrates up any kind of virtual interaction even beyond QED. In that case, a level of experimental and theoretical precision has been reached where new physics might enter the last of the 8 significant digits [5].) Still, neutrino exchange is considered to be the leading term. In this context, a theorem has been proved by Schechter and Valle telling that models which allow 0νββ-decay necessarily imply the existence of Majorana neutrinos with finite masses [67]. These arguments demonstrate that 0νββ-decay is involved in many open questions which cannot be answered simultaneously by a single decay rate. Its discovery would rather open a door into a new and wide territory.

Until recent years, all experiments on different nuclides have reported only upper limits of $\Gamma_{0\nu\beta\beta}$ and corresponding $m_{ee}$-values. The Heidelberg-Moscow experiment on double β-decay of $^{76}$Ge, which is considered the most sensitive one, reported a result in 2001 [68]

$$m_{ee} < 0.35 \text{ eV} \quad \text{at } 90\% \ C.L.. \tag{14}$$

Subsequently, a subgroup of this collaborative grouping published a refined analysis which revealed a signal at the right decay energy with a significance of $3\sigma$ above background [69]. Additional data and a new calibration have strengthened the signal up to $4.2\sigma$; it corresponds to a decay rate $\Gamma_{0\nu\beta} = 0.84 \times 10^{-26}/y$ from which the authors derive an effective neutrino mass in the limits [70]

$$0.1 \leq m_{ee}/\text{eV} \leq 0.9 \ . \tag{15}$$

The result (15) lies above the hierarchic limit (7) and would classify neutrino masses as being degenerate.

These analyses with positive results for 0νββ were received with scepticism by parts of the community (e.g. [71,72] for direct criticisms to [67]). The significance could be enhanced to $6\sigma$ by pulse shape selection [73]. This data cut rejects most of the γ-background as giving rise to pulses which are diluted in time due to a delocalized, so called multi-site energy deposition, whereas ββ0ν-events deposit their energy locally (single-site event). Still, both types of events populate rather broad classes of pulse shapes with some overlap which cannot be strictly quantified. Hence this latest result may be regarded as enhanced significance for a *lower limit* rather than an improved *value* of the decay rate. However, the scepticism felt by part of the community still remained, although no stringent counter-argument could falsify the claim.

This situation clearly calls for experimental clarification by independent experiments with enhanced sensitivity on $^{76}$Ge and other suitable nuclides. In the future, several experiments on 0νββ-decay (for a recent review see [74]) aim at reaching a sensitivity limit in the range of the limit (7) which would be sufficient to discover the case of inverted hierarchy (provided the phase $\alpha_2$ is constructive) (see figure 4).

*2.5 Neutrino masses from kinematics*

Besides nuclear β-decay, which is treated extensively in later chapters, several other laboratory experiments, as well the Supernova event 1987A, have provided upper limits of absolute neutrino masses from the analysis of neutrino kinematics. Before the discovery of neutrino oscillations, it was necessary to address each neutrino flavour separately in this context. In the following, we will briefly summarise the results.

*2.5.1. m($\nu_\mu$) from pion decay*

As compared to β-decay, the rest mass of neutrinos which are associated with the production or decay of a muon or a tau is much harder to observe, since these neutrinos carry much higher energies. The muon neutrino mass $m(\nu_\mu)$ has been investigated in the two-body decay of pions at rest:

$$\pi^+ \to \mu^+ + \nu_\mu \text{ or } \pi^- \to \mu^- + \bar{\nu}_\mu . \tag{16}$$

Energy and momentum conservation result in sharp momenta $p(\mu) = p(\nu)$ from which follows:

$$m^2(\nu_\mu) = \left(m^2(\pi) + m^2(\mu)\right) - 2m(\pi)\sqrt{m^2(\mu) + p^2(\mu)} \tag{17}$$

according to the relativistic invariant for total energy $E_{tot}$ and momentum $p$

$$m^2 = E_{tot}^2 - p^2 . \tag{18}$$

In a dedicated precision experiment at the Paul Scherrer Institute (Zurich) [75] the muon momentum has been determined to be $p(\mu) = 29.791998(110)$ MeV. Using input parameters $m(\mu) = 105.6583568(52)$ MeV and $m(\pi) = 139.570180(350)$ MeV [5] the authors have obtained from (17) a mass squared value

$$m^2(\nu_\mu) = -(0.016 \pm 0.023) \text{ MeV}^2 \tag{19}$$

from which an upper limit on the muon neutrino mass itself can be derived [5] (We will clarify the definition of m($\nu_\mu$) in the context of neutrino mixing at the end of the following subsection.)

$$m(\nu_\mu) < 190 \text{ keV } (90\% \text{ C. L. }). \tag{20}$$

Why is this limit so much higher than the uncertainties of the input masses and of the muon momentum from which it is derived? As already indicated above, this is a trivial consequence of relativistic kinematics, namely, of the quadratic form of the energy-momentum relation (18). Therefore, the given input uncertainties of neutrino energy and -momentum $\Delta E_\nu$ and $\Delta p_\nu$ have to be scaled up with the full energy $E_{\text{tot }\nu}$ and momentum $p_\nu$ respectively, when calculating the uncertainty of the derived neutrino mass squared:

$$\Delta m_\nu^2 \approx \Delta E_{\text{tot }\nu}^2 + \Delta p_\nu^2 \approx 2 E_{\text{tot }\nu} \Delta E_{\text{tot }\nu} + 2 p_\nu \Delta p_\nu. \tag{21}$$

In any search for a kinetic neutrino mass, neutrino energy should, therefore, be as small as possible; otherwise relativity hides the mass! This argument favours the search for $m(\nu_e)$ in low energy nuclear β-decay by many orders of magnitude as compared to the case of the other neutrino flavours. On the other hand, any decay rate into neutrinos shrinks with their phase space density, and hence with their energy squared! In-between these two poles little space is left, and tremendous effort will be required to devise new experiments which might extend decisively the existing limits of kinetic mass measurements.

*2.5.2. m(ν$_\tau$) from tau decay*

The most sensitive direct information on the mass of the tau neutrino $m(\nu_\tau)$ comes from the investigation of tau pairs which are produced at electron-positron colliders and decay into pions. Decays into a maximum of pions (19 and 20) yield the highest sensitivity on $m(\nu_\tau)$ because they restrict the available phase space of the neutrino, however, at the expense of the branching ratio. The quantity looked at is the invariant mass of the multiple pions $M_\pi$. In the rest frame of the decaying tau the respective relations are:

$$M_\pi^2 = \left( \sum_j E_{\text{tot }j}(\pi), \sum_j \boldsymbol{p}_j(\pi) \right)^2 = \left( m(\tau) - E_{\text{tot}}(\nu_\tau), -\boldsymbol{p}(\nu_\tau) \right)^2 \leq \left( m(\tau) - m(\nu_\tau) \right)^2. \tag{22}$$

The most precise result has been obtained by the ALEPH collaboration at LEP. A 2-dimensional analysis in the $\left( M_\pi, \sum_j E_{\text{tot }j,\text{lab}}(\pi) \right)$- plane restricts the tau neutrino mass to [76]

$$m(\nu_\tau) < 18.2 \text{ MeV} \quad (95\% \text{ C..L.}). \tag{23}$$

The existence of neutrino mixing tells us that the flavour neutrino eigenstates $\nu_e$, $\nu_\mu$ and $\nu_\tau$ are certain superpositions of mass eigenstates $m_i$. How do we then interpret the results of the pion and tau decay experiments? Due to the limited experimental resolution, the tiny splitting between the squared neutrino mass eigenvalues were far from being resolved. This holds, too, for all neutrino mass searches in nuclear β-decay. Therefore, the obtained upper limits for $m(\nu_e)$, $m(\nu_\mu)$ and $m(\nu_\tau)$ (see equations (1), (19), and (23)) correspond to the weighted average of the neutrino mass eigenstates contributing to the given flavour:

$$m^2(\nu_e) = \sum_{i=1}^{3} |U_{ei}^2|^2 m_i^2 \qquad m^2(\nu_\mu) = \sum_{i=1}^{3} |U_{\mu i}^2|^2 m_i^2 \qquad m^2(\nu_\tau) = \sum_{i=1}^{3} |U_{\tau i}^2|^2 m_i^2 \qquad (24)$$

Although questioned by the KARMEN experiment [77], and later excluded by the MiniBooNE experiment [78], the evidence for neutrino oscillation by the LSND experiment [79] would require a third squared neutrino mass difference $\Delta m^2_{ij}$ in addition to $\Delta m^2_{12}$ and $\Delta m^2_{23}$ which explain the findings of solar and atmospheric neutrino experiments. The question was, therefore, raised as to whether the neutrino mixing matrix should be extended to more than 3 neutrino states. These additional neutrinos cannot couple to the $Z^0$- and $W^\pm$-bosons, as we know from the measurement of the $Z^0$-pole width, there are exactly three active neutrino states [5]. Good candidates for these so-called "sterile neutrinos" are the right-handed neutrinos discussed in the context of Majorana neutrino mass terms (see section 2.5.2). Although they do not couple to the weak gauge bosons, sterile neutrinos become visible by their mixing to the neutrino mass states. In a measurement of weak decay kinematics, e.g. of a β-decay, we would then have to extend our neutrino mass formula by $n_s$ sterile neutrino mass states; however, this would not affect the shape of the β-spectrum, as long as the different mass states are not resolved:

$$m^2(\nu_e) = \sum_{i=1}^{3+n_s} |U_{ei}^2|^2 m_i^2 \qquad (25)$$

*2.5.3. Neutrinos from supernova 1987*

In section 2.5.1 we argued that in order to achieve access to the neutrino mass, we have to restrict the available neutrino phase space as much as possible in order to allow the neutrinos to exhibit their mass when they are not fully relativistic. The investigation of μ- and τ-decays, cannot, therefore, compete with β-decays measurements.

Although we cannot bypass the arguments of relativistic kinematics, one can still compensate for this disadvantage in that the neutrino mass uncertainty $\Delta m^2_\nu$ scales with the total neutrino energy $E_{tot\,\nu}$ (e.g. equation (21)) by the determination of the neutrino energy with extremely small uncertainty $\Delta E_{tot\,\nu}$. Astrophysics offers extremely bright neutrino sources, like, for example, a Type II supernova, which could provide a neutrino time-of-flight measurement. A reasonable compromise between a flight path long enough to obtain the necessary small uncertainty $\Delta E_{\text{tot }\nu}$ and a still acceptable neutrino detection rate is provided by Type II supernova explosions in our own, or in a neighbouring galaxy. These nearby supernovae can be expected to occur a few times per century.

In 1987, underground detectors at Baksan, Caucas [80], in the Morton Thiokol mine, Ohio [81], and in the Kamioka mine, Japan [82], observed a burst of neutrinos which, undoubtedly, had arrived from the well-known supernova 1987A in the Large Magellanic Cloud, close to our galaxy, after a time of flight of about $t \approx 5 \times 10^{12}$ s . Altogether about two dozen events occurred within a time interval of about $\Delta t \approx 10$ s at energies between $E_{\text{tot min}} \approx 10 \, \text{MeV}$ and $E_{\text{tot max}} \approx 40 \, \text{MeV}$ with the more energetic particles tending to arrive earlier. If one ascribed this effect to a rest mass, one can arrive at a rough estimate quite easily using $\beta_{\max} \approx 1 \approx \beta_{\min}$ and $E_{\text{tot max}}^2 >> E_{\text{tot min}}^2$:

$$m^2 = E_{tot}^2 - p^2 = E_{tot}^2 \left(1 - \beta^2\right) = E_{tot}^2 (1-\beta)(1+\beta) \approx 2 E_{tot}^2 (1-\beta)$$

$$\Rightarrow \beta \approx 1 - \frac{m^2}{2 E_{tot}^2}$$

$$\frac{\Delta t}{2t} = \frac{L/\beta_{\min} - L/\beta_{\max}}{L/\beta_{\min} + L/\beta_{\max}} = \frac{\beta_{\max} - \beta_{\min}}{\beta_{\max} + \beta_{\min}} \approx \frac{\beta_{\max} - \beta_{\min}}{2} = \frac{1}{2}\left(-\frac{m^2}{2 E_{tot\,\max}^2} + \frac{m^2}{2 E_{tot\,\min}^2}\right) \approx \frac{m^2}{4 E_{tot\,\min}^2}$$

$$\Rightarrow m \approx E_{\min\,tot} \sqrt{\frac{2\Delta t}{t}} = 10 \, \text{MeV} \sqrt{\frac{20 \, \text{s}}{5 \cdot 10^{12} \, \text{s}}} = 20 \, \text{eV}$$

(26)

The latest detailed analysis identifies an upper limit of $m(\nu_e) < 5.7$ eV based on input, which also includes the dynamics of realistic supernova models [83]. Relation (26) is another example of the consequence of how relativity uncompromisingly reduces enormous relative accuracy in the raw data of $10^{-12}$ down to its square-root in the final result. Nonetheless, this mass limit (though being model dependent) matches the order of magnitude of the best laboratory results. If neutrinos from another nearby supernova arrive on earth at some time in the future, they will encounter a set of sensitive neutrino detectors, well prepared to learn more neutrino- as well as supernova physics from the event.

*2.6 Complementarity of the different methods*

We want to remark upon the important difference between the composition of the masses $m(\nu_e)$ (equation (24)) and $m_{ee}$ (equation (13)) observed in single and neutrinoless double β-decay, respectively: In the former case we measure an incoherent and unresolved sum of *β*-spectra each leading with probability $|U_{ei}|^2$ to a mass eigenstate $m_i$ in the latter case a coherent sum of these masses with unknown phases. The present neutrino mass limit of $m(\nu_e) < 2\,\text{eV}/c^2$ from tritium β-decay (1) does not provide a critical check of $m_{ee}$ in the claimed range of (15). But the follow-up experiment KATRIN, designed to reach a sensitivity of $0.2\,\text{eV}$ fits into it well. If non-zero neutrino masses $m(\nu_e)$ and $m_{ee}$ will be found both in single and in neutrinoless double β-decay, respectively, their difference could be used to gain information on the Majorana phases $\alpha_2, \alpha_3$, which are otherwise not accessible [84].

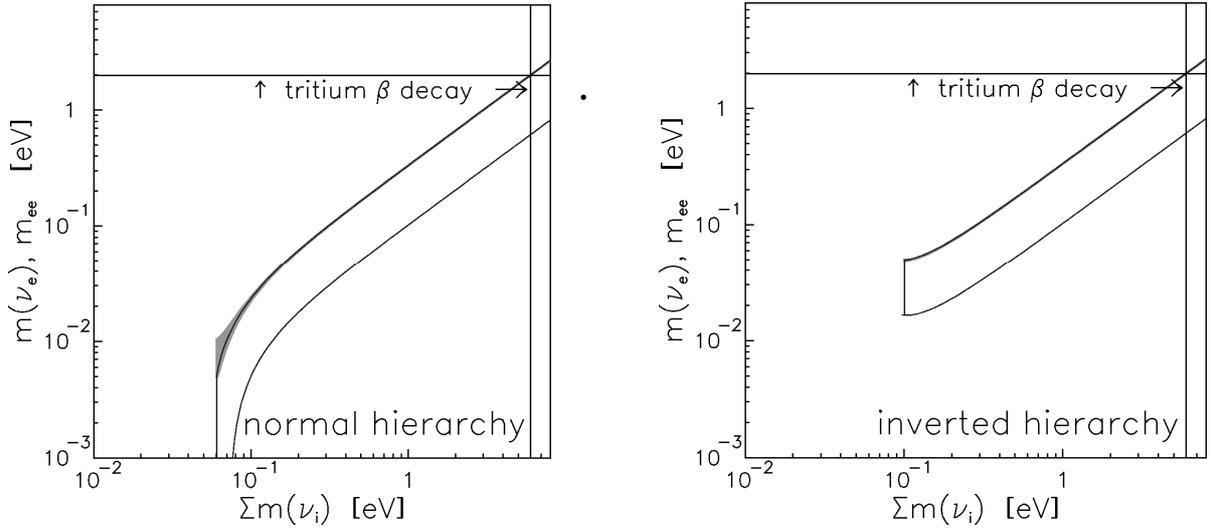

**Figure 4**: Observables of neutrinoless double β-decay $m_{ee}$ (open band) and of direct neutrino mass determination by single β-decay $m(\nu_e)$ (thin gray area sitting at the upper end of the $m_{ee}$ band) versus the cosmological relevant sum of neutrino mass eigenvalues $\Sigma m(\nu_i)$ for the case of normal hierarchy (left) and of inverted hierarchy (right). The width of the bands/areas is caused by the experimental uncertainties of the neutrino mixing angles (6) and in the case of $m_{ee}$ also by the completely unknown Majorana- and CP-phases $\alpha_2, (\alpha_3 + 2\varphi)$ (13). Uncertainties of the nuclear matrix elements, which enter $m_{ee}$, are not considered.

As equation (24) is free of undetermined interference terms, it would answer the question about absolute neutrino masses unambiguously. Figure 4 illustrates that there is a clear correlation of $m(\nu_e)$ with the cosmological relevant neutrino mass sum $\sum m(\nu_i)$. On the other hand, the unknown CP- and Majorana-phases $\alpha_2, (\alpha_3 + 2\varphi)$ in the coherent sum (13) do not permit neutrinoless double β-decay a precise determination of $\sum m(\nu_i)$ nor of the mass scale $m(\nu_e)$ (see figure 4). But we would like to note that the importance of the search for neutrinoless double β-decay does not only relate to the value of the neutrino mass. The search for neutrinoless double β-decay is the only method to establish a possible Majorana character of neutrinos, which is of great importance for particle physics.

In summary, recent years have witnessed enormous progress in our knowledge of neutrino properties, which is giving decisive input and drive to theoretical particle physics. But we are still only halfway; great efforts are being made to master the remaining half as well, in order to arrive at a conclusive experimental and theoretical picture of neutrino physics.

## 3. Nuclear β-decay and neutrino mass

*3.1. Suitable candidates for neutrino mass search*

As previously mentioned, the most sensitive upper limits on the mass of the electron neutrino $m(\nu_e)$ (1)) have been achieved by investigating the $\beta^-$ -spectrum from nuclear beta decay. The phase space region of low energy neutrinos, where the highest sensitivity to the neutrino mass is achieved according to (21), corresponds to the very upper end of the ß-spectrum. This extremely tiny part of the spectrum can be emphasized with respect to the, with regard to the neutrino mass, embarrassing bulk by choosing β-emitters with very low endpoint energy like $^{187}$Re with $E_0 \approx 2.6$ keV, or tritium with $E_0 \approx 18.6$ keV, respectively. Although $^{187}$Re has a 7.5 times lower endpoint energy compared to tritium, there are quite a few arguments which favour tritium as the ideal isotope for a neutrino mass search experiment:

1. Tritium β-decay is a super-allowed decay with a rather short half-life of 12.3 y, whereas the uniquely forbidden $^{187}$Re β-decay has a half-life of $4.3 \cdot 10^{10}$ y. Therefore, tritium is the only β-emitter, which allows a specific activity large enough for an experimental set up with a β-source and a separated β-electron spectrometer.

2. Only (super-) allowed decays have a nuclear matrix element, which does not show any dependence on the energy of the β-electron.

3. $T_2$ is the simplest molecule allowing quantitative calculation of its final state spectrum (s. sec. 3.4).

The arguments 1. and 3. against $^{187}$Re β-decay can in principle be overcome, however, by using cryogenic bolometers, which act at the same time as a β-source and as a detector (sec. 5.2).

*3.2. Q-value and endpoint of β-spectrum*

The candidates selected above are $\beta^-$-emitters characterized by

$$(A,Z) \rightarrow (A,Z+1)^+ + e^- + \bar{\nu}_e + Q. \qquad (27)$$

They release surplus energy $Q$ which is shared between the kinetic energy of the ß-particle ($E$), the total energy of the neutrino ($E_{tot\,\nu}$), the minute recoil energy ($E_{rec}$) carried by the much heavier daughter, and the excitation of the daughter to a final state of energy $V_j$:

$$Q = E + E_{tot\,\nu} + E_{rec} + V_j = E_0 + E_{rec} = \Delta M - \Delta E_B. \qquad (28)$$

We will briefly discuss the different terms in (28). In case of zero neutrino mass and the daughter being produced in its very groundstate ($V_j=V_0=0$), the β-spectrum would terminate at the so called endpoint

$$E_0 = Q - E_{rec}. \qquad (29)$$

Figure 5 demonstrates that $Q$ is given by the mass difference 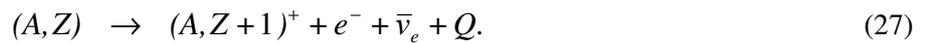 $\Delta M = M(A,Z) - M(A,Z+1)$ between the (by definition) *neutral* mother and daughter atoms, corrected for the difference in electronic binding energy $\Delta E_B$ between the *atomic mother/daughter pair* and the *actual mother/daughter systems* of the experiment, for instance, a neutral molecule and a molecular ion, respectively. This correction may be calculated, for example, from a combination of ionisation energies ($E_{ion}$) and

molecular dissociation energies ($E_D$), as illustrated in figure 5. For the decay of gaseous $T_2$ into the groundstate of the molecular daughter ion $(^3HeT)^+$, the correction is [85,86]

$$\Delta M = M(T) - M(^3He) = Q + E_D(T_2) - E_D(^3HeT^+) + E_{ion}(T)$$
$$= Q + 4.59\,eV - 1.90\,eV + 13.60\,eV = Q + 16.29\,eV = Q + \Delta E_B \quad (30)$$

With $E_{rec} = 1.72\,eV$ (see (40)) one calculates from the most recent $\Delta M$-value [30] using (29) and (30) for this decay the endpoint energy of

$$E_0(T_2) \;=\; 18571.8 \pm 1.2 \;\; eV. \quad (31)$$

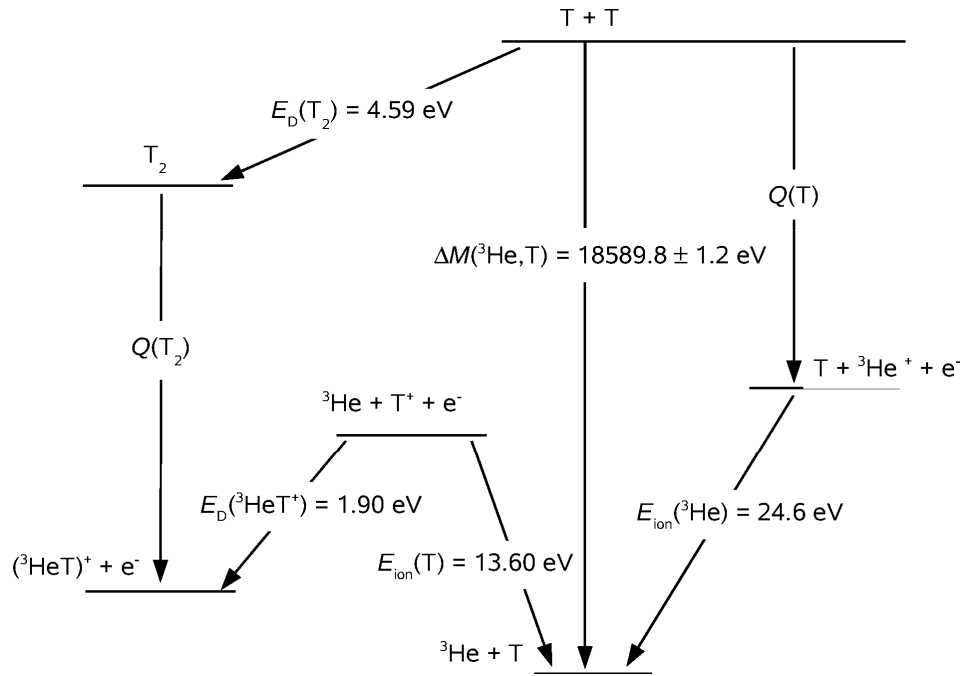

**Figure 5**: Level diagram illustrating the relation between mass difference $\Delta M(^3He,T)$ measured by cyclotron resonance in a Penning trap [30] and the Q-values of molecular and atomic tritium decay.

The endpoint of the decay of atomic tritium $E_0(T)$ is lower by 10.0 eV (This includes the reduction by the twice as large recoil energy $E_{rec} = 3.44$ eV of the daughter). In section 3.4, we will argue that the rotation-vibration excitation of the ($^3$HeT)$^+$- daughter will lower the *average* endpoint of molecular tritium β-decay by another 1.72 eV, such that the *average total energy given to nuclear motion* is the same as in atomic tritium decay.

.

*3.3. Nuclear β-spectrum*

The decay probability $\Gamma$ of a state $|i\rangle$ (given in energy units due to the suppression of a factor $1/\hbar$) is $2\pi$ times the square of the transition matrix element $M_{fi}$ summed and integrated over all possible discrete and continuum state $|f\rangle$ of the daughter system and the emitted particles (Fermi's golden rule)

$$\Gamma = 2\pi \sum \int |M_{fi}|^2 df . \tag{32}$$

The matrix element in (32) is factorized into a leptonic part ($M_{lep}$) and a nuclear part ($M_{nuc}$), multiplied by the universal Fermi coupling constant $G_F = 1.16637(1) \cdot 10^{-5}$ (GeV)$^{-2}$ [5] (suppressing a dimensional factor $(\hbar c)^3$) and its projection $|V_{ud}| = \cos\Theta_C = 0.97377(27)$ [5] onto (u,d)-coupling by the Cabbibo angle $\Theta_C$:

$$M = G_F \cos\Theta_C M_{lep} M_{nuc} . \tag{33}$$

According to Fermi's original ansatz of a point-like weak interaction, the leptonic part $|M_{lep}|^2$ essentially results in the probability of the two leptons to being found at the nucleus. In case of an allowed, or super-allowed, decay like tritium where no orbital angular momentum is involved, this probability is $1/V$ for a plane neutrino wave emitted into the box normalisation volume $V$. For the electron, this density is modified by the Fermi function

$$F(E, Z+1) = \frac{2\pi\eta}{1 - \exp(-2\pi\eta)} . \tag{34}$$

It accounts for the final electromagnetic interaction of the outgoing electron with the daughter nucleus; $\eta = \alpha(Z+1)/\beta$ is the Sommerfeld parameter, $\alpha =$ fine structure constant, $\beta = v/c$. The leptonic matrix element – reduced to its kinetic part – can now be written as

$$|M_{lep}|^2 = F(E, Z+1)/V^2.$$
(35)

Its spin structure and coupling to the nuclear spin, as well as its (β,ν)-angular correlation, is usually contracted into the nuclear matrix element.

For an allowed or super-allowed transition, the nuclear matrix element $M_{nuc}$ is independent of the electron energy. Generally, this matrix element can be divided into a vector current, or Fermi part, and into an axial current, or Gamow-Teller part. In the former case, electron- and neutrino spin couple to $S = 0$, in the latter to $S = 1$. Summing over spin states and averaging over the (β, ν)-angular correlation factor

$$1 + a(\boldsymbol{\beta} \cdot \boldsymbol{\beta}_\nu)$$
(36)

(with the electron velocity vector $\boldsymbol{\beta} = \boldsymbol{v}/c$ and the neutrino velocity vector $\boldsymbol{\beta}_\nu = \boldsymbol{v}_\nu/c$) [87], the hadronic matrix element for tritium decay is [47]

$$|M_{nuc}|^2 = 5.55.$$
(37)

Next, we address the continuum of the outgoing β and ν which forms the β-spectrum. The number of states d$n$ in a phase space element d$\Phi$ into which the β is emitted (suppressing a factor $\hbar^{-3}$) is

$$\mathrm{d}n = \mathrm{d}\Phi = \frac{V \cdot p^2 \mathrm{d}p\, \mathrm{d}\Omega}{(2\pi)^3} = \frac{V \cdot p E_{tot} \mathrm{d}E_{tot}\, \mathrm{d}\Omega}{(2\pi)^3} = \frac{V \cdot E_{tot} \sqrt{E_{tot}^2 - m^2}\, \mathrm{d}E_{tot}\, \mathrm{d}\Omega}{(2\pi)^3}.$$
(38)

where $V$ is again the box normalisation volume. On the right, we have transformed momentum into energy space by help of (18). Then, the two particle phase space density of β *and* ν is given by the product

$$\rho_2(E_{tot}, E_{tot\,\nu}) = \frac{\mathrm{d}n}{\mathrm{d}E_{tot}\, \mathrm{d}\Omega} \frac{\mathrm{d}n_\nu}{\mathrm{d}E_{tot\,\nu}\, \mathrm{d}\Omega_\nu} = \frac{V^2 \cdot E_{tot} \sqrt{E_{tot}^2 - m^2} \cdot E_{tot\,\nu} \sqrt{E_{tot\,\nu}^2 - m_\nu^2}}{(2\pi)^6}$$
(39)

where the indexed variables stand for the $\nu$. The recoiling daughter has no phase space of its own, since its momentum is fixed to $-(\boldsymbol{p} + \boldsymbol{p}_\nu)$ which yields the recoil energy in the limits

$$0 \leq E_{\text{rec}} = (\boldsymbol{p} + \boldsymbol{p}_\nu)^2 / 2m_{\text{daught}} \leq E_{\text{rec max}} = p_{\text{max}}^2 / 2m_{\text{daught}} = (E_0^2 + 2E_0 m)/(2m_{\text{daught}}). \quad (40)$$

At the β-endpoint $E_{\text{rec}}$ reaches its maximum value given on the right of (40) (neglecting $m_\nu$); for a tritium ion this amounts to 3.44 eV, for the centre of mass motion of the daughter molecule ($^3$HeT)$^+$ to 1.72 eV.

The phase space density (39) is distributed over a surface in the two-particle phase space which is defined by a $\delta$-function conserving the decay energy. With this prescription, we can integrate (32) over the continuum states and get the partial decay rate into a single channel; for instance, the ground state of the daughter system with probability $P_0$:

$$\Gamma_0 = P_0 \iint_{E_t, E_{t\nu}, \Omega, \Omega_\nu} \left\{ \frac{G_F^2 \cos^2 \Theta_C |M_{\text{nuc}}|^2}{32\pi^5} F E_{\text{tot}} \sqrt{E_{\text{tot}}^2 - m^2} E_{\text{tot}\nu} \sqrt{E_{\text{tot}\nu}^2 - m_\nu^2} \right.$$
$$\left. (1 + a(\boldsymbol{\beta} \cdot \boldsymbol{\beta}_\nu)) \delta(Q + m - E_{\text{tot}} - E_{\text{rec}} - E_{\text{tot}\nu}) \right\} dE_{\text{tot}} dE_{\text{tot}\nu} d\Omega d\Omega_\nu. \quad (41)$$

A correct integration over the unobserved neutrino variables in (41) has to respect the (β,ν)-angular correlation factor (36) which enters the recoil energy $E_{\text{rec}}$ (40). The variation of $E_{\text{rec}}$ near the endpoint is tiny [88]. Even for the most sensitive tritium β-decay experiment, the up-coming KATRIN experiment, the variation of $E_{\text{rec}}$ over the energy interval of investigation (the last 25eV below the endpoint) can be neglected and replaced by a constant value $E_{\text{rec}} = 1.72$ eV yielding a fixed endpoint $E_0 = Q - E_{\text{rec}}$ [89]. We can, then, integrate over $E_{\text{tot}\nu}$ simply by fixing it through the $\delta$-function to the missing energy $E_{\text{tot}\nu} = (E_0 - E)$: , the difference between endpoint energy $E_0$ and kinetic energy of the β. Further integration over the angles yields through (36) an averaged nuclear matrix element, as mentioned above. Besides integrating over the (β, ν)-continuum, we have to sum over all other final states. It is a double sum, one over the 3 neutrino mass eigenstates $m_i$ with probabilities $|U_{ei}|^2$, the other over all of the electronic final states of the daughter system with probabilities $P_j$ and excitation energies $V_j$. The latter give rise to shifted endpoint energies $(E_0 - V_j)$. Introducing the definition

$$\varepsilon := (E_0 - E) \tag{42}$$

the total neutrino energy now amounts to $E_{\text{tot }\nu} = \varepsilon - V_j$. Rather than in the total decay rate, we are interested in its energy spectrum $\gamma = d\Gamma/dE$, which we can read directly from (41) without performing the second integration over the β-energy. Written in terms of $\varepsilon$ and summed up over the final states it reads

$$\gamma(\varepsilon) = \frac{G_F \cos^2 \Theta_C}{2\pi^3} |M_{\text{nuc}}|^2 F \cdot (E_0 + mc^2 - \varepsilon) \sqrt{(E_0 + mc^2 - \varepsilon)^2 - m^2} \times \\ \times \sum_{i,j} |U_{ei}|^2 P_j \cdot (\varepsilon - V_j) \sqrt{(\varepsilon - V_j)^2 - m_i^2} \cdot \Theta(\varepsilon - V_j - m_i^2). \tag{43}$$

The $\Theta$-function confines the spectral components to the physical sector $\varepsilon - V_j - m_i > 0$. This causes a technical difficulty in fitting mass values smaller than the sensitivity limit of the data, as statistical fluctuations of the measured spectrum might occur which cannot longer be fitted within the allowed physical parameter space. Therefore, one has to define a reasonable mathematical continuation of the spectrum into the region $m_i^2 < 0$ which leads to parabolic $\chi^2$-parabolas around $m_i^2 = 0$ (see e. g. [90]). But one may equally well use formulas describing a physical model with the signature of a spectrum stretching beyond $E_0$ like tachyonic neutrinos [91] (with the caution, of course, that one should not to jump to spectacular conclusions from significant fit values $m_i^2 < 0$ instead of carefully searching for systematic errors in the data).

Furthermore, one may apply radiative corrections to the spectrum [92,93]. However, they are quite small and would influence the result on $m^2(v_e)$ only by few percent of its present systematic uncertainty. One may also raise the point of whether possible contributions from right handed currents might not lead to measurable spectral anomalies [94,95]. It has been checked that the present limits on the corresponding right handed boson mass [5,85] rule out a sizeable contribution within present experimental uncertainties. Even the forthcoming KATRIN experiment will hardly be sensitive to this problem [96].

*3.4. Final state spectrum*

Neutrino masses enter the β-spectrum by their square in the second square root term of (43) which represents the neutrino momentum. There the mass competes with the total neutrino energy $E_{\text{tot }\nu} = \varepsilon - V_j$. Hence, a sensitive and correct mass measurement requires not only, restriction to small $E_{\text{tot }\nu}$, but also precise knowledge of the final state spectrum $(P_j, V_j)$. In the case of a free atom, this concerns the final excitation of its electronic shell by the decay; in the case of a gaseous molecular source, the nuclear recoil excites, in addition, a dense spectrum of vibration and rotation states of the daughter molecule. For a condensed source even electronic excitation of next neighbours is not a rare process, but has to be considered. At first sight, the dynamics of these final state interactions seem by far too complicated to be amenable to precise calculation. Fortunately, this is not the case for simple systems: as the β-particle is very fast, it rarely interacts directly with the shell of its mother atom or molecule, represented by a wave function $|\Psi_M\rangle$ and energy $E_M$, but suddenly leaves behind a nucleus with different charge and different eigenstates and eigenvalues of its electron shell $(|\Psi_{Dj}^+\rangle, E_{Dj}^+)$ which have to emerge as final states of the daughter ion. Theory solves this problem by the sudden approximation [97] which yields an expansion of the original state in terms of the $|\Psi_{Dj}^+\rangle$ with coefficients $P_j$ and excitation energies $V_j$ (with respect to the mother)

$$P_j = \left|\langle \Psi_{Dj}^+ | \Psi_M \rangle\right|^2 \quad V_j = E_{Dj}^+ - E_M \ . \tag{44}$$

For the simplest case of a tritium atom decaying into $^3\text{He}^+$ the evaluation of (44) is obviously a straightforward task. Quite precise results are still obtained for light atoms and molecules by numerical means, in particular for $T_2$ and its daughter $(^3\text{He T})^+$ [98]. However, these results were questioned when early data from Mainz [88] confirmed the large negative $m^2(\nu_e)$ values of the Los Alamos [40] and Livermore [35] experiments when being analysed over the same spectral region, namely the last 500eV below the endpoint $E_0$. Furthermore, a new feature was observed. The large negative values of $m^2(\nu_e)$ disappeared when analysing much shorter intervals below $E_0$. This effect, which could only be investigated by the high luminosity and resolution of the MAC-E filters, pointed towards an underestimated energy loss process in the source, or a missing component in the final state spectrum at $V_{extra} \approx 70$ eV, $P_{extra} \approx 4\%$, seemingly present in all experiments. The only common feature of the various experiments seemed to be their analysis with the same theoretical final state spectrum. Consequently, the different theory groups started to check these calculations in more detail. The expansion was calculated to one order further and new interesting insight into this problem was obtained, but no significant changes were found (see [99]) and references therein): Using the improved spectrum, instead of the earlier one, would shift $m^2(\nu_e)$ by <1 eV$^2$ in the analysis of typical data sets from Mainz or Troitzk. This is still within the limits of the present uncertainties. As to the missing component, however, Mainz found, later, a non- trivial experimental explanation through an unexpected effect, namely de-wetting of the T$_2$-film from the substrate [100,101,22] (see section 4.4.).

The calculation by Saenz et al. yields for the expansion (44) a unitarity check of $\sum P_j = 99.83\%$ giving high confidence in the result. Their spectrum of final states is used in recent analyses of T$_2$-decay and is shown in figure 6. The first, restricted group, concerns rotation and vibration excitation of ($^3$HeT)$^+$ in the electronic ground state; this spectrum comprises a fraction of $P_g$=57.41% of the total decay rate. It stretches to somewhat more than 4 eV; that is into the unbound $^3$He + T$^+$ continuum which starts at 1,9 eV. Its mean excitation energy is about 1.72 eV for a ß-energy close to the endpoint. The same amount of recoil energy is given to the centre of mass motion of the daughter system and is already considered here in the value $E_0 = Q - 1.72\,\text{eV}$. In solid T$_2$, the recoil may additionally excite some phonons. But in sudden approximation, which is quite valid here, the mean overall recoil energy will, even then, remain at the energy a nucleus would receive in classical mechanics from a sudden impact in the moment of its decay, irrespective of its being bound to a neighbour. Hence the sum of the mean excitation energy of the ground state (1.72 eV in the case of T$_2$, see figure 6) and the recoil energy to the molecule ($E_{rec}(T_2) = 1.72\,\text{eV}$ in the case of T$_2$) equals the recoil energy of the single tritium atom ($E_{rec}(T) = 3.44\,\text{eV}$); this also holds for the other hydrogen molecules containing tritium, HT and DT [84]. Thus, the effect on $m^2(\nu_e)$ due to contamination of a T$_2$ source by HT and DT is rather small [100],

After this first, so-called, elastic group, we observe an important gap in the spectrum up to the first excited electronic state of ($^3$HeT)$^+$ at 24 eV. This gap could, in principle, be filled by a ($^3$He,T$^+$)-continuum, which starts at the dissociation energy of 1.9 eV. But dissociation at the cost of ß-energy is strongly disfavoured in the Born-Oppenheimer approximation. The spectrum of electronic excitations of ($^3$HeT)$^+$ sums up to $P_{ex} \approx 27\%$; it consists of broad resonances decaying into the continua ($^3$He,T$^+$) and ($^3$He$^+$,T) according to their excitation energy, respectively. Flat continuum states ($^3$He$^+$,T$^+$,e), ($^3$He$^{++}$,T$^+$, 2e) are populated with about 14% probability, the rest falls to Rydberg states. Quite recently Doss and Tennyson [102] have, again, addressed the problem of the excited ($^3$HeT)$^+$ states with *R*-Matrix theory; they have confirmed the results of Saenz et al. in detail. Hence, the final state spectrum of the β-decay of gaseous T$_2$ is not a matter of concern in evaluating the neutrino mass.

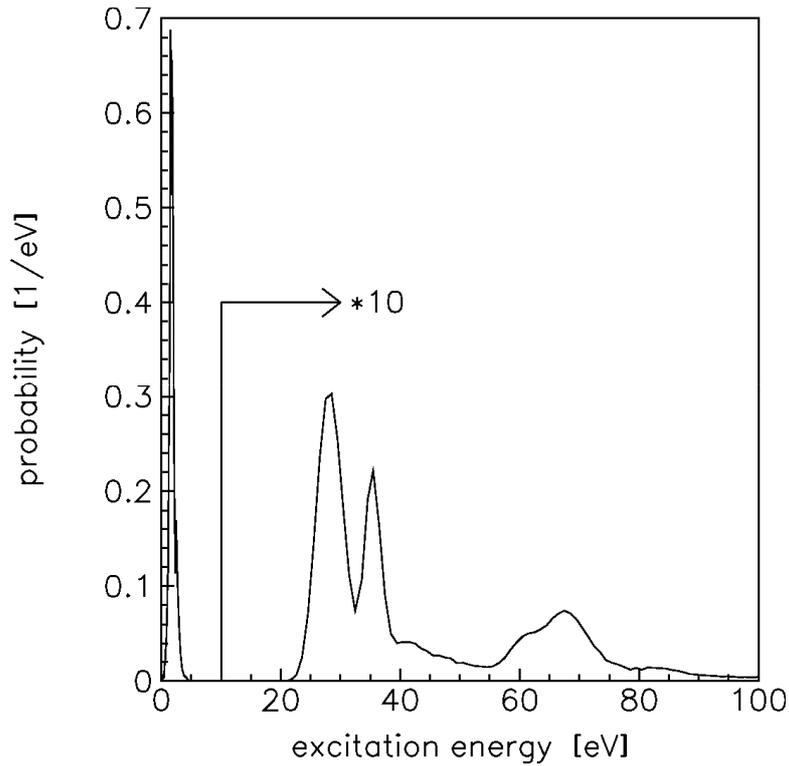

**Figure 6**: Excitation spectrum of the daughter $(^3\text{HeT})^+$ in β-decay of molecular tritium (reprinted from ref [22]). The first peak below 5eV comprises the rotation and vibration excitation of $(^3\text{HeT})^+$ in the electronic ground state. The excitations above 24eV are electronic excitations of the $(^3\text{HeT})^+$ ions combined with ro-vibrational excitations of the molecule. Above the ionisation limit of 45eV the continuum starts overlapped with some autoionizing states (the electronic excitation energy is high enough to ionise the molecular ion ones more, but the excitation energy is shared between 2 electrons, thus avoiding the direct ionisation. But, finally, the electronic shell will rearrange by emitting an electron.).

In solid $T_2$ the sudden appearance of an additional nuclear charge may also excite neighbouring molecules by the need to find a new configuration of local equilibrium. Kolos et al. [104] have calculated the chance of this spectator excitation to be approximately 5.9%, which is taken into account (with some modification) in the analysis of the Mainz experiment [22, 105]. Moreover, the endpoint shifts by +0.88 eV due to the polarisation of the lattice by the charged daughter. Apart from this general shift, the final state spectrum of $(^3\text{HeT})^+$ is said not to be changed significantly in the solid phase [101], although the ultimate accuracy, which has been obtained meanwhile in calculations for $T_2$ in the gaseous phase, cannot be claimed for $T_2$ in the solid phase.

*3.5 $m^2(\nu_e)$-sensitivity of the ß-spectrum as measured by an electrostatic filter*

Figure 7 shows the last 40 eV of the $T_2$-ß-spectra for vanishing neutrino mass graphically in the form of dotted and the dashed lines. The dotted line includes transitions to electronic excitations of the daughter; the dashed line shows only the ground state fraction with $P_g = 57.41\%$. The solid line shows the effect of degenerate neutrino masses $m_i = m(\nu_e) = 10$ eV on the ground state fraction. For this arbitrary choice, the missing decay rate in the last 10 eV would amount only to $2 \times 10^{-10}$ of the total decay rate, scaling as $m^3(\nu_e)$. For the actual limit of 2 eV this fraction shrinks, therefore, by two more decades, and by another three decades for the sensitivity limit of 0.2 eV aspired to by KATRIN [39].

We learn from these numbers that the minute useful high energy end of the spectrum carries enormous ballast at lower energies. However, this can be rejected safely by an electrostatic filter which can be passed only by electrons with kinetic energy $E$ larger than a potential barrier $qU$ to be climbed. Any momentum analysing, e.g. magnetic spectrometer, cannot guarantee this strict rejection as scattering events may introduce tails to both sides of the resolution function. Energy sensitive detectors of semiconductor type or bolometers, on the other hand, suffer from pile-up events at the necessary total decay rates of an order of $10^9$/s.

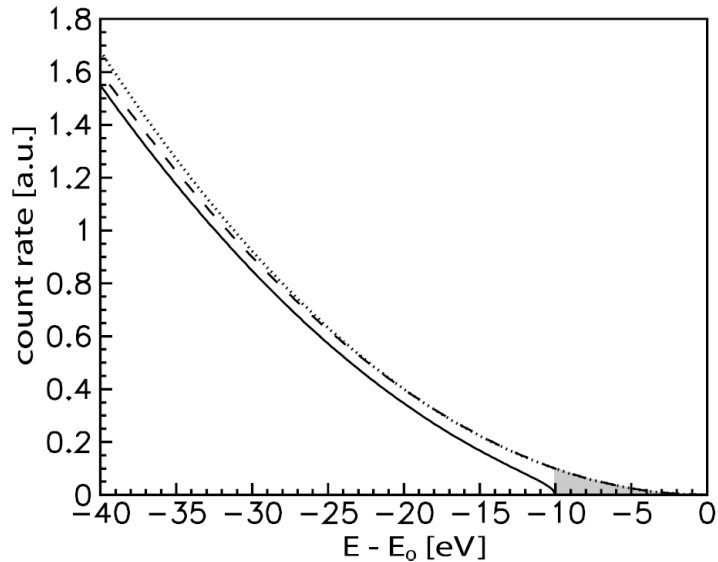

**Figure 7:** $T_2$- β-spectrum close to the endpoint $E_0$. The dotted and the dashed lines correspond to $m(\nu_e) = 0$, the solid line to $m(\nu_e) = 10$ eV. In the case of the dashed and the solid line only the decay into the electronic ground state of the daughter is considered. For $m(\nu_e) = 10$ eV, the missing decay rate in the last 10 eV below $E_0$ (shaded region) is a fraction of $2 \cdot 10^{-10}$ of the total decay rate, scaling as $m^3(\nu_e)$ (reprinted from ref [22]).

Actually, the relevant signature of $m^2(\nu_e)$ extends further below the shaded triangle of the missing rate in figure 7, namely into the region where it causes an asymptotically constant offset (see figure 7 and below). Let us apply a *sharp* filter in this region which integrates the spectrum for energies $E = E_0 - \varepsilon > qU$. For small missing energy $\varepsilon$, we may treat, for the purpose of the following discussion, all factors in front of the sum in (43) as a constant $C_\gamma$. Moreover, we take into consideration within this short interval, for the sake of clarity, only the dominant decay mode into the electronic ground state (figure 7). The narrow recoil spectrum is replaced by its mean. Then, from a source of $N_{\mathrm{nuc}}$ nuclei, observed with a solid angle $\Delta\Omega$ the integral count rate $s$ and a background rate $e$ is obtained. These should be independent of the filter setting

$$r(\varepsilon) = N_{\mathrm{nuc}} \frac{\Delta\Omega}{4\pi} C_\gamma \int_0^\varepsilon \gamma(\varepsilon')\mathrm{d}\varepsilon' + b = A \sum_i |U_{ei}|^2 (\varepsilon^2 - m_i^2)^{3/2} + b = s + b \qquad (45)$$

where the energy independent factors on the left have been contracted on the right to the characteristic amplitude $A$. Under practical conditions, the signal rate $s$ integrated over the measurement time $t$ separates from the background noise $\sqrt{bt}$ only at missing energies $\varepsilon \gg m_s$ where $m_s$ is the sensitivity limit on the mass. There we may develop (45) to first order

$$r(\varepsilon) = A\left(\varepsilon^3 - \frac{3}{2}\varepsilon \sum_i |U_{ei}|^2 m_i^2\right) + b . \qquad (46)$$

Next to the leading $\varepsilon^3$-term this approximate integral spectrum displays a mass dependent signal as the product of the integration interval $\varepsilon$ and the weighted squared mass $m^2(\nu_e) = \sum_i |U_{ei}|^2 m_i^2$ ; this is the signature of the neutrino mass in realistic β-decay measurements as already anticipated in (24).

The statistical noise on the number of counts $(s+b)t$ after a measuring time $t$ will be dominated near $E_0$ by the background and, further below, by the cubic term. The noise of the latter rises like $\varepsilon^{3/2}$ and hence, faster than the mass dependent signal. In-between there must be point $\varepsilon_{\mathrm{opt}}$ with optimal sensitivity on $m^2(\nu_e)$; it is found at

$$s(\varepsilon_{\mathrm{opt}}) = 2b . \qquad (47)$$

Aided by (46), one can calculate [106] that measuring at that setting for a time $t$ would yield a statistical uncertainty

$$\delta m^2(\nu_e) = \frac{2^{2/3}}{\sqrt{3}} A^{-2/3} b^{1/6} / \sqrt{t} \ . \tag{48}$$

We see that for this optimal choice, the dependence on the background rate is, fortunately, much weaker than that on the characteristic amplitude $A$. For an accepted β-flux of the order $6 \cdot 10^8$/s and $b = 0.015$ cts/s which are typical numbers for the Mainz and Troitzk experiments, one finds optimal sensitivity at $\varepsilon_{opt} \approx 15$ eV and for the uncertainty (48) a value of the order $10^3 (\text{eV}^2) / \sqrt{t/s}$ which within a period of 10 days, would even drop to 1 eV². In an actual experiment, of course, one needs quite a number of measuring points within a reasonable interval, in order to check the spectral shape and also to fix the other parameters $A$, $E_0$, $b$ by a $\chi^2$-fit. The multi-parameter fit enhances the uncertainties considerably by correlations. Full-fledged simulations of the statistics of a realistic experimental procedure show that (48) underestimates the necessary measurement time by about one order of magnitude; but they confirm (47) and also the functional dependence of (48). Hence, these equations provide valuable guide lines for an experiment.

At a particular measuring point $\varepsilon$, an endpoint uncertainty $\delta E_0$ correlates to $\delta m^2(\nu_e)$ according to (46) as

$$0 = \delta r = \frac{\partial r}{\partial m^2(\nu_e)} \delta m^2(\nu_e) + \frac{\partial r}{\partial E_0} \delta E_0 \quad \Rightarrow \quad \delta m^2(\nu_e) \approx 2\varepsilon \cdot \delta E_0 . \tag{49}$$

Considering that, in the chosen approximation, $\varepsilon$ is the total neutrino energy $E_{\text{tot }\nu}$, the correlation (49) just repeats the general relation (21) and underlines, yet again, the necessity of measuring the neutrino mass close to the ß-endpoint. Figure 8 shows that a fitting of simulated experimental data reproduces (49) fairly well.

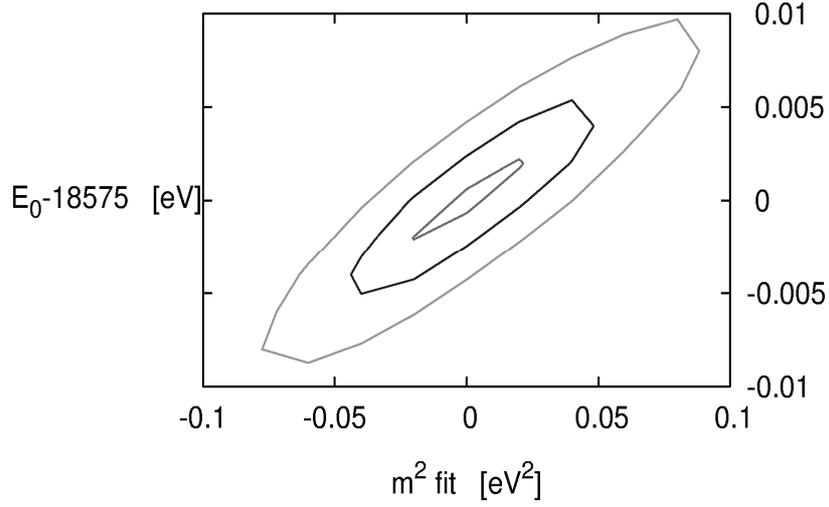

**Figure 8:** $\chi^2$ contour plot to illustrate the correlation between the fitted endpoint $E_0$ and the fitted neutrino mass squared $m^2(\nu_e)$ from Monte Carlo simulations with conditions similar to the KATRIN experiment [39] for a measurement interval of 25eV below the endpoint. Here the mass sensitivity peaks around $\varepsilon_{opt} \approx 5eV$ (47). The ellipses correspond to $1\sigma$, $2\sigma$ and $3\sigma$ contours. The $2\sigma$ uncertainty of $m^2(\nu_e)$ amounts to about $\pm 0.05$ eV$^2$ and corresponds to an endpoint uncertainty of $\Delta E_0 = \pm 5$meV reproducing fairly well (49) for $\varepsilon = \varepsilon_{opt}$ (reprinted from [106].

Instead of fitting $E_0$ together with the other parameters from the data, one could consider using the known mass difference of $\Delta M(T, ^3He) = 18589.8(1,2)$ eV [30]. For gaseous $T_2$, one calculates from this result an endpoint energy of $E_0 = (18571.8 \pm 1.2)$ eV (31), which is in good agreement to the measurement from a solid $T_2$-film yielding $E_0 = (18572.6 \pm 3.0)$ eV [88] (including corrections for polarisation shift in the $T_2$ film [101] and the electrical potentials of the Mainz experiment).
The combined uncertainty, however, would cause in the most sensitive region of present experiments, i.e. around $\varepsilon_{opt} = 15$ eV, through (49) an uncertainty in $m^2(\nu_e)$ of about 100eV$^2$. This is far beyond present values obtained from inclusive fits. Regarding voltage measurement, the latter is sensitive only to the easily measured small *voltage differences* in the scan, rather than to the *absolute energy scale*.

In the meantime, mass measurements with Penning traps have passed the 10$^{-10}$ level [107], and proposals to reach even the 10$^{-11}$ level of relative mass accuracy are under discussion [108]. Whether absolute measurements of the analysing potential will be able to cope with the corresponding 30 meV precision in the future is questionable; because this problem not only concerns the high voltage control, but more seriously, even the control of source charging by the escaping β's and of the work function of the electrode materials on which the vacuum potential depends. On the other hand, an improved external $\Delta M(T, ^3He)$-value will be of great help in uncovering, understanding and solving experimental problems; such procedures form a constituent part of the culture of precision experimentation [109].

Given the fact that $E_0$ came to be fixed from a multi-parameter fit of the spectrum, we look again at the correlation (49): Concerning the determination of $E_0$ somewhat larger $\varepsilon$-values should be included, because the uncertainty $\delta E_0$ will de-correlate from $\delta m^2(\nu_e)$ as $1/2\varepsilon$. To sum up, there are, in principle, three spectral regions from which the basic parameters $b$, $m^2(\nu_e)$, $E_0$, $A$ are fitted most sensitively and with a minimum of cross-talk:

 (i) A region beyond $E_0$ fixing $b$,
 (ii) A region around $\varepsilon_{opt}$ fixing $m^2(\nu_e)$,
 (iii) A region at somewhat larger $\varepsilon$ fixing $E_0$, $A$.

In the region (iii), however, the inelastic components of the spectrum and their uncertainties start to matter, which come finally to dominate the systematic error. This concerns predominantly the external energy loss which the β-particles suffer in crossing source thicknesses of order $10^{17}$ atoms/cm$^2$ (the internal excitations were said to be quite precisely calculable by sudden approximation). In this situation, we expect there to be an optimal length of the measuring interval at which a proper balance between the systematic and statistical uncertainty of the result is reached. In view of the non-trivial energy loss spectrum, this optimum has to be found by numerical simulations. Still, we can get some analytical insight into the correlation between the uncertainties of endpoint and energy loss by means of a rough model which sums up this spectrum into a single component at an average energy loss $\overline{E}_{loss}$ with a relative amplitude $\overline{a}$. In the region under discussion $\varepsilon > \overline{E}_{loss}$ we may neglect background and mass terms and write (46) as

$$r(\varepsilon) = A\left(\overline{a}(\varepsilon - \overline{E}_{loss})^3 + (1-\overline{a})\varepsilon^3\right). \tag{50}$$

In first approximation $\overline{a} \ll 1$ we then get the following correlation of uncertainties:

$$\delta E_0 \approx \frac{1}{3}\varepsilon\left(1 - \left(\frac{\varepsilon - \overline{E}_{loss}}{\varepsilon}\right)^3\right)\delta\overline{a} + \overline{a}\left(\frac{\varepsilon - \overline{E}_{loss}}{\varepsilon}\right)^2 \delta\overline{E}_{loss}. \tag{51}$$

We see that under (over)-estimation of the mean energy loss $\overline{E}_{loss}$ and/or its amplitude $\overline{a}$ leads to a false lowering (lifting) of the endpoint $E_0$ which transfers through (49) further to $m^2(\nu_e)$ with the same sign. This mechanism was at the bottom of the unreasonable high $m^2(\nu_e)$-value claimed by Ljubimov et al [25], as well as of the tendency towards unphysical, negative $m^2(\nu_e)$-values as a function of $\varepsilon$ which was observed in the early Mainz result [88]. For given uncertainties $\delta\overline{a}, \delta\overline{E}_{loss}$ the only possibility of reducing their influence on $m^2(\nu_e)$ lies with an improvement of the signal to background ratio allowing a shorter $\varepsilon$-interval.

Finally, we point to another consequence of the quadratic relations (21), (49) relevant for experiments with electrostatic filters. It concerns a spreading $\delta E$ of the analysing energy around its average $\langle E \rangle$, resulting from its instrumental resolution, a ripple on its voltage, etc. The spreading causes a *negative* shift of $m^2(\nu_e)$ with respect to the value which is calculated from the average $\langle E \rangle$ [47]:

$$\langle \delta m^2 \rangle = \langle 2(E_0 - E)\delta E \rangle = \langle 2(E_0 - (\langle E \rangle + \delta E))\delta E \rangle = -2\langle \delta E^2 \rangle = -2\sigma_E^2 \qquad (52)$$

Hence energy spreading has to be carefully controlled in order to avoid a systematic error from this source.

## 4. Search for neutrino mass by β-spectroscopy with MAC-E-Filters

Section 3 has made clear that sensitive and uncontaminated search for the neutrino mass in β-spectra requires measuring close to a low endpoint with high luminosity as well as with high resolution. Half a century ago, Hamilton et al. already recognised that an electrostatic filter is well suited to fulfil these particular requirements [110]. We start this section by recalling the principal ideas underpinning their method, since it has formal analogy to MAC-E-Filters: Let us consider a β-emitter placed onto a small half-spherical substrate of radius $r$, concentrically surrounded by a half-spherical grid with much larger radius $R$ which provided the analysing or filter voltage $U$ (figure 9). The transmitted current is measured by an analogue circuit as a function of $U$. The concentric arrangement allows for a wide solid angle, whereas the large ratio $R/r$ guarantees high resolution according to the following elementary calculation: The spherical symmetry of the electric field conserves the angular momentum of the β-particle which takes the value $L(\vartheta) = rp \sin \vartheta$ for starting angle $\vartheta$ with respect to $r$. When it passes the grid $G_2$, its corresponding transverse energy has been reduced from its starting value, therefore, by the square of the radius ratio:

$$E_\perp(\vartheta, R) = \frac{L^2}{2mR^2} = \frac{r^2 p^2 \sin^2 \vartheta}{2mR^2} = \left(\frac{r}{R}\right)^2 \sin^2 \vartheta \cdot E. \qquad (53)$$

The relative width of the filter is then

$$\Delta E / E = (r/R)^2. \qquad (54)$$

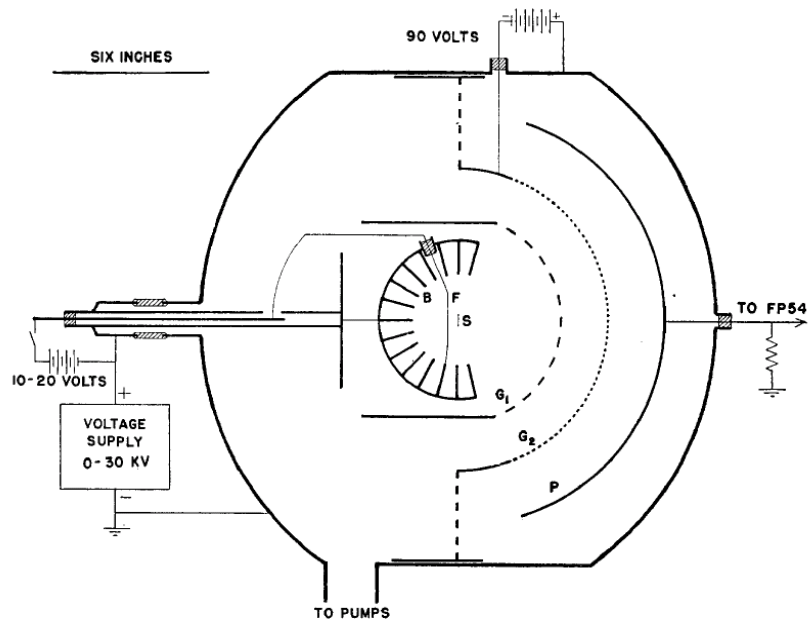

**Figure 9:** Hamilton's spherical electrostatic β-spectrograph showing collector *P*, grids $G_1$ and $G_2$, source *S*, decharging filament *F* and electron backstop *B* (reprinted from ref [107] with kind permission). For practical reasons the source is a small disk instead of a sphere and the intermediate grid $G_1$ is at source potential; still (54) remains valid.

Within that width the transmission drops from 1 at $qU = E - \Delta E$ to 0 at $qU = E$, according to the residual transverse energy (53) as function of $\vartheta$. Realised on a modest scale, the experiment still reached a resolution of 0.7% and yielded an upper limit on the neutrino mass m($\nu_e$) of 500 eV [112] which was competitive at the time.

*4.1. Magnetic adiabatic collimation applied to an electrostatic filter (MAC-E-Filter)*
The salient point of Hamilton's simple concept was the following: Collimation of the particles along the repelling electrical field lines was achieved without any solid angle limiting means of electron optics, but just by letting them expand to a larger (spherical) cross section at which their momentum space is squeezed accordingly. A similar overall collimation effect for an electrostatic filter is reached, in a more flexible and more powerful manner, by letting β-particles adiabatically expand into a decreasing magnetic field *B* before they reach the full analysing potential. The magnetic field *B* has to be strong enough at any point in order to confine the particle track to a spiral motion at the cyclotron frequency $\omega_c$ around a particular *B*-line, called the guiding centre. This new type of spectrometer is based on early work by Kruit [113] and was later re-developed for the purpose of tritium beta spectroscopy independently at Troitzk and Mainz [36, 37, 88,114]. The main features of a MAC-E-Filter are illustrated in figure.10
:

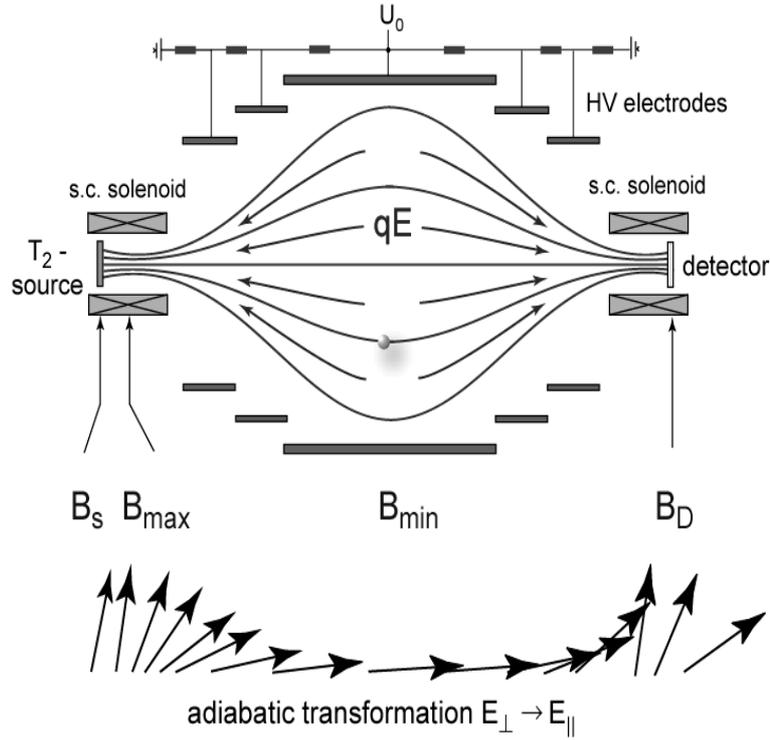

**Figure 10:** Principle of the MAC-E-Filter: Two superconducting solenoids produce a magnetic guiding field. The tritium source is placed in the left solenoid, whereas the detector sits in the right solenoid. Between the 2 solenoids there is an electrostatic retardation system consisting of several cylindrical electrodes. In the so-called analysing plane, the minimum of the magnetic field $B_{min}$ coincides with the maximum of the electric retarding potential $U_0$. The vectors in the lower part of the picture illustrate the aligning of the momentum vector along $B$ in the region of low magnetic field by the adiabatic transformation (58) (plotted without electrostatic retardation).

Two superconducting solenoids provide the magnetic guiding field. β-particles, which are emitted from the tritium source in the strong field $B_S$ inside the left solenoid into the forward hemisphere with a starting angle $\vartheta_S$ with respect to $B$, are guided magnetically into the spectrometer along the $B$-line at which they are born. Those with sufficient energy pass the analysing potential inside the spectrometer and are magnetically refocused onto the detector in the right solenoid. In principle, a solid angle of nearly $2\pi$ can be transported and, moreover, it can be analysed sharply. The latter statement follows from phase space arguments which act here as follows: On their way into the spectrometer the $B$-field drops down by several orders of magnitude to a saddle point minimum $B_{min} = B_a$, the field in the analysing centre. Therefore, the magnetic gradient force transforms most of the primary cyclotron energy $E_{\perp S}$ at the source into longitudinal motion along $B$. This is illustrated at the bottom of figure 10 by a momentum vector. In adiabatic approximation, which holds under the condition

$$\left| \frac{1}{B} \frac{d\mathbf{B}}{dt} \right| \ll \omega_c = \frac{eB}{\gamma m} \tag{55}$$

(with the cyclotron frequency $\omega_c$), the amount of the angular momentum and hence the product of the orbital magnetic moment

$$\mu = \frac{e \omega_c r^2}{2} \tag{56}$$

around **B** and the relativistic factor $\gamma$ is a constant of motion [115]:

$$\gamma \mu = \gamma \frac{e \omega_c r^2}{2} = \frac{p_\perp^2}{2mB} = const. \tag{57}$$

In non-relativistic approximation, which is reasonably valid in tritium β-spectroscopy, (57) simplifies to

$$\mu = \frac{E_\perp}{B} = const. \tag{58}$$

The conservation of the orbital magnetic moment $\mu = E_\perp / B$ can be understood by the conservation of the magnetic flux $\Phi$ through the particle orbit

$$\Phi = \pi \rho^2 B = \pi \left( \frac{m v_\perp}{eB} \right)^2 B \propto \frac{m v_\perp^2}{2B} = \frac{E_\perp}{B} \tag{59}$$

Equation. (59) means, that the transverse, non-analysable fraction of the particle energy is reduced on the way into the analysing centre in proportion to the magnetic field:

$$E_{\perp a} = \frac{B_a}{B_S} E_{\perp S} = \frac{B_a}{B_S} \sin^2 \vartheta_S \cdot E = \frac{R_S^2}{R_a^2} \sin^2 \vartheta_S \cdot E . \tag{60}$$

On the right of (60) we have replaced the ratio of the magnetic fields by the squared ratio of the corresponding radii of the total magnetic flux $\Phi = \pi R^2 B$ (which is conserved as well as the flux through the particle orbit) showing the identity with (53); this applies also to the relative width of the filter. In terms of the fields it reads

$$\frac{\Delta E}{E} = \frac{B_a}{B_S} . \tag{61}$$

Note that the width can be easily adjusted to the desired value by choosing the appropriate field ratio. For an isotropically emitting source, a simple analytic transmission function is derived from (60)

$$T(E,U) = 1 - \sqrt{1 - \frac{E - qU}{E} \frac{B_s}{B_a}} \qquad (62)$$

(defined in the interval $E \cdot (1 - B_a / B_s) \leq qU \leq E$; outside, the transmission is 1 or 0, respectively). Thus the transmission function (62) describes an energy high-pass filter. In practise (62) is modified somewhat by putting an additional scanning voltage on the source and by providing downstream a field maximum $B_{max} > B_S$ [22,110]. According to (60), it rejects particles by the magnetic mirror effect which have been emitted at angles beyond a certain pinch angle

$$\vartheta_{max} = \arcsin\left(\sqrt{\frac{B_s}{B_{max}}}\right) \qquad (63)$$

and which have, therefore, suffered too much energy loss in the source.

Figure 11 shows an early example of the N32-conversion lines from a frozen [83m]Kr-source [116] scanned at a resolution of $\Delta E / E = 2 \times 10^{-4}$. It shows essentially the transmission function with some rounded edges, which correspond to the Lorentzian width of the unresolved N2/3 doublet. The analysis which considered a number of corrections for chemical shifts, work functions etc. yielded the nuclear transition energy to be $E_\gamma = 32151.5(1.1)\,\text{eV}$. The uncertainty is dominated by systematics (the statistical uncertainty of the line position from the fit is much smaller). The result agrees well with the recent direct re-measurement of the γ-energy yielding $E_\gamma = 32151.5(0.5)\,\text{eV}$ [117]. The experiment on [83m]Kr-conversion lines was meant as a forerunner for the planned $T_2$-β-spectroscopy. As such it checked the feasibility of MAC-E-Filters, including the technique of utilising solid sources from volatile gases at low temperature; and also, it has proved to be able to achieve an absolute accuracy of the order of 1 eV.

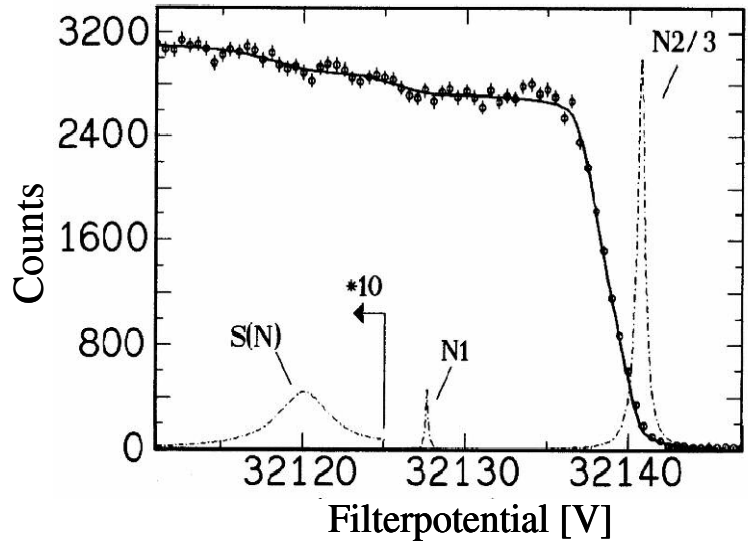

**Figure 11:** Scan of the N-conversion electron lines of $^{83m}$Kr by an integrating electrostatic filter of MAC-E type. The full line shows the convolution of a sum of Lorentzians with the transmission function, fitted to the data. The Lorentzian components found by the fit are shown by *dotted lines*. The elastic peak of the unresolved N2/3-doublet dominates. S(N) represents qualitatively shake up/shake off events (reprinted from [112]).

Before entering the discussion of T$_2$-β-spectroscopy with MAC-E-Filters, we will point out a few more basic properties of these instruments. Thanks to strong magnetic guidance (i.e. dominance of Lorentz- over electric force) the function of the filter is, fortunately, insensitive to any misalignment between magnetic and electric field vectors ***B***, ***E***. This is seen from the adiabatic equations of motion [118]. In first approximation the particle follows the encircled field line and its longitudinal energy is controlled only by the local electric potential of the filter and the size of the magnetic field;

$$E_\parallel(\mathbf{r}) = E - qU(\mathbf{r}) - \mu B(\mathbf{r}) . \tag{64}$$

they define the point of reflection at $E_\parallel(r) = 0$. Instead of exactly configured vector fields, which are required for precise momentum analysis, this energy analysing instrument only demands that $U(r)$ and, to a lesser extent, also $B(r)$ form flat saddles in the analysing centre of the spectrometer. The instruments at Mainz and Troitzk show, in the symmetry plane, radial variations of the order $\Delta U/U \, 10^{-4}$ and $\Delta B/B \lesssim 10\%$, respectively; the radial dependencies have to be convoluted into the transmission function.

In next order transverse electric acceleration is no longer averaged out by the fast cyclotron motion, but leads to a distortion of the cyclotron orbits, which integrates up to a transverse drift $u$ of the guiding centre, known as magnetron drift. Also, the curvature of ***B*** produces transverse drift terms in second order. Summing up all drift terms we get

$$\boldsymbol{u}_\perp = \left( \frac{\boldsymbol{E}}{B^2} + \frac{2E_\parallel + E_\perp}{e} \frac{\nabla_\perp B}{B^3} \right) \times \boldsymbol{B} . \qquad (65)$$

The energy contained in the transverse drift turns out to be negligible in the analysing centre of the spectrometer, since all the relevant parameters – electric field, kinetic energies, and relative magnetic field gradient – attain minimal values there. Nonetheless, one has to watch $E_\parallel$. If the particle moves too fast through a region of low $B$-field, either because it has too much surplus energy above the analysing potential, or because this potential does not stretch far enough into the high $B$-field region, then the general condition of adiabatic motion (55) is violated, and the particle might even be lost from the guiding magnetic flux tube . Setting the Mainz spectrometer to a central field value $B_a = 5.7 \times 10^{-4}$ T ( used also for T$_2$-runs), one has observed that the transmission of the high energy conversion lines of $^{83m}$Kr around 32 keV started to decrease at surplus energies >500 eV above the filter setting [110]. Hence, integration intervals have to be kept sufficiently short.

The MAC-E-Filter may be converted from its usual function as a high pass filter into a narrow band filter by running it in a time of flight mode (MAC-E-TOF filter) [119]. The salient point is that the time of flight is dominated by the passage through the long central section at full analysing potential where particles are slow. This takes, for example, several μs for electrons with longitudinal energy of 10 eV,. The high energy cut-off of the transmitted band is then determined by requiring a minimum time of flight. This is controlled by pulsing the total transmission (here easily achieved by applying an offset voltage to the electron source, which is rapidly switched on and off) and by measuring the time of arrival at the detector. Both slopes have equal width given by (61). In a test experiment at the Mainz spectrometer the K32-line of $^{83m}$Kr at 17830 eV was measured in this mode with resulting FWHM of only 6 eV and time-averaged transmission of ≈25% in the peak [115].

*4.2. Set-ups at Mainz and Troitsk*
Ideas to build MAC-E-Filters for T$_2$-β-spectroscopy arose independently at Troitsk and Mainz in the 1980's, with the aim of checking the positive ITEP claim [25] with an independent method which promised higher sensitivity and resolution. A proposal from Troitsk was published by Lobashew and Spivak in 1985 [36]. The Mainz group published a technical paper [110] and first physics results with the set up on $^{83m}$Kr-spectroscopy [112] in 1992. Figures 12 and 13 show the set ups in Troitsk [37] and Mainz (the latter in the improved version of phase II [22]). Although the instruments differ significantly in concept and size, both came up finally with quite similar signal- and background rates.

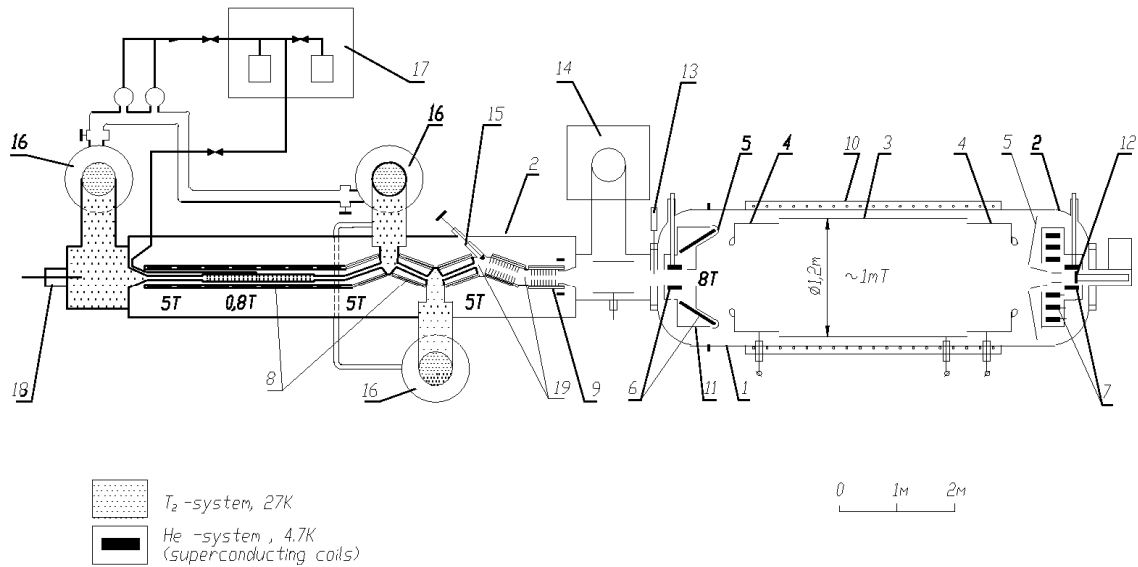

**Figure 12:** Tritium-Beta-Spectrometer at Troitsk: (l), (2) vacuum tank; (3), (4) electrostatic analyzer; (5) grounded electrode; (6), (7), (8), (9) superconducting solenoids; (10) warm coil; (11) liquid-N2 jacket; (12) detector; (13) fast shutter; (14) Ti-pump; (15) cold valve; (16) Hg diffusion pump;(17) T, purification system; (18) electron gun; (19) argon pump (reprinted from ref. [37] with kind permission).

The main difference lies with the sources. Troitsk chose the ambitious concept of a gaseous $T_2$-source which had been pioneered at Los Alamos [34]. $T_2$-gas at 27K enters the centre of a 3m long source section and is pumped out at both ends and re-circulated. The through-put is reduced by narrowing the outlets. The magnetic flux has to be contracted there accordingly by raising $B$ from 0.8 T to 5 T, thereby also trapping β's with pitch angles $\vartheta > 23.6°$ magnetically. Downstream, another differential pumping section and a cryo-trap prevent $T_2$ from entering the spectrometer. The beam tube has several bends to reduce the tritium flow into the spectrometer, whereas the β-electrons are guided by a series of superconducting solenoids. Inside the spectrometer (ø =1.5 m, $L$ =7 m) conical, superconducting coils perform an adiabatic field transition from $B_{max} \approx 8\,T$ to $B_a \approx 1\,mT$ in the analysing central volume. The analysing potential was provided by a single central electrode whose fringe field formed a sufficiently smooth slope. Ultra high vacuum is mandatory inside the spectrometer because the strong electromagnetic fields support Penning-like discharges down to very low pressure. The superconducting coils prevented the spectrometer from being baked on the one hand, but on the other hand, they served as cryo-pumps at LHe-temperature, maintaining a vacuum of $10^{-9}$ mbar. The experiment ran at a source column density of about $10^{17}$ molecules/cm$^2$, being analysed by a luminosity (= source area × $\Delta\Omega/4\pi$) of 0.27 cm$^2$ at resolving power of 5300, typically.

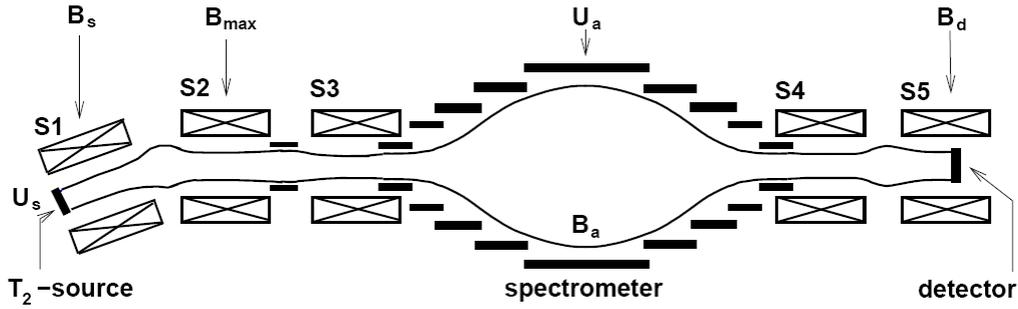

**Figure 13:** Mainz MAC-E-Filter for $T_2$-β-spectroscopy. From left to right: Frozen $T_2$ source; spectrometer with altogether 27 electrodes; PIN-diode detector; guiding solenoids S1 – S5 provide the magnetic flux tube within which β-particles are guided (reprinted from ref. [22]).

In comparison, the Mainz experiment (figure 13) is a fairly small set up, limited by laboratory dimensions to a total flight path of 6 m. The source is a film of up to 140 mono-layers $T_2$, shock-condensed onto a substrate of highly oriented pyrolytic graphite at 1.8 K. A cryo-trap in a single bend of the beam protected the spectrometer from evaporating $T_2$. The spectrometer (ø =1 m, $L$ = 4m) is shown in more detail in the lower part of figure 13. Inside, the magnetic field is shaped essentially by the fringe fields of the entrance and the exit solenoids. The central field $B_a$ was adjusted by auxiliary coils outside. The steep field slopes in front of the solenoids and the short overall length of the spectrometer rendered adiabaticity conditions somewhat difficult. It was decided, therefore, to install 2 sets of electrodes with potentials decreasing towards the solenoids in order to remove continuously kinetic energy in accordance with the transformation (64). The price paid for this precaution was an increased sensitivity for starting Penning discharges which prohibited running at the full magnetic design field, in spite of excellent vacuum conditions (<$10^{-10}$ mbar). For tritium runs the spectrometer was usually operated at a combination of high luminosity (up to 0.24 cm$^2$, reached at $\vartheta_{max} \approx 61.6°$) and reduced resolving power (≈ 4000) which together optimized the sensitivity to $m^2(v_e)$ [22].

*4.3. Background events in MAC-E-Filters*

Understanding and suppressing background at MAC-E-Filters are non-trivial problems. Under good running conditions, Mainz and Troitsk have achieved background rates down to the order of $10^{-2}$ cts/s. This sounds quite satisfactory, considering source strengths of order $10^9$ Bq. But the source contributed only indirectly to the background (see below) – if at all. Figure 13 shows a typical background spectrum from the final phase II of the Mainz experiment, as measured by a silicon diode detector segmented into 5 rings of 1 cm$^2$ area each. Whether the source was closed off mechanically by a valve or electrically by setting the analysing energy beyond the endpoint, did not make any difference in the rate.

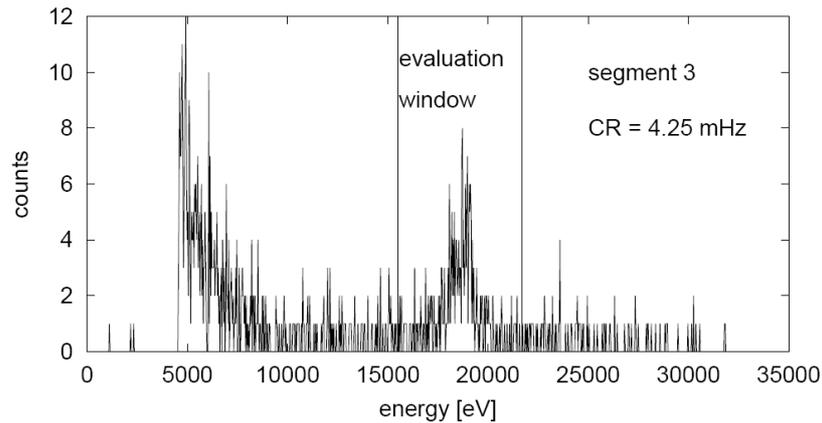

**Figure 14:** Spectrometer background spectrum, collected for 13 h on detector segment 3. The perpendicular lines frame the accepted energy window for evaluation (reprinted from ref. [22]).

The dominant peak happens to coincide precisely with the analysing potential within narrow error bars. Hence it falls right into the evaluation window of accepted events and cannot be discriminated energetically from the signal. The position of the peak reveals its origin: It has to stem from electrons which are produced at low kinetic energy within the large central section of the spectrometer at full analysing potential, and are released from there towards the detector. Consider, for instance, ionisation events of residual gas in the analysing centre. The emerging electrons have average kinetic energies around 30 eV, and the probability of the transverse fraction being smaller than the energy resolution ($E_{a\perp} < \Delta E$) is high. In this case the electrons can pass the magnetic mirror in front of the detector immediately and can be accelerated towards the detector.

The ionising particles can be, for example, recoil- or secondary ions from the source which are magnetically guided and finally accelerated into the spectrometer. They caused a background rate of about 1ct/s in the first attempt to measure a tritium spectrum at Mainz. It was easily eliminated by putting some negative voltage on the source which retained them [88]. Ions may also enter the sensitive flux tube from some local Penning discharge in the spectrometer or field emission from its electrodes. It is well known that traps can be purged of stored particles by heating them up with rf-pulses. This card was successfully played at the Mainz spectrometer for the purpose of suppressing this kind of background source – yet in a trial and error approach, lacking knowledge about their precise position and characteristics [102]., Troitsk has observed chains of correlated of background events, obviously caused by $T_2$-molecules decaying within the spectrometer. The high energy β-particle has a good chance of being magnetically trapped in the centre and to perform successive ionisations. At low β-signal rate close to the endpoint these chains could be discriminated by their signature of excessive rate within the chain [37].

Numerous slow electrons at the full potential emerge from the surface of the large central electrodes which are hit by cosmic and local radioactivity. But they are born outside the magnetic flux tube which crosses the detector; hence they are guided adiabatically past the detector. This decisive magnetic shielding effect was investigated at Mainz with an external γ-source, as well as by coincidence, with traversing cosmic muons; a magnetic shielding factor of around $10^5$ was measured [120]. Furthermore similar checks at Troitsk pointed to 10 times better shielding [121], which probably results from the better adiabaticity conditions of this larger instrument. In case the axial symmetry of the electromagnetic field configuration is broken (e. g. by stray fields) the transverse drift (65) will have a radial component, and it will be all the faster the weaker the guiding field. This drift can transport slow electrons from the surface into the inner, sensitive flux tube within which they are accelerated onto the detector. The effect is probably present at Mainz [122]. After finishing tritium measurements in 2001, Mainz developed electrostatic solutions which strengthened shielding of surface electrons by an additional factor ≈ 10. This was achieved by covering the electrodes with negatively biased grids built from thin wires (Mainz/Münster collaboration [123]). Such grids are now under construction at Münster for the KATRIN-spectrometer also. This measure (in addition to improved adiabaticity) will contribute decisively to keeping the background rate from this much larger instrument down to the design level of $10^{-2}$ cts/s [39].

### 4.4. Measurements of $T_2$-spectra and discussion of results
### 4.4.1. Early results from Mainz experiment

The Mainz team took data in the period 1991 to 2001 in a rhythm of measuring and R&D phases. First results were published in 1993 from a source of 40 monolayers solid $T_2$, kept at 2.8 K during running [88]. The analysing potential was periodically scanned between 500 V below and 200V above endpoint. Figure 15 shows the results of $m^2(v_e)$ and $E_0$ as a function of the lower limit of the fit interval $E_l$ (bottom scale) or of the interval $\varepsilon_{max} = E_0 - E_l$ (top scale) which has been included in the fit. Both observables show an identical down-sloping structure as a function of $\varepsilon_{max}$ which proves their close correlation according to (49) with a ratio $\delta m^2(v_e)/\delta E_0 \approx 60 \, \mathrm{eV}$. The corresponding $\varepsilon$-value of 30 eV agrees pretty well with $\varepsilon_{opt} \approx 33 \, \mathrm{eV}$, read from the measured spectrum, as the point at which the signal is twice the background (see (47)) and where the sensitivity on $m^2(v_e)$ peaks. We see that with increasing $\varepsilon_{max}$ a systematic error in $E_0$ develops which drives $m^2(v_e)$ into the unphysical negative sector. As shown by the approximate, analytic correlation (51), this signature points to underestimation of inelastic events. Simulations show that it would be removed by assuming an extra energy loss component at ≈ 75 eV with amplitude ≈ 4%, both parameters being strongly correlated [88,103]. Inserting this numerical result tentatively into (51) yields for the widest evaluation interval down to $\varepsilon_{max} = 500 \, \mathrm{eV}$ a downshift of the endpoint by $-2.5 \, \mathrm{eV}$, which is about the number one reads from the plot in figure 15. The corresponding unphysical $m^2(v_e)$-value of about $-120\,\text{eV}^2/\text{c}^4$ happens to match within uncertainty limits the Los Alamos result of $(-147 \pm 68 \pm 41)\,\text{eV}^2$ obtained from the same spectral interval [40]. This coincidence aroused suspicion about the theoretical final state spectrum in T$_2$-decay which has in the meanwhile been fully cleaned out (see section 3.2). Hence, all the experiments with negative $m^2(v_e)$-values shown in Figure 1 apparently seemed to carry some hidden external source of energy loss.

In view of this situation, the Mainz group decided, first of all, to discard the contaminated results at large $\varepsilon$ and to restrict the interval of accepted data to $\varepsilon \leq 137$ eV (see figure 15) where the observables $m^2(v_e)$ and $E_l$ do not depend on the fit interval $\varepsilon_{max}$ and the uncertainty of the respective result $m^2(v_e) = (-39 \pm 34_{stat} \pm 15_{syst})\,\text{eV}^2$ still touches the physical sector. Following the Bayesian approach one derives from this value an upper limit of $m(v_e) < 7.2$ eV at 95% C. L.

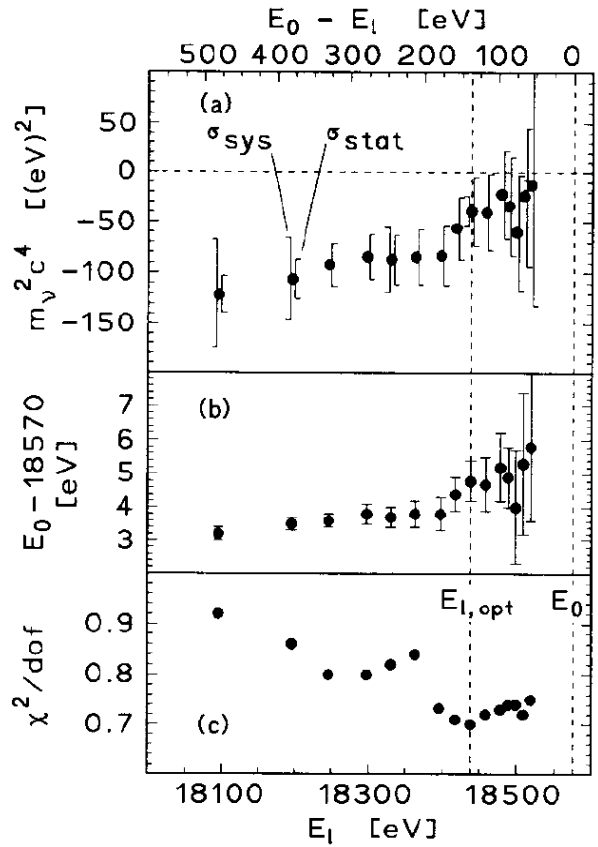

**Figure 15**: First results from T$_2$-β-spectroscopy with a MAC-E-Filter. The observables, mass squared (a) as well as endpoint (b) still suffer from a dependence from the lower limit $E_l$ of the fit interval. The left broken line indicates the optimum fit interval at which the total mass squared uncertainty is still independent of the fit interval and still includes zero. (The correction of the electric potential (-2.2eV) was not applied to the endpoint energy $E_0$ plotted here) (Reprinted from ref. [88]).

Calibration of the instruments measuring the retarding potential of the spectrometer allowed an absolute determination of the endpoint energy being $E_0 = (18574.8 \pm 0.6)\,eV - 2.2\,eV = (18572.6 \pm 0.6)\,eV$ (see figure 15) and (by using (29) and (30)) of a mass difference being $m(T) - m(^3He) = (18591 \pm 3)\,eV/c^2$ in agreement with current values [29,30,33,40]. On the other hand, the insight gained into the interplay between parameters indicated clear tasks to be tackled:

- Search for hidden sources of energy loss
- Re-measure the energy loss spectrum in molecular hydrogen in order to reduce its uncertainty
- Improve the signal rate to enable determination of $E_0$ from shorter $\varepsilon$-intervals and to reduce thereby its dangerous correlation (51) to energy loss uncertainties.
- Improve signal to background ratio in order to shift the region of maximum $m^2(\nu_e)$-sensitivity closer to $E_0$ and to reduce thereby the correlation (49) to endpoint uncertainty.

*4.4.2. Early results from Troitsk experiment*

The Troitsk group was the first to achieve really high statistics data in 1994 [37], collected with a signal rate roughly 10 times that of the early runs at Mainz [88]. The integral spectrum was measured with settings of the analysing energy from 400 eV below to 100 eV above endpoint. Data points close to the endpoint are shown in figure 16. A background of 0.015 cts/s is subtracted. The analysis showed that the data collected at $\varepsilon \gtrsim 250\,eV$ were also contaminated by some excessive energy loss which could be represented by an extra component between 100 eV and 150 eV with amplitude of several percent. Hence these data were not considered further. The analysis of shorter integration intervals down to $\varepsilon_{max} = 55\,eV$ yielded pretty stable fit values, varying by only $\pm 0.3\,eV$ for $E_0$ and $\pm 5\,eV^2$ for $m^2(\nu_e)$; however, the central value of the latter at about $-25\,eV^2$ still falls significantly into the negative sector. This fact correlates to a small step-like shoulder in the spectrum of figure 16. If one introduces step position and size as additional free parameters, the fit places it at $\varepsilon \approx 7\,eV$ with amplitude of $0.0025\,cts/s$; moreover, the negative $m^2(\nu_e)$ vanishes within the error bars. For a data interval down to $\varepsilon \approx 220\,eV$ the total uncertainty minimizes yielding the result $m^2(\nu_e) = (-4.1 \pm 10.9)\,eV^2$ from which the authors derive an upper limit of $m(\nu_e) < 4.35\,eV$. Of course, the step parameters do strongly correlate to $m^2(\nu_e)$ and enlarge the uncertainty of the latter. Note also that due to the integrating transmission function of a MAC-E-Filter such a step has the signature of a small, mono-energetic line in the original spectrum for which, however, no simple and reasonable origin could be traced. The higher signal rate at Troitsk has led to a significantly improved mass limit, but it still carries a tiny residual spectral distortion of unknown origin, which has to be eliminated by the fit in a phenomenological manner which is not wholly satisfactory.

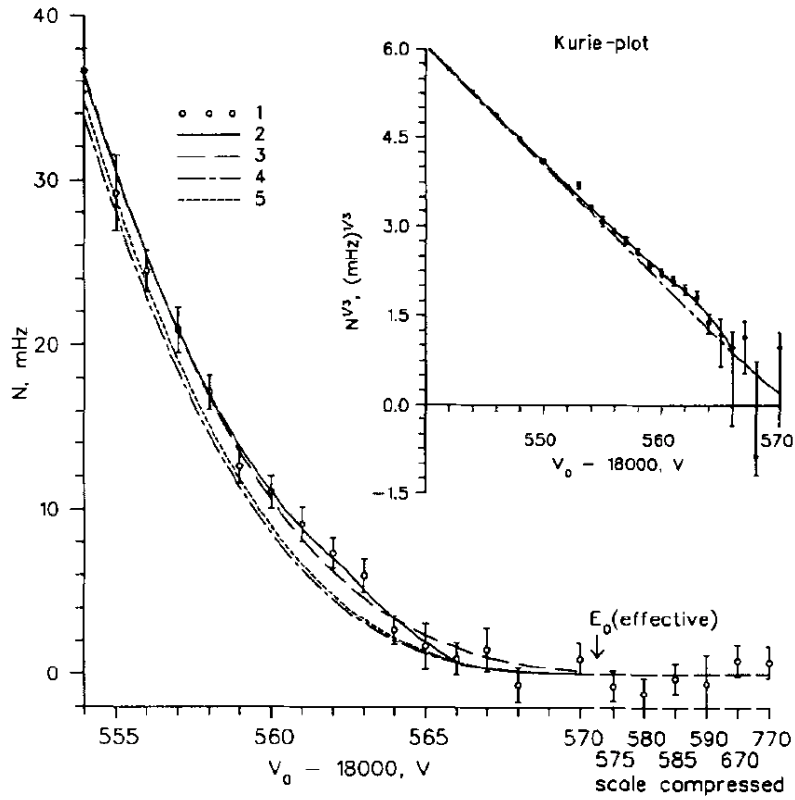

**Figure 16**: Integral $T_2$-$\beta$-spectrum close to the endpoint measured by the Troitsk experiment (linear plot (left), linearized Kurie-plot (inset) (b)). *V* is the spectrometer retarding voltage; The open circles (1) represent the experimental spectrum. The lines (2-5) describe fitted curves using different variables: The solid line (2) includes step position and size as free variables, the dashed line (3) has no step, the dashed-dotted line (4) is the solid line (2) with the fitted step function subtracted and the dotted line (5) is equal to the fit without step (compare to line (3)) but with neutrino mass fixed to zero (reprinted from ref. [37] with kind permission).

*4.4.3. Side experiments and improvements*

On the one hand, the important gain in sensitivity and resolution achieved with MAC-E-Filters already led in the first round of experiments to a reduction of statistical uncertainties of $m^2(v_e)$ by a large factor; on the other hand, it revealed residual systematic problems of the experiments on a finer, so far inaccessible scale. The latter had to be tackled in the next round.

As the calculations of final state excitation in $T_2$-decay have been proven to be exact (see section 3.4.) the origin of the extra energy loss components observed at Mainz and Troitsk has to be sought in the sources themselves. The Mainz group got the decisive hint that solid hydrogen films may dewet from their substrate even at LHe-temperature and contract from a shock-condensed, amorphous film into small crystallized islands [124]. The crystals grow up to size of (0.5 – 1) μm [125] at which double scattering of the β-particles prevails, each with an average energy loss of (34.4 ± 3) eV [126]; this observation may explain the position of the extra component at 75 eV qualitatively[88]. Mainz was forced, therefore, to investigate the problem systematically as a function of temperature and isotopes. This was accomplished with the (already in place) laser diagnostics for ellipsometric control of the average film thickness, but at this point looking for the scattered instead of the reflected light. Scattering appears when the crystal size approaches the optical wavelength. Films were condensed on various substrates at temperatures down to 1.6 K (achieved by a new cryostat); thereafter they were warmed up to a fixed temperature, at which stage dewetting was observed as a function of time through the signature of increasing Rayleigh/Mie-scattering from the growing crystals. Dewetting occurred with any of the chosen substrates at $T \gtrsim 3$ K [98]. The observed dewetting dynamics (see the example of $D_2$ in figure 17) could be quantitatively analysed and fitted in terms of a surface diffusion model with an activation energy of 45 $k_B$ K for $T_2$ [99]. From this number one can derive a dewetting time constant > 1y at the operating temperature of 1.8 K used in later runs. Indeed, the extra component no longer appeared [22,102].

An explanation of the extra energy loss component observed at (100 - 150) eV in the Troitsk experiment has been given in [127]. β-particles emitted at pitch angle $\vartheta > \vartheta_{max,s} = 23.6^o$ are magnetically trapped (63) within the gaseous source ($B_s = 0.8\,T$) by the neck of the magnetic flux ($B_{max,s} = 5\,T$) at both ends (see Figure 12). They may, however, have a certain probability to be scattered into the transmitted solid angle $\vartheta < \vartheta_{max} = 18.4^o$ defined by the field ($B_{max} = 8\,T$) at the spectrometer entrance; then they leave the source with enhanced energy loss according to the extra flight path performed in the source.

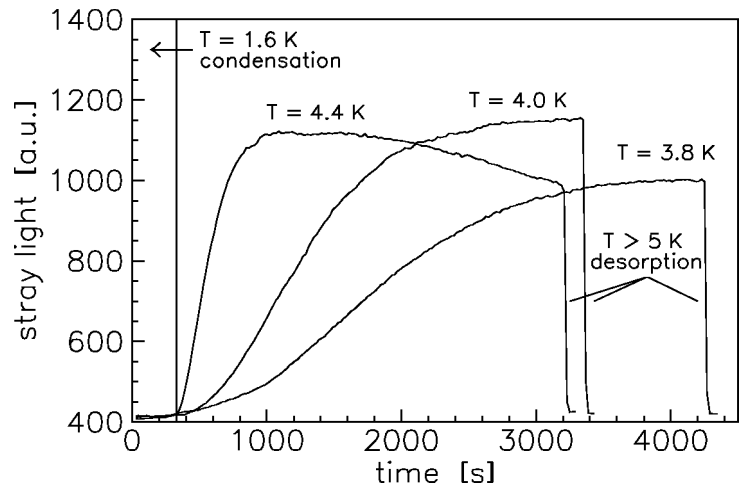

**Figure 17**: Temporal rise of stray light intensity from dewetting and hence roughening, frozen $D_2$-films measured at different temperatures 3.8 K, 4.0 K, 4.4 K. The temperature was raised from 1.6 K to the dewetting temperature at 330 s and raised further to a desorption temperature $T > 5$ K later. The temperature dependence of the dewetting rate follows Arrhenius' law. At 4.4 K slow desorption already competes with dewetting (reprinted from ref [99].

The first results from Mainz and Troitsk have been evaluated using literature data on energy loss spectra of energetic electrons in gaseous hydrogen. It was felt that further improvement should include re-investigating energy loss in the sources. To that end, dedicated experiments were performed at both spectrometers of which the results have been published in a joint paper [122]. Troitsk passed a mono-energetic electron beam from a gun through the gaseous $T_2$-source at energy somewhat above $E_0$ such that the β-spectrum did not interfere. Figure18 shows the transmitted beam for various source pressures as a function of analysing energy. With a lowering of the analysing energy (filter potential, retarding energy), the spectrometer starts to transmit electrons at about 18,644 keV (17.828 keV) in the Troitsk (Mainz) case. About 10 eV below this onset of transmission, the count rate stabilizes at a plateau representing those electrons, which have not undergone inelastic scattering ("no loss fraction") The second rise integrates the energy loss spectrum up to the second plateau at full transmission which is practically reached at 100 eV energy loss. At given column density of the source, the ratio of the 2 plateaux is a measure of the total inelastic cross section. Corresponding data for condensed sources were taken at Mainz in a similar fashion, except that instead of an electron gun the 17.83 keV conversion line of $^{83m}$Kr served as source (see figure 18); it was condensed on a graphite substrate and then covered by a shock condensed $D_2$-film [122] (Warming up this film to about 4 K provoked dewetting and thus allowed the enhanced energy loss in the dewetted phase of the film to be observed directly [99]).

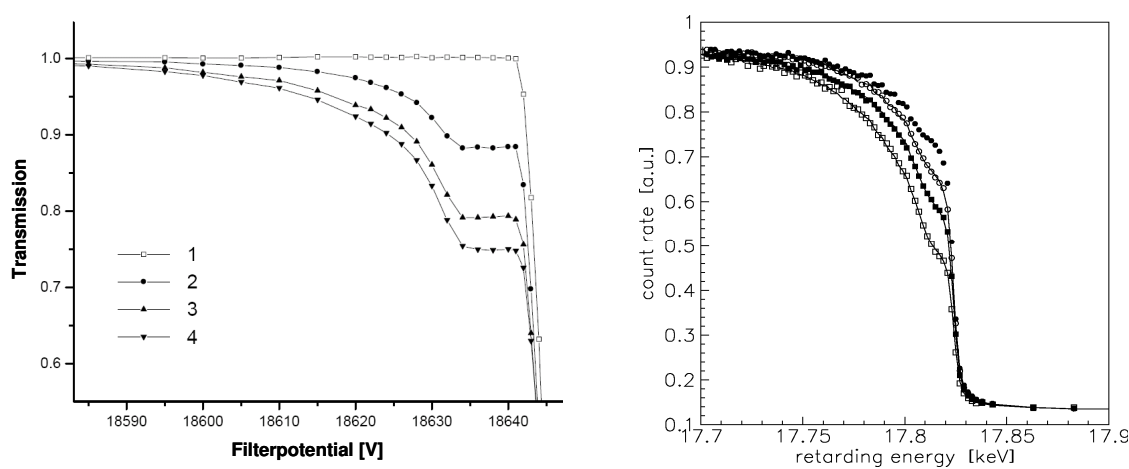

**Figure 18**: Integral energy loss spectra of monochromatic electrons passing the gaseous $T_2$-source of the Troitsk experiment (left) and of $^{83m}$Kr conversion electrons passing a quench condensed $D_2$-film of the Mainz experiment (right) as measured by the MAC-E-Filters as a function of the filter potential energy. For the Troitsk spectra $T_2$-pressures of (0, 0.3, 0.48 and 0.6) torr were used for the spectra 1 – 4 respectively. At Mainz the $D_2$-film thicknesses were determined by laser ellipsometry to be zero (filled circles), 3.9 molecules/Å$^{-2}$ (open circles), 7.7 molecules/Å$^{-2}$ (filled squares) and 14.3 molecules/Å$^{-2}$ (open squares), respectively (reprinted from ref [122]).

To generate an energy loss spectrum from these data by de-convolution would be a somewhat ill-posed problem in the region of sharp rise due to competition by the width of the transmission function as well as by the shake-up/off satellites of the $^{83m}$Kr conversion electron line. Instead, the spectrum was composed of trial functions – one for the peak of molecular excitation, the other for the tail of ionisation (figure 19) – whose parameters were least squares fitted to the measured transmission curve. Troitsk and Mainz used somewhat different trial functions to account for the modifications due to the solid state effects in case of the Mainz quench-condensed $D_2$ film (see below). Their details, like the unphysical kink at the matching point in one of them, for example, don't really matter, once they are folded with the transmission function and the continuous β-spectrum. What does matter, however, are their mean position and width, and their total area. Here we observe significant differences between the gaseous and the condensed phase. The peak of molecular excitation shifts from $(12.6\pm0.3)\,\text{eV}$ to $(14.1\pm0.7)\,\text{eV}$, and the total inelastic cross section from $(3.40\pm0.07)\times10^{-18}\,\text{cm}^2$ to $(2.98\pm0.16)\times10^{-18}\,\text{cm}^2$. These findings agree well with quantum chemical calculations, reported and comprehensively discussed in [122]. The principal reason for these shifts is partial Pauli blocking of phase space of the excited molecular wave functions by its neighbour molecules.

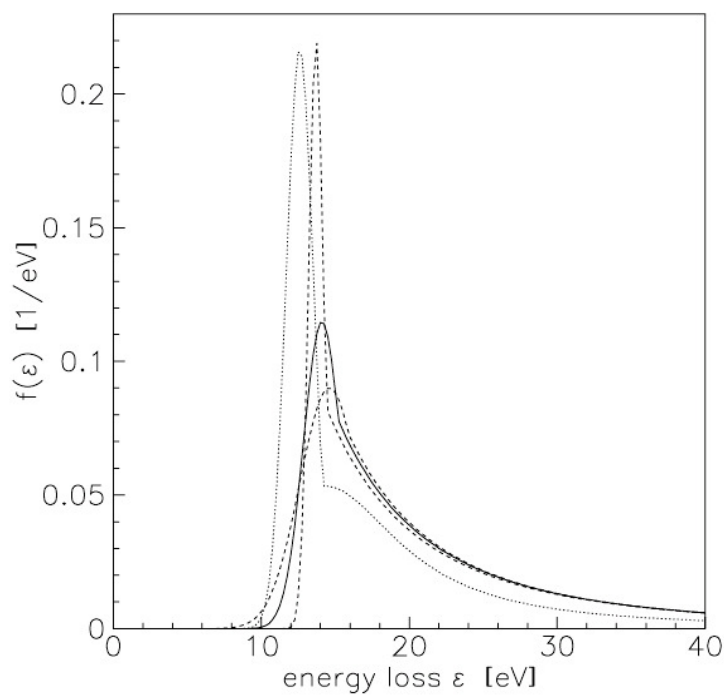

**Figure 19**: Normalised energy loss functions of 18 keV electrons in gaseous $T_2$ (dotted line) and condensed $D_2$ (solid line) fitted to integral energy loss spectra measured by the MAC-E-Filters at Troitsk and Mainz. Dashed lines correspond to $\pm1$ standard deviation of the fitted line width of the principal energy loss peak in the solid phase (reprinted from ref [122]).

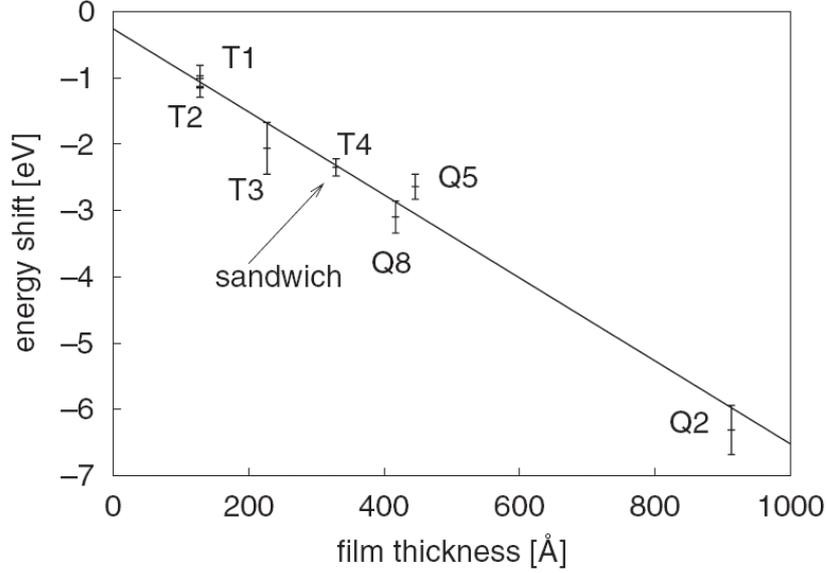

**Figure 20**: Self-charging of frozen $T_2$-films by β-decay. Shown is the energy shift of the 17.83 keV conversion line of $^{83m}$Kr condensed on top of a tritium film as function of its thickness. The labels mark results from dedicated test experiments ($T_1$-$T_4$) and from old films ($Q_2$, $Q_5$, $Q_8$) taken at the end of long term β-spectroscopy runs, respectively (reprinted from ref. [125]).

Yet another, so far unknown, feature of solid $T_2$-films had to be cleared up in a side experiment, before the final phase of β-spectroscopy was resumed. It concerns a downshift of the endpoint, which was observed at Mainz initially for a very thick $T_2$-film of 284 monolayers. But it was definitely not caused by excessive energy loss; rather it reflected a positive self-charging of the insulating $T_2$-film by β-decay [131,132]. The phenomenon could be investigated quantitatively by spectroscopy of the 17.83 keV conversion line of $^{83m}$Kr. To that end $^{83m}$Kr was condensed on top of $T_2$-films of various thicknesses, or sandwiched somewhere inside the film, and its energy measured by the MAC-E-Filter.

Figure 20 shows an essentially linear downshift of the $^{83}$Kr-conversion line as a function of the distance of the source from the conducting graphite substrate corresponding to a constant field strength of about $|E| \approx 20$ V/monolayer as for a co-planar capacitor. This observation lends itself to an explanation in terms of a 'hopping model' of charges, trapped at their lattice site by a potential $\phi_0$ which is degraded in one and upgraded in the other direction by the electric field. Starting from Frenkel's law for the hopping frequency $\nu$ over a trapping potential $\phi_0$

$$\nu = \nu_0 \exp\left(\frac{-\Phi_0}{k_B T}\right) \qquad (66)$$

one derives that the hopping frequencies in opposite directions to an applied electric field *E* differ by

$$\nu^+ - \nu^- = \nu_0 \left( \exp\left( \frac{-\phi_0 + e|\mathbf{E}|g/2}{k_B T} \right) - \exp\left( \frac{-\phi_0 - e|\mathbf{E}|g/2}{k_B T} \right) \right), \qquad (67)$$

where $\nu_0 \approx 10^{12}$ Hz is the vibrational frequency at a lattice site and *g* is the lattice constant. At $T \approx 2$ K the dependence of the first exponential of (67) on $|E|$ gets extremely steep when the exponent approaches zero. This defines a kind of breakthrough field at which the film becomes conductive and up to which it charges asymptotically. Also, the dynamics of this model have been worked out and found to agree with the observed time dependence of self-charging. The trapping potential was determined to be $\varphi_0 = (175 \pm 8)k_B K$. Backed by this knowledge, the effect of self-charging could be folded into the fit function of β-spectra which allowed the use of $T_2$-films 4 times thicker than in the first round of experiments [22].

Apart from the new, upgraded source section, the Mainz set up was altered in several respects concerning background reduction. As for the gaseous sources, a cold bend was introduced into the transport line (see figure 13) serving as a baffle where $T_2$ evaporating from the source was retained on amorphous graphite at liquid helium temperature. This measure removed any source-related background. Furthermore, the electrode system was redesigned in part and its tendency to destabilize background conditions by traps and/or field emission repressed. Automation and an alert system allowed the experiment to be run essentially in a stand alone mode for several weeks [22].

*4.4.4. Phase II of $T_2$-spectroscopy at Mainz and Troitsk*
The upgraded Mainz experiment came up with signal and background rates quite similar to those at Troitsk. It took data in the years 1997 – 2001. The gain in sensitivity is evident when comparing the two spectra in figure 21 taken before and after improvement, respectively [102]. In particular, Mainz was now also sensitive to the minute step-like anomaly which continued to occur in the Troitsk spectra close to the endpoint.

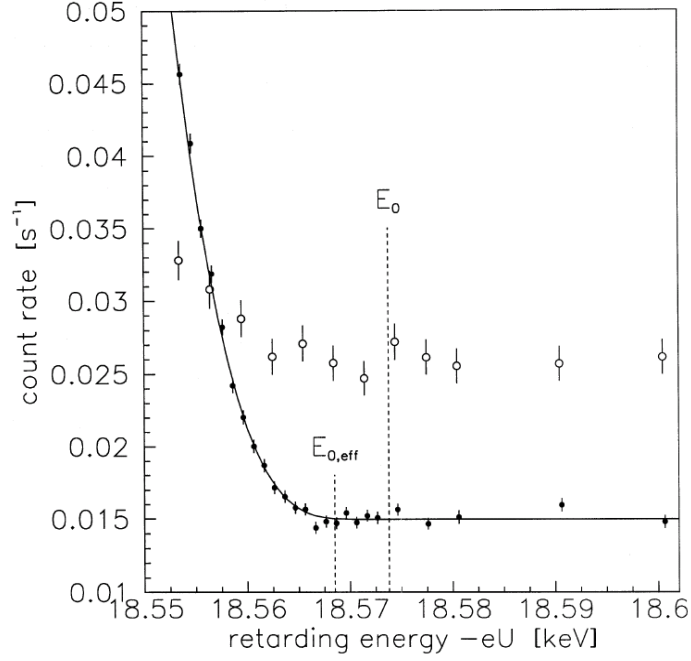

**Figure 21**: Integral $T_2$-$\beta$-spectrum near the endpoint $E_0$ from the first high statistics run with the improved Mainz set up (dots) compared with earlier data from 1994 (circles). The effective endpoint $E_{0,\text{eff}}$ considers the convolution with the response function (72) and the rotation-vibration spectrum of the electronic ground state of the $^3\text{HeT}^+$ daughter molecule. The line shows a fit to the data for $m^2(\nu_e) = 0$ over the interval shown (reprinted from ref [102]).

In 1999 both groups published in parallel results from their enhanced data sets:

$$m^2(\nu_e) = (-1.9 \pm 3.4_{\text{stat}} \pm 2.2_{\text{syst}})\,\text{eV}^2 \quad \Rightarrow m(\nu_e) < 2.5\,\text{eV} \qquad \text{(Troitsk [23])} \qquad (68)$$

$$m^2(\nu_e) = (-3.7 \pm 5.3_{\text{stat}} \pm 2.1_{\text{syst}})\,\text{eV}^2 \quad \Rightarrow m(\nu_e) < 2.8\,\text{eV} \qquad \text{(Mainz [102])}. \qquad (69)$$

Figure 22 shows results of $m^2(\nu_e)$ as a function of the evaluated interval for the sum of Troitsk data for the period 1994 – 1998 [23] (left panel) and for the single runs Q4 and Q5 in 1998 from Mainz [102] (right panel).

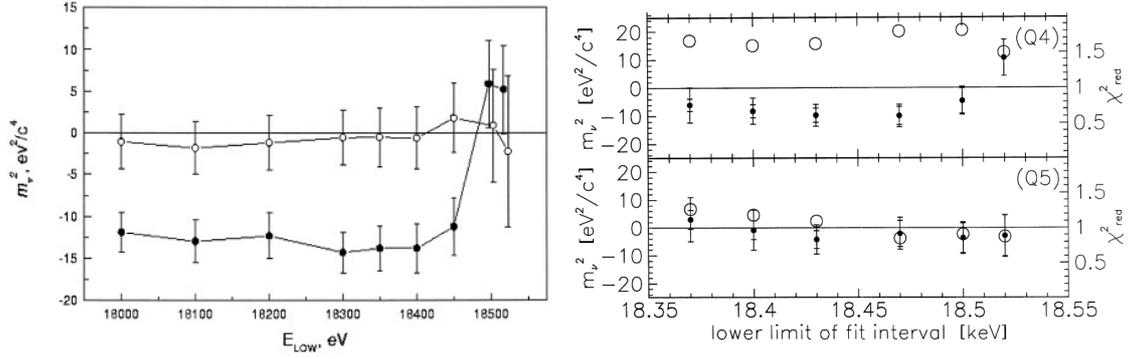

**Figure 22**: Dependence of $m^2(\nu_e)$ from lower limit of fitted data interval for the sum of runs 94, 96,97/2, 98 from Troitsk [23] (left panel) and runs Q4, Q5 from phase II of running at Mainz [102] (right panel). The Troitsk data have been fitted once without step function (dots), once with additional parameters for step size and position, which were fitted to the data run-by-run (circles). In case of the Mainz data dots indicate $m^2(\nu_e)$-values (left scale) and circles $\chi^2_{red}$ -values of the fit (right scale) (reprinted from [102] and [23] with kind permission).

The before mentioned effect of enhanced energy loss by trapping and scattering in the source has now been considered in the fit of the Troitsk data. One can see from the plots that this measure has stabilized the fit results for long fit intervals as well. In case of the full dots, a step-like anomaly in the Troitsk data was ignored. Then – in the *average* over all runs – a significant negative value around $-12\,\mathrm{eV}^2$ is observed for all evaluation intervals with relevant statistical accuracy. As in 1994 this down shift into the unphysical sector can be removed when the fit of *individual* runs is performed with 2 more parameters for position and size of the step. Note that position and size of the anomaly vary from run to run. The problems connected with this correction are discussed in [23]. It is clear that the step parameters strongly correlate to $m^2(\nu_e)$ as demonstrated by the two plots in figure 22 (left). The correlation enlarges significantly the uncertainty of $m^2(\nu_e)$ resulting from the fit. It is argued by the Troitsk group that this enlargement also covers the systematic uncertainty which could possibly stem from the fact that structure and origin of the anomaly are *a priori* unknown. For runs exhibiting a step quite close to $E_0$ the fit could not discriminate the signature of $m^2(\nu_e)$ from that of the step within uncertainty limits and the respective fits do not converge. These runs had to be discarded from the evaluation.

Results from the improved Mainz set up no longer showed the strong drive towards negative $m^2(\nu_e)$, although source films 4 to 7 times thicker than before were used; this proves that energy loss and self-charging are now well under control. Figure 22 (right) shows $m^2(\nu_e)$-plots as a function of the evaluation interval from two long production runs performed at Mainz in 1998 [102]. In addition, the open circles on the right scale show the corresponding $\chi^2_{red}$ -values of the fit. For run Q5 one finds $m^2(v_e)$ to be close to 0 and $\chi^2_{red}$ close to 1 at all evaluation intervals. For run Q4, however, $m^2(v_e)$ falls still slightly into the unphysical sector, comparable to the uncorrected Troitsk values of figure 22 (left). Also $\chi^2_{red}$-values of up to 1.8 are not satisfactory considering the number of degrees of freedom of about 30. In fact, the residues evidence a form of step-like deviation from the fit function with parameters in the range also observed at Troitsk. The data from the earlier runs in Q2 and Q3 also showed similar residual problems, yet these residues did not exhibit a clear step-like signature. Mainz tried different methods to account for the residual anomalies in the fits; they led to $m^2(v_e)$-values close to zero within narrow error bars [102]. However, there was still the "clean" result from run Q5. This run differed from the previous ones in that it was the first during which the background was reduced and stabilized by expelling trapped particles through an rf-pulse every 20 s (see section 4.4.3.). Working under the hypothesis that the anomaly is an effect of the apparatus which can be suppressed by certain measures, Mainz took the purist decision to discard contaminated data sets rather than to correct them. Hence the above quoted Mainz result (69) was based just on the Q5 run. The runs Q6, Q7, Q8, following run Q5, were performed under the same conditions regarding preparation of the set up before the measurement (out-baking and conditioning) and during the measurement (applying rf-pulses in short measurement interruptions every 20s to expel trapped particles). Additionally, objective criteria (like $\chi^2$) were used to reject suspicious sub-runs. Altogether the results of these runs were completely stable in the critical observables $m^2(v_e)$ and $E_0$ with respect to the fit interval and exhibited good $\chi^2$ [22] (see figure 23).

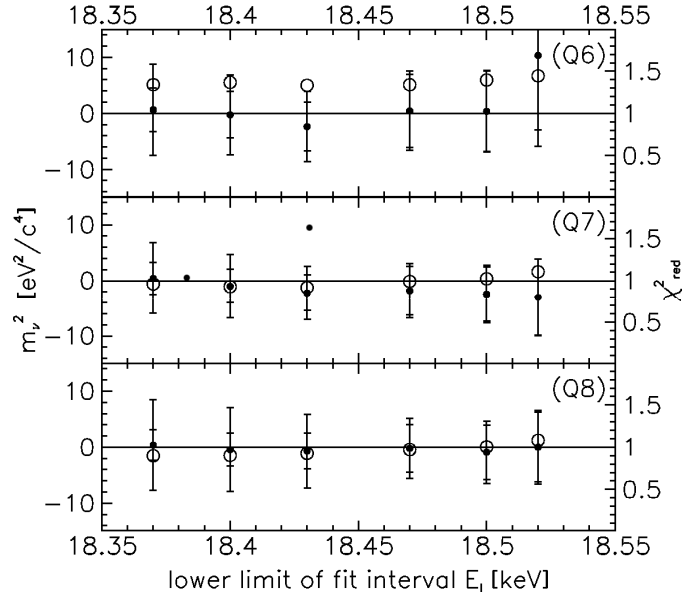

**Figure 23**: Dependence of $m^2(v_e)$ (filled circles) and $\chi^2$ (open circles) from lower limit of fitted data interval for the three Mainz runs Q6, Q7, Q8 of 1999 which were performed after the same careful out-baking and conditioning and with the same rf-pulsing in measurement pauses every 20s as Q5. The inner error bar indicates the statistical uncertainty only, whereas the outer error bar shows the total uncertainty (reprinted from [133]).

*4.4.5. Troitsk-anomaly*

It was always stated by authors and observers that the residual anomaly in the Troitsk spectrum is likely to be an apparatus effect. However, its origin is anything but easy to identify at a level of $10^{-10}$ of the full beam intensity which enters the spectrometer. As mentioned above, Mainz has introduced an active measure to repress such anomalies (intermediate rf-pulses which expel trapped charges from the spectrometer). In the last running period in 2001, it was found that runs Q11 and Q12 which were preceded by quite extensive baking and conditioning of the spectrometer showed a very low and stable background of 0.012 cts/s ; this residual background rate could no longer be manipulated by rf-pulses. Also in these cases the anomaly was no longer present [22]. Although the mechanism of the excess background rate has not been cleared up in any detail, two decisive practical lessons have been learned for the Mainz set up: (i) the source of the effect are slowly charging Penning traps; (ii) the Penning plasma does not develop after the spectrometer has been properly baked and conditioned.

Tentatively taking the effect as a new physics phenomenon, on the other hand, a spectral line on the β-spectrum can only be produced by capture from a surrounding, dense neutrino cloud. A line position below the endpoint would indicate negative neutrino energy due to some long range weak potential. A model, not being in conflict with existing knowledge, was worked out by Stephenson et al. [134]. It was mandatory to check this possibility in spite of its meagre prospect of becoming true. As the effect, as measured in Troitsk, apparently happens to fluctuate strongly in time, Mainz could have missed it during its "clean" runs. The judgement in favour of an apparatus effect was brought about in December 2000 by running in parallel at Mainz and Troitsk. The result of the step search at both locations is shown in figure 24 [22,135]. In order not to miss small steps by correlated fit uncertainties, the number of fit parameters is minimized, i.e. $m^2(v_e)$ is fixed to 0 (for 2 out of the 3 fits shown) and fits are performed for a chosen set of fixed step positions. One then looks for a significant reduction of $\chi^2$ as function of the chosen step position. The Troitsk result in plot a) shows a very significant $\chi^2$ drop of up to 10 units at 18553 eV with a correspondingly large fit value of the step amplitude of 0.012 cts/s (plot b)). However, the $\chi^2$-plot c) of the parallel data set from Mainz displays only statistical fluctuations. Therefore, the Troitsk anomaly seems to be an experimental artefact of the set up, whose origin is not yet fully understood .

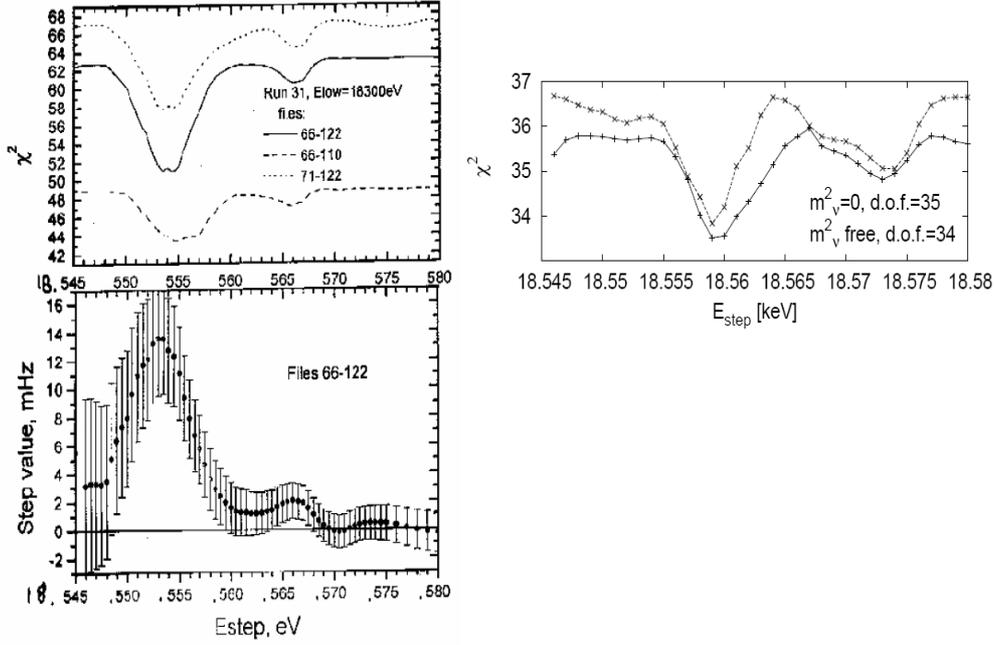

**Figure 24**: Analysis of a search for a step-like anomaly in parallel measurements in December 2000 at Troitsk (left) and Mainz (right). Each point in the upper left and right results from a fit to the data with a step position fixed to the respective energy. File 66-122 (left) from Troitsk shows a very significant drop $\chi^2$ by 10 units at 18.554 keV with correspondingly large fitted step amplitude of 14 mHz (lower left). In contrast the $\chi^2$-plot of the Mainz data is quite flat at that energy, and the overall fluctuations by $\pm 1$ unit correspond to a normal statistical behaviour (reprinted from [22]).

*4.4.6. Final results*

From the 11 production runs performed with the improved Mainz set up, 6 were selected for the final evaluation. The selection criteria were mainly temporal stability, and self-consistency of the runs, as some runs suffered from unfavourable experimental and unstable background conditions [22]. These selected data sets represent 217 days of running. The data from each run were fitted separately to a function $F(U)$ which is a convolution of a spectral function $\gamma'(\varepsilon)$ and the response function $T'(E,U) = T'(E_0 - \varepsilon, U)$ plus constant background rate $b$:

$$F(U) = \int_{eU}^{E_0} \gamma'(\varepsilon) T'(E_0 - \varepsilon, U) \mathrm{d}\varepsilon + b. \qquad (70)$$

The spectral function is essentially the β-spectrum (43), simplified by contracting all constant fore factors into the amplitude $A$ and the neutrino masses into the mean squared mass $m^2(\nu_e)$:

$$\gamma'(\varepsilon) = A\, F(E, Z+1)\, (E_0 + m - \varepsilon)\, \sqrt{(E_0 + m - \varepsilon)^2 - m^2} \times \qquad (71)$$
$$\times \sum_j P_j \cdot (\varepsilon - V_j) \sqrt{(\varepsilon - V_j)^2 - m^2(v_e)} \cdot \Theta(\varepsilon - V_j - m^2(v_e)).$$

The response function is a convolution of the exact transmission function of the spectrometer [110] (approximately given by (62)) with 4 correction functions for energy loss, source charging, backscattering from the substrate, and energy dependence of the detection efficiency

$$T' = T \otimes f_{\text{loss}} \otimes f_{\text{charge}} \otimes f_{\text{back}} \otimes f_{\text{det}}. \qquad (72)$$

Each carries its particular systematic uncertainty. Since $f_{\text{loss}}$ is the dominating correction, $T'$ is quite similar to the transmission curves shown in figure 18.

Issues of statistical and systematic uncertainties are discussed in detail in [22]. Figure 25 shows on the left the parabolas of $m^2(v_e)$ versus $\chi^2$ which result from fitting each individual run. From the width of their sum (which is narrowed by summation), the statistical uncertainty of the final result was determined. On the right, all relevant uncertainties are plotted as a function of the evaluation interval. Whereas the statistical uncertainty decreases with the interval length, all systematic uncertainties increase. Uncertainties of energy loss dominate. Open squares result from the uncertainties of the total inelastic cross section (5.4%) and of the determination of film thickness by ellipsometry (3%). In addition to its geometrical thickness, a density reduction of the amorphous film with respect to the crystal phase by 6.8% has been determined by ellipsometry via a corresponding drop of refraction index. This is considered in calculating energy loss. Stars represent the energy loss uncertainty stemming from an additional $H_2$-coverage which condenses in the course of running at a rate of about 0.3 monolayer/day from the rest gas on top of the $T_2$-film. This effect was also traced by ellipsometry and was found to accord to a slight down sloping of fitted $m^2(v_e)$-values with increasing source age. The respective correction has also been entered into the uncertainty budget in a conservative manner at full size.

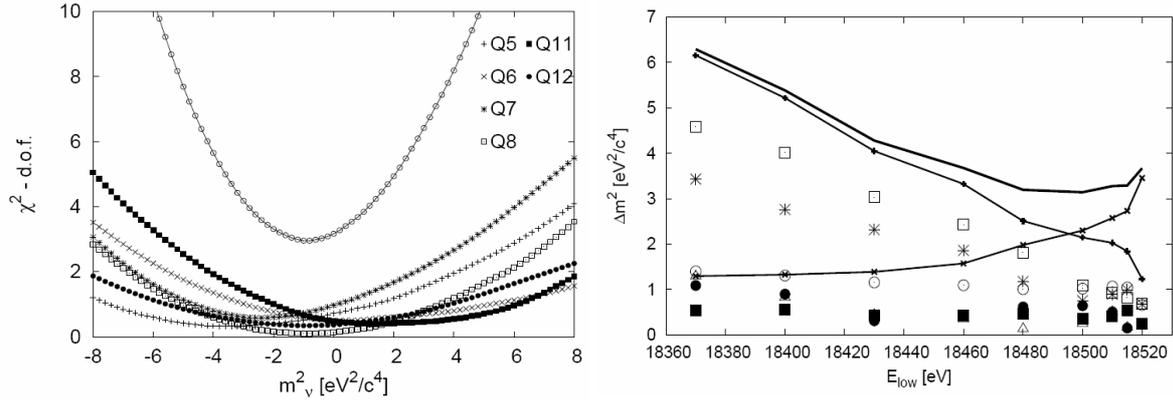

**Figure 25**: $m^2(\nu_e)$-uncertainty of the final Mainz result. On the left, the $\chi^2$-parabolas of $m^2(\nu_e)$-fits for each individual run, as well as their sum (circles), is shown. The fit intervals are restricted to a lower limit of 75 eV below $E_0$. On the right, the individual and the geometric sum of uncertainties are shown for the joint data set, calculated for different lengths of the data interval. The line with crosses shows the geometric sum of systematic uncertainties, the line with stars gives the statistical uncertainty showing opposite slope. The upper most line shows the total geometric sum of uncertainties which attains a minimum at 18500 eV (75 eV below $E_0$) (reprinted from ref [22]).

Particular attention has been paid to prompt neighbour excitation (open circles) which already has been mentioned in section 3.4. . Kolos' calculation of $P_{ne} = 5.9\%$ excitation probability is certainly based on good semi-empirical grounds; but a check of this number by quantum chemistry calculations is still missing. If one considers, in a qualitative manner, the porosity of the film as well as the reduction of excitation by Pauli-blocking (section 3.4.), $P_{ne}$ shifts down to 4.6%. On the other hand, an independent determination from the data itself was tried by fitting $P_{ne}$ as an additional *free* parameter from the *full* data set down to $\varepsilon = 170\,\mathrm{eV}$ which has maximum sensitivity to energy loss. The resulting contour plot of $\chi^2$ in the ($P_{ne}$, $m^2(\nu_e)$)-plane (see figure 26) yielded the very satisfactory result $P_{ne} = 5\% \pm 1.6\%$, $m^2(\nu_e) = (0 \pm 3)\,\mathrm{eV}^2$. This $P_{ne}$-value was then used as a *fixed* parameter in fitting restricted intervals where energy loss contributes much less, taking consistently into account its uncertainty and its correlation to the energy loss parameters during the fits. This operation had only a marginal influence on the final result; primarily though it reinforced confidence in assigning adequate systematic uncertainties to this effect.

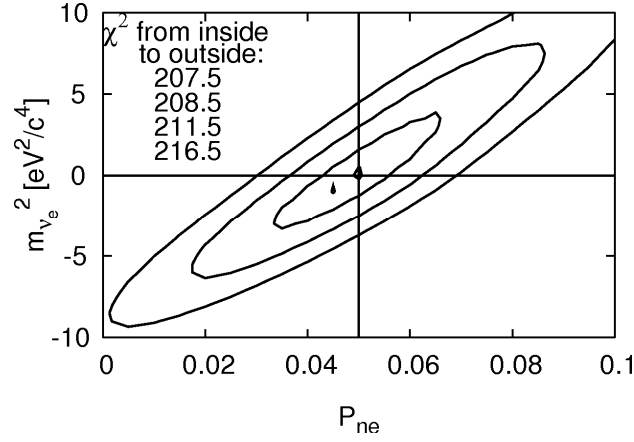

**Figure 26**: Shown are contour plots of $\chi^2(P_{ne}, m^2(\nu_e))$ at 1σ, 2σ and 3σ around its minimum value at (0.05, 0eV$^2$) (reprinted from [22]).

The excitation spectrum of the daughter molecule ($^3$HeT)$^+$ shifts upwards slightly in solid T$_2$ with respect to the gaseous phase for the same reasons as discussed above for neutral T$_2$. Saenz has estimated shifts of 0.8 eV for the second and 1.4 V for the third group of excited states; the first group hardly shifts [136]. These shifts have been considered and have also been fully entered into the uncertainty budget.

The geometric sum of all uncertainties, i.e. the total uncertainty of the final Mainz result, reaches a minimum for a data interval stretching to $\varepsilon \approx 75\,\text{eV}$. From that choice one obtains [22]

$$m^2(\nu_e) = (-0.6 \pm 2.2_{\text{stat}} \pm 2.1_{\text{syst}})\,\text{eV}^2 \quad \Rightarrow m(\nu_e) \leq 2.3\,\text{eV at 95\% C.L.} \tag{73}$$

The observable $m^2(\nu_e)$ is fully compatible with the physical limit 0. The upper limit is calculated from these numbers according to the unified approach.

The latest results from Troitsk were communicated in 2003 [137]. The problem with step-like anomalies has persisted. Runs showing a step closer than 8 eV to the endpoint were excluded because the correlation between step amplitude and $m^2(\nu_e)$ was too strong as mentioned above. Rather large steps of around 20 eV below $E_0$ have been observed in the latest runs; these runs have also been rejected. In total about half of the data was included in the final evaluation.

The following systematic uncertainties are of $m^2(\nu_e)$ are quoted:

| | |
|---|---|
| Energy loss | 1.2 eV$^2$ |
| Transmission function | 0.5 eV$^2$ |
| Final states of daughter | 0.7 eV² |
| Possible space charge in the source | 1.0 eV$^2$. |

Together with some other small items the geometric sum up is 2.0 eV$^2$. The presence of steps enlarges through correlations the uncertainty in fitting $m^2(\nu_e)$ by 1.5 eV$^2$ which is included in the total fit uncertainty of 2.5 eV$^2$. Again it is argued that this enlargement also covers the systematic uncertainty which could possibly stem from the fact that structure and origin of the anomaly are *a priori* unknown. This statement may apply qualitatively; but it cannot apply in general and its validity in the given case has not been demonstrated. The final result from Troitzk is communicated to be [129]

$$m^2(\nu_e) = (-2.3 \pm 2.5_{stat} \pm 2.0_{syst})\,\text{eV}^2 \quad \Rightarrow m(\nu_e) \leq 2.05\,\text{eV at 95\% C.L.} \quad (74)$$

The central value of (74) lies in the negative sector at -0.7σ (with $\sigma = \sqrt{\sigma_{stat}^2 + \sigma_{syst}^2}$ as usual). The upper mass limit is lowered somewhat by this circumstance.

Alternatively, the $m^2(\nu_e)$-results may also be interpreted in terms of the so called sensitivity limit, defined simply as the square root of the 2σ-value of the experimental $m^2(\nu_e)$-value, irrespective of its position. This limit is calculated from (73) and (74) to be 2.46 eV and 2.53 eV, respectively.

The Particle Data Group published in 2006 [5] an upper limit of $m(\nu_e) < 2$ eV as we have stated in the introduction (1). The limit is said to be based on the final result (73) from Mainz in 2005 and from the Troitsk result (68) in 1999 (which is the latest published as a regular journal paper). From the physics point of view, there is no need to discuss the subtle differences in calculating and presenting these upper limits any further. Here the primary results on the observable $m^2(\nu_e)$, (73) from Mainz and (74) from Troitsk, have to be emphasized. It is remarkable how close they come to each other, in spite of the quite different source concepts upon which they are based and the experimental problems connected to them. It is also clear that both experiments have virtually exhausted their potential at this point with respect to statistical accuracy.

## 5. Preview on forthcoming experiments.

The Troitsk group is reconstructing its apparatus in part in the meantime [138]. A larger spectrometer vessel of 2.2 m in diameter will improve resolution. A major aim of this next phase of the experiment is tracing and eventually eliminating the sources of the step-like anomaly. A measure from which its result would certainly profit.

At a meeting in Erice in 1997 preliminary ideas for next generation experiments on tritium β-decay in search for the absolute neutrino mass were presented by Troitsk [139] and by Mainz [124]. More details on the latter have been published by Bonn et al. [115]. With the discovery of neutrino oscillations in 1998 [11] the discussion gained momentum. Motivated by a long record in neutrino physics through the GALLEX- and KARMEN-Experiments [14,75] and backed by the presence of a dedicated tritium laboratory on site, the Forschungszentrum Karlsruhe decided to get involved in the plans for a new neutrino mass experiment. It was named KATRIN and is described in the following section.

*5.1. The Karlsruhe Tritium experiment on the neutrino mass (KATRIN)*

From a workshop in Bad Liebenzell in 2001 a letter of intent for the KATRIN-Experiment [38] emerged from close collaboration of group members from the earlier neutrino mass experiments at Los Alamos (now at University of Washington, Seattle), Mainz, and Troitsk with Forschungszentrum Karlsruhe. A design report [39] was approved in 2004. Construction of the experiment is under way and expected to be completed in 2010. The experiment aims at an improvement of the sensitivity limit by an order of magnitude down to $m(v_e) < 0.2$ eV to check the cosmologically relevant neutrino mass range and to distinguish degenerate neutrino mass scenarios from hierarchical ones. Furthermore, Majorana neutrinos sufficiently massive to cause the double β-decay rate of $^{76}$Ge which part of the Heidelberg Moscow collaboration claims to have observed [68] would be observable in the KATRIN experiment in a model independent way. The true challenge becomes clear by drawing attention to the experimental observable $m^2(v_e)$ whose uncertainties have then to be lowered by two orders of magnitude.

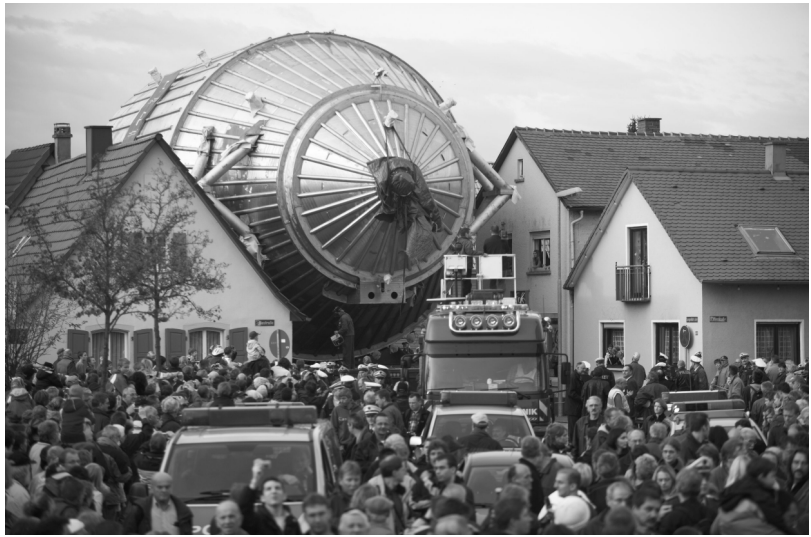

**Figure 27**: The KATRIN main spectrometer passes through the village Leopoldshafen on its way from the river Rhine to the Forschungszentrum Karlsruhe on November 25, 2006 (printed with kind permission from Forschungszentrum Karlsruhe).

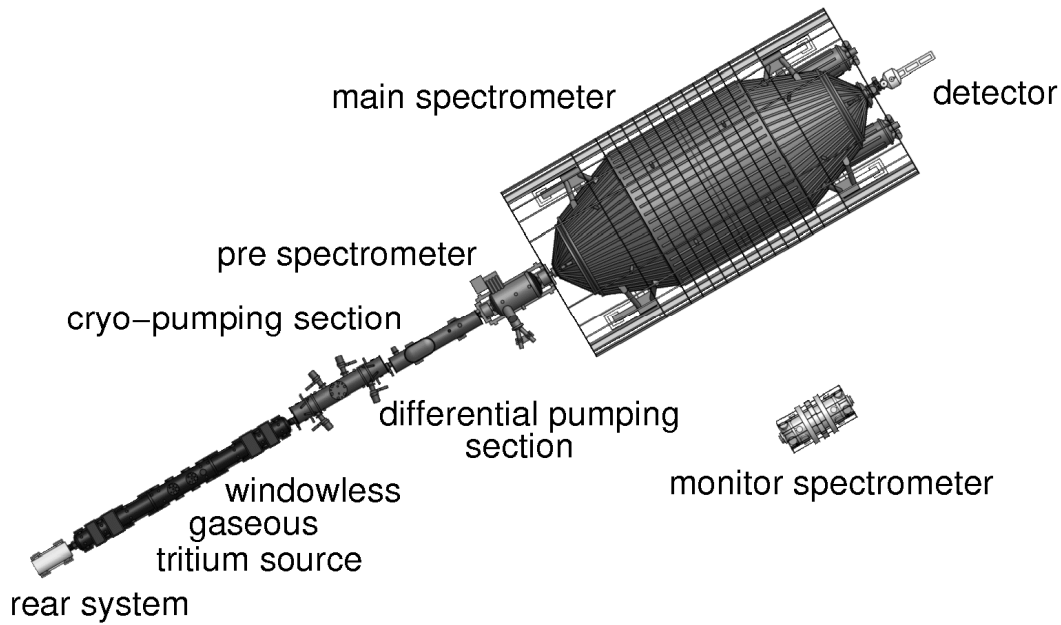

**Figure 28**: Schematic view of the 70 m long KATRIN experiment consisting of calibration and monitor rear system, windowless gaseous $T_2$-source, differential pumping and cryo-trapping section, small pre-spectrometer and large main spectrometer, segmented PIN-diode detector and separate monitor spectrometer.

Improving tritium β-spectroscopy by a factor of 100 evidently requires brute force, based on proven experimental concepts. It was decided, therefore, to build a MAC-E-Filter with a diameter of 10 m, corresponding to a 100 times larger analysing plane as compared to the pilot instruments at Mainz and Troitsk. Accordingly one gains a factor of 100 in quality factor which we may define as the product of accepted cross section of the source times resolving power $E/\Delta E$ for the emitted β-particles. Figure 27 shows the spectrometer tank of KATRIN on its way to Forschungszentrum Karlsruhe, Figure 28 depicts a schematic plan of the whole, 70 m long set up. Meantime the spectrometer has been set up and has reached its design out-gasing rate in the range of $10^{-12}$ mbar l s$^{-1}$ cm$^{-2}$.

A decay rate of the order $10^{11}$ Bq is aimed for from a source with diameter ≈ 9 cm. The choice between a gaseous and a frozen $T_2$-source was not an easy one. The latter could have been built with quite modest effort; but the sought-after activity requires a column density of $5\cdot10^{17}$ molecules/cm$^2$, corresponding to a thickness of 500 monolayers, at which it would charge up to a voltage of 10 V across the frozen film (compare sect. 4.4.3); this would spoil the energy resolution. Since a solution to this problem has yet to be found, preference was given to a windowless gaseous $T_2$-source (WGTS) in spite of its extraordinary demands in terms of size and cryo-techniques, which would be required to handle the flux of $10^{19}$ $T_2$-molecules/s safely. $T_2$ is injected at the midpoint of a 10m long source tube kept at a temperature of 27 K by a 2-phase liquid neon bath. The integral column density of the source of $5\cdot10^{17}$ molecules/cm$^2$ has to be stabilized within 0.1%. For background reasons, the $T_2$-flux entering the spectrometer should not exceed $10^5$ $T_2$-molecules/s. This will be achieved by differential pumping sections (DPS), followed by cryo-pumping sections (CPS) which trap residual $T_2$ on argon frost at 4 K. Each system reduces the throughput by $10^7$, which has been demonstrated for the cryo-pumping section by a dedicated experiment at Forschungszentrum Karlsruhe. The $T_2$-gas collected by the DPS-pumps will be purified and recycled. A pre-spectrometer will transmit only the uppermost end of the β-spectrum into the main spectrometer in order to reduce the rate of background producing ionisation events therein. The entire pre- and main spectrometer vessels will each be put on their respective analysing potentials, which are shifted inside by about $-200\,\text{eV}$, however, due to the installation of a background reducing inner screen grid system (figure 29). A ratio of the maximum magnetic field in the pinch magnet to the minimum magnetic field in the central analysing plane of the main spectrometer of 20000 provides an energy resolution of $\Delta E$=0.93 eV near the tritium endpoint $E_0$.

The residual inhomogeneities of the electric retarding potential and the magnetic fields in the analysing plane will be corrected by the spatial information from a 148 pixel PIN diode detector. Active and passive shields will minimize the background rate in the detector. Additional post-acceleration will reduce the background rate within the energy window of interest. Special care has to be taken to stabilize and to measure the retarding voltage. Therefore, the spectrometer of the former Mainz Neutrino Mass Experiment will be operated at KATRIN as a high voltage monitor spectrometer which continuously measures the position of the $^{83m}$Kr-K32 conversion electron line at 17.8 keV, parallel to the retarding energy of the main spectrometer . To that end its energy resolution has been refined to $\Delta E = 1$ eV

The β-particles will be guided from the source through the spectrometer to the detector within a magnetic flux tube of 191 Tcm$^2$, which is provided by a series of superconducting solenoids. This tight transverse confinement by the Lorentz force applies also to the $10^{11}$/s daughter ions, emerging from β-decay in the source tube, as well as to the $10^{12}$/s electron ion pairs produced therein by the β-flux through ionisation of $T_2$ molecules. . The strong magnetic field of 3.5 Tesla within the source is confining this plasma strictly in the transverse direction such that charged particles cannot diffuse to the conducting wall of the source tube for getting neutralized. The question, how the plasma in the source becomes neutralized then or at which potential it might charge up eventually, has been raised and dealt with only recently [140].The salient point is, however, that the longitudinal mobility is not influenced by the magnetic field. Hence the resulting high longitudinal conductance of the plasma will stabilize the potential along a magnetic field line to that value which this field line meets at the point where it crosses a rear wall. This provides a lever to control the plasma potential. Meanwhile the Troitsk group has performed a first experiment on the problem [141]. They have mixed $^{83m}$Kr into their gaseous $T_2$ and searched for a broadening of the L3-conversion line at 30.47 keV which might be due to an inhomogeneous source potential. Their data fit is compatible with a possible broadening of 0.2 eV, which would not affect their results but suggests further investigation at KATRIN.

The sensitivity limit of KATRIN on $m^2(v_e)$ has been simulated (see below) on the basis of a background rate of $10^{-2}$ cts/s, observed at Mainz and Troitsk. Whether this small number can also be reached at the so much larger KATRIN-instrument – or even be lowered – has yet to be proven. On the one side, the large dimensions of the main spectrometer are helpful, as they improve straight adiabatic motion due to reduced field gradients. On the other hand, the central flux tube faces a 100 times larger electrode surface at the analysing potential from which secondary electrons might sneak in. As already discussed in sect. 4.3, this background source can be cut back by two orders of magnitude by locating a repelling grid in front of the electrode [119]. Such grids with two wire layers will also be installed at KATRIN (see figure 29).

A simulated spectrum covering 3 years of data taking at KATRIN and is shown in figure 30; a measured spectrum from Mainz is added for comparison. Due to the gain in the signal to background ratio, the region of optimal mass sensitivity around $\varepsilon_{opt}$ (compare (47)) has moved much closer to the endpoint and one already notices on first sight a marked mass effect for $m(v_e) = 0.5\,\text{eV}$.

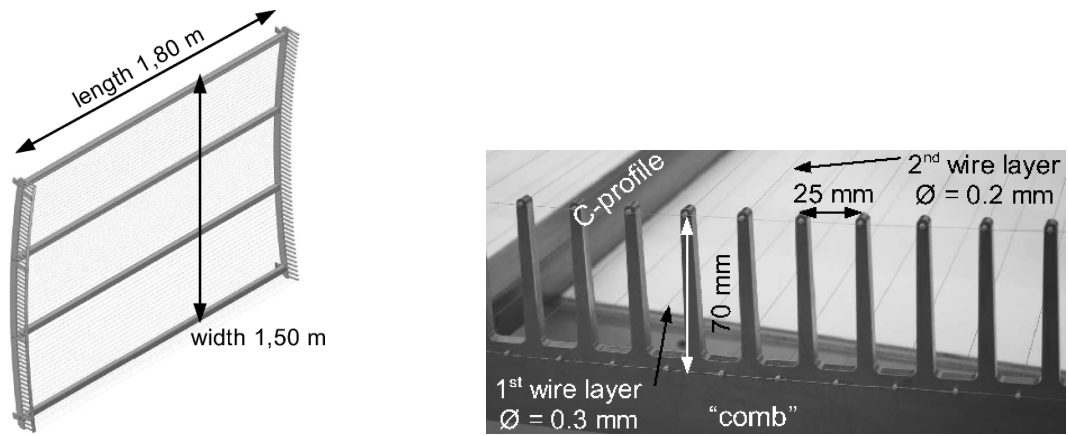

**Figure 29**: Prototype of one of the 248 modules of the double-layer wire electrode system for the KATRIN main spectrometer. Wires with a diameter of 300µm (200µm) are used for the outer (inner) layer. The wires are mounted by precision ceramic holders onto a frame consisting of ``combs'' and C-profiles and keep their relative distance along their length within a few tenths of a mm. Materials are chosen to be non-magnetic and bakable in order to reach the required low outgassing rate of $10^{-12}$ mbar l s$^{-1}$ cm$^{-2}$ (reprinted from [142]).

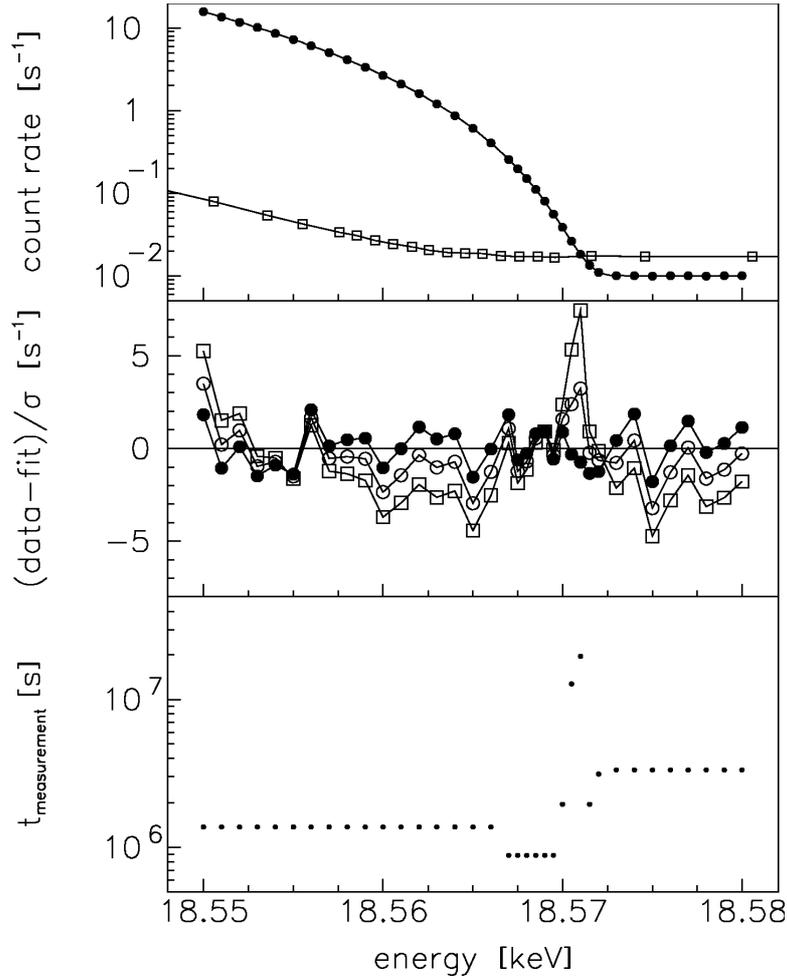

**Figure 30**: Top: Simulated integral β-spectra (assuming $m(\nu_e) = 0$ eV and $E_0 = 18.575$ keV) resulting from 3 years of KATRIN-running under KATRIN design conditions (filled circles) and from phase 2 of the Mainz experiment for comparison (open squares). Middle: Difference of data and fit normalized to the statistical uncertainty for $m(\nu_e)$ fixed in the fit to 0eV (filled circles), 0.35eV (open circles) and 0.5 eV (open squares); this difference peaks at 3.5 eV below the endpoint and thus shows the setting with highest mass sensitivity. Bottom: Distribution of measuring points, optimized in position and measuring time.

One also notices that the typical third power rise of the integral spectrum below $E_0$ (see (45), (46)) is delayed. This is mainly due to ro-vibrational excitations of the daughter molecule which centre at $\varepsilon = 1.72$ eV and stretch up to more than $\varepsilon \approx 4$ eV with a width of $\sigma_{\text{ro-vib}} = \pm 0.42$ eV (figure 6). This width diminishes the mass sensitivity as compared to an atomic source with a sharp endpoint. At KATRIN this effect will be felt for the first time, but still amounts to only 5.5% sensitivity loss on $m^2(\nu_e)$, according to a simulation with KATRIN design parameters. An atomic source would only be worthwhile, if in the future the background could be dramatically suppressed, so that the optimum sensitivity according to (47) (which applies to sharp endpoint and sharp spectrometer) would shift down into a region of $\varepsilon_{opt} \lesssim \sigma_{ro-vib}$. At a background rate of $b = 10^{-4}$ cts/s, for instance, an atomic source would gain about 40% sensitivity on $m^2(v_e)$ with respect to a molecular one. A thorough discussion of an atomic source would also have to consider questions of final states, energy loss, molecular contamination *etc*. In any case, the development of a sufficiently strong atomic source would require an enormous amount of R&D work.

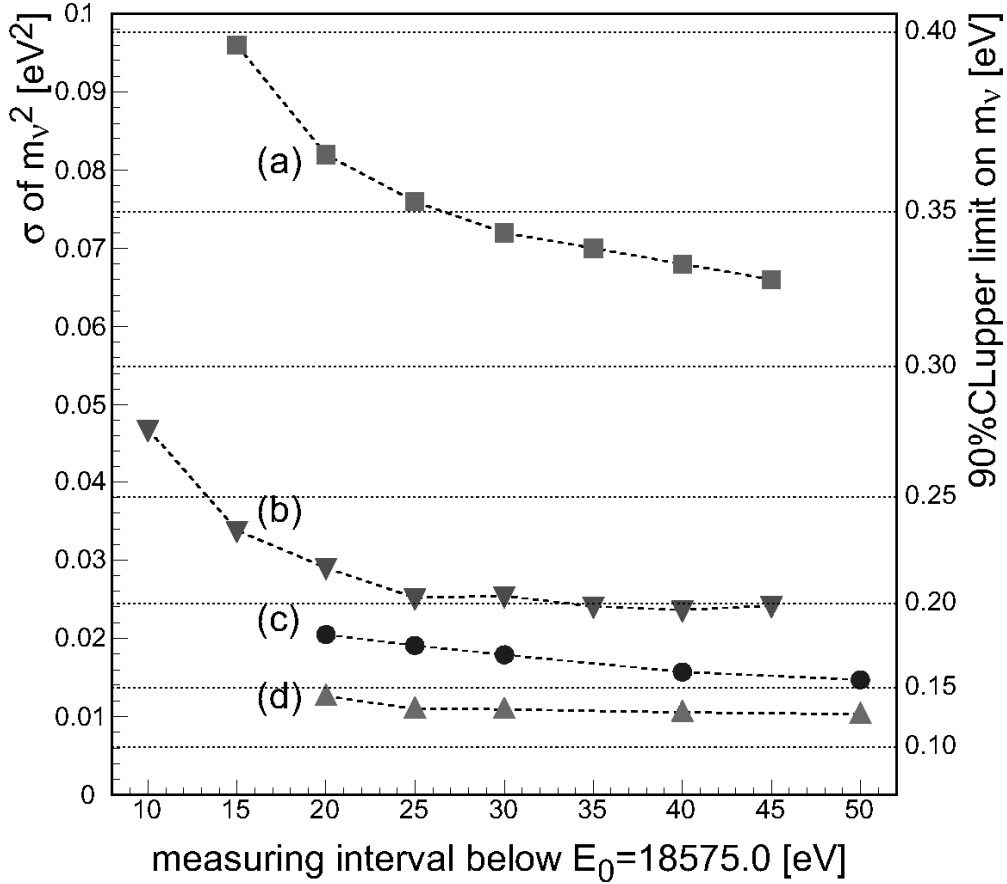

**Figure 31**: Simulations of statistical neutrino mass squared uncertainty expected at KATRIN after 3 years of running, calculated in dependence on the fit interval under following conditions. Spectrometer diameter = 7 m as originally proposed [38]: (a); final 10m design [39]: (b, c, d); background = $10^{-2}$ counts/s: (a, b, c); background = $10^{-3}$ counts/s: (d); equidistant measuring point distribution: (a, b); measuring point distribution optimized according to local mass sensitivity: (c, d) (reprinted from ref. [39]).

But already, with the present molecular source, KATRIN´s sensitivity would profit from a substantial reduction of the background rate below the design value (see figure 31d). This goal might be realised in the future through the deployment of a detector array consisting of highly-resolving cryogenic bolometers (compare section 5.2) instead of silicon PIN diodes. Of course, the development and the operation of such a cryo-detector array would challenge present technology to the limits, but it would reduce KATRIN´s background decisively in at least 2 respects:

1) The excellent energy resolution of cryogenic bolometers would allow the reduction of the energy window of interest by up to 2 orders of magnitude.
2) If, even with the installation of the two-layer wire electrode system (see figure 29), the main spectrometer is still not completely 'quiet' as far as background electrons, born at any of the tank or electrode surfaces are concerned (background noise), most of these residual electrons would be rejected by the excellent energy resolution capacity of the cryo-bolometers.

Figure 31 shows simulations of the statistical uncertainty of the observable $m^2(v_e)$ and corresponding upper mass limits (without systematic uncertainties) which are expected from the KATRIN-Experiment after 3 years of data taking at background rates of $10^{-2}$ cts/s and $10^{-3}$ cts/s, respectively. They are plotted as a function of the width of the spectral interval, as measured with equidistant or optimized distribution of settings for analysing potential as well as for measuring time. The dependence on the interval length is pretty flat, in particular for the lower choice of background. For the reference value $b = 10^{-2}$ cts/s one expects to reach a total uncertainty somewhat below 0.02 eV². Fortunately, the improved signal to noise ratio is very helpful with regard to the systematic uncertainties, as it reduces the decisive correlated uncertainties (49) through lowering $\varepsilon_{opt}$ and shortening the measured spectral interval $\varepsilon_{max}$. In addition, if the measurement interval $\varepsilon_{max}$ drops below energy thresholds of inelastic processes like the first electronic excitation of the $(^3HeT)^+$-ion at 24eV (see figure 6) and the minimum energy loss of inelastic scattering on $T_2$-molecules of 10eV (see figure 19), the corresponding uncertainties vanish completely. From figure 31 it is clear that KATRIN aims at measuring intervals of about $\varepsilon_{max} \approx 25$eV, for which the following systematic uncertainties and the corresponding counter-measures play a role:

- Uncertainty of the energy dependent cross section of inelastic scattering of β-electrons on $T_2$ in the windowless gaseous tritium source.
  Counter-measures: energy loss measurements with an e-gun as done in Troitsk (see figure 18) analysed by special deconvolution methods.

- Fluctuations of the $T_2$ column density in the windowless gaseous tritium source.
  Counter-measures: temperature and pressure control of the tritium source to the $10^{-3}$ level, laser Raman spectroscopy to monitor the $T_2$ concentration compared to HT, DT, $H_2$, $D_2$ and HD.

- Spatial inhomogeneity of the transmission function.
  Counter-measures: spatially resolved measurements with an e-gun or, alternatively, with an $^{83m}$Kr source.

- Stability of retardation voltage.
  Counter-measures: a) measurement of HV with ppm-precision by a HV-divider and a voltage standard; b) applying the retarding voltage also to the monitor spectrometer, which continuously measures $^{83m}$Kr conversion electron lines.
- Electric potential inhomogeneities in the WGTS due to plasma effects.
  Counter-measures: potential-defining plate at the rear exit of the WGTS; monitoring of potential within WGTS possible by special runs with $^{83m}$Kr/T$_2$-mixtures.

Each systematic uncertainty contributes to the uncertainty of $m^2(v_e)$ with less than 0.0075 eV$^2$, resulting in a total systematic uncertainty of $\Delta m^2(v_e) = 0.017 eV^2$. The improvement on the observable $m^2(v_e)$ will be two orders of magnitude compared to present. The total uncertainty will allow a sensitivity on $m(v_e)$ of 0.2 eV to be reached. If no neutrino mass is observed, this sensitivity corresponds to an upper limit on $m(v_e)$ of 0.2 eV at 90% C.L, or, otherwise, to evidence for (discovery of) a non-zero neutrino mass value at $m(v_e) = 0.3$eV (0.35eV) with 3σ (5σ) significance. For more details we refer to the KATRIN Design Report [39].

*5.2. Calorimetric β-spectroscopy of $^{187}$Re and the MARE proposal*

As stated, the search for the neutrino mass by β-spectroscopy requires (i) low β-endpoint and (ii) full control over the energy which is expended in exciting the daughter and other molecules in the source. Condition (ii) can be relaxed if the source is imbedded into some calorimetric type of energy detector which also sum up this expended energy, in addition to the resulting β-energy. Early in the history of β-decay, the decisive proof for the (at that time still mysterious) missing energy was already achieved by calorimetric measurements in an *analogue* manner [143]. Modern cryogenic micro-calorimeters are capable of measuring the heat released by an *individual* event and have achieved an energy resolution better than 5 eV for soft X-rays. Applied to β-decay, one measures (by taking the difference to the endpoint) the spectrum of the *total missing* energy $E_{tot\nu}$ carried off by the neutrino, irrespective of how the *detected* energy is split in the decay between the β-particle and the final states $V_j$ of the daughter (compare (41), (43)). Close to the endpoint, moreover, the shape (not,however, the amplitude) of the neutrino spectrum is determined to the first order by the neutrino phase space $\propto E_{tot\,\nu}\sqrt{E_{tot\,\nu}^2 - m^2(v_e)}$ alone. Only further up in the neutrino spectrum is the influence of the excitation energy $V_j$ felt in the spectral shape by a decrease of the corresponding β-phase space. Close to the endpoint, one can, therefore, ignore the final state spectrum in first approximation, which eases the case of high-Z elements decisively. Hence, there are good reasons to consider the β-decay of $^{187}$Re with primordial half-life

$$^{187}\text{Re}(I=5/2^+) \xrightarrow{T_{1/2}=43.2\times10^9 \text{ y}} {}^{187}\text{Os}(I=1/2^-) + e^- + \bar{\nu} + 2.47\text{keV} . \qquad (75)$$

It exhibits extraordinary low *Q*-value [144] and long lifetime [145]; its natural isotopic abundance is 62.8%. As the decay is uniquely forbidden, another known, weakly varying shape factor *S*(*E*) enters the spectrum (43). The low *Q*-value off pays twice as compared to tritium: (i) The required energy resolution $\Delta E$ can be achieved at 7.5 times smaller resolving power $E/\Delta E$ of the detector; (ii) the burden of useless events occurring outside the tiny investigated window close to the endpoint is 400 times smaller. Both factors are essential in order to address the disadvantages of calorimetric spectroscopy with respect to MAC-E-Filters, namely: (i) Limitation of resolving power; (ii) Limitation of source strength due to the pile-up of low energy events in the rather slow calorimeter.

Calorimetric β-spectroscopy of $^{187}$Re was piloted first by the MANU experiment at Genoa [146,148]. The experimental scheme is shown in figure 32. A single crystal of 1.5 mg of metallic Re is cooled down to 60 mK into a superconducting phase at which its heat capacity is minimal; its activity is around 1 Bq. Within less than 1 ms, the decay heat equilibrates between the Re-source and a glued-on germanium thermistor, whose temperature change is measured by a sensitive circuit by means of the corresponding resistance change
.

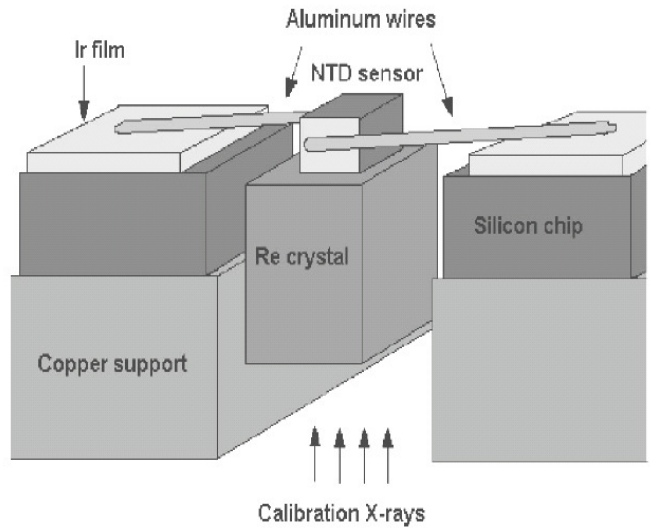

**Figure 32**: Set up of the MANU experiment, Genoa, for calorimetric β-spectroscopy at 60 mK from a metallic, superconducting Re-crystal (reprinted from ref. [144] with kind permission).

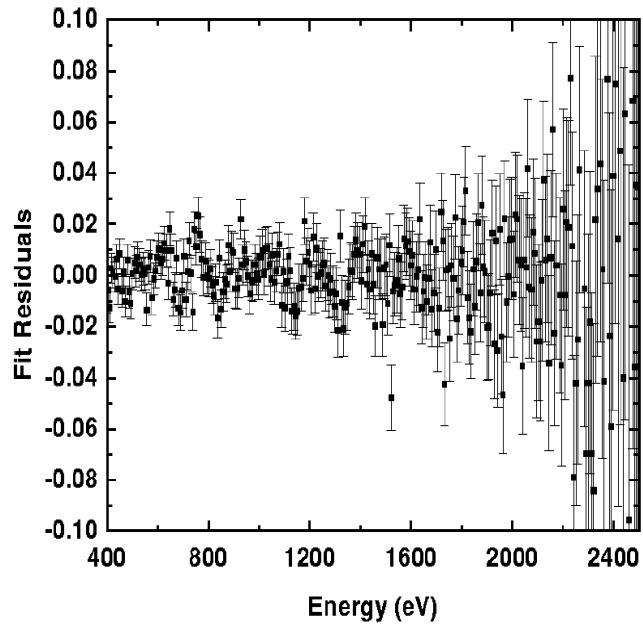

**Figure 33**: Residuals of the best fit of the theoretical to the experimental Re-β-spectrum, obtained in the MANU experiment. The quite significant modulation at low energy is an interference effect of the β-wave in the crystal lattice (reprinted from ref. [144] with kind permission).

The heat pulse is carried off to a sink by the connecting wires with a time constant of several tens of ms which only allows for a quite slow count rate. X-ray spectra are used for energy calibration of the output signals. The energy resolution is characterized by a Gaussian line shape with around 80 eV FWHM. Figure 33 shows the residuals from fitting a theoretical β-spectrum to the data. It shows a very peculiar oscillation superimposed onto a β-spectrum, which the authors have identified as beta environmental fine structure (BEFS) [151]; it is an interference effect of the particle wave in the crystal lattice which is also observed for photo electron cross sections in that energy range and known there as EXAFS (**E**xtended **X**-ray **A**bsorption **F**ine **S**tructure). In β-spectroscopy, the effect was already observed in 1985 as a slight anomaly at the lower end of the spectrum of tritium implanted into a silicon detector. But it was misinterpreted as a spectral component ascribed to the admixture of a heavy neutrino of mass 17 keV [152], which attracted great attention in subsequent years. The correct interpretation in analogy to EXAFS was given in 1991 [153]. Apart from this effect, the fit has yielded, for the first time, a precise endpoint energy of $E_0(^{187}Re) = 2470(4)$ eV and an upper limit of the neutrino mass of $m(v_e) < 26$ eV [154].

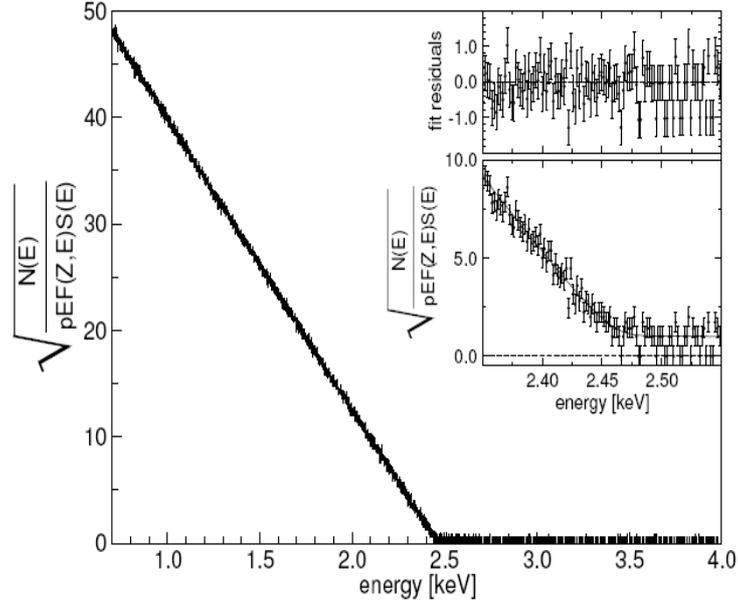

**Figure 34:** Kurie plot of Re-β-spectrum containing $6.2 \times 10^6$ counts obtained by the MIBETA array of microbolometers (reprinted from ref. [145] with kind permission).

Obviously a β-source of 1 Bq cannot lead to a competitive neutrino mass limit at this present time. Because of the 'pile-up' problem, a way out of the rate problem cannot be found by scaling up just the size of micro-calorimeters by large factors, if one does not succeed in improving the time resolution accordingly; however, the micro-calorimeters can be scaled up in number. First steps in this direction have been taken by the Milano/Como-Collaboration MIBETA [145,156]. They prepared an array of micro-bolometers, consisting of 10 AgReO$_4$ mono-crystals, weighing about 300 μg each. Doped silicon thermistors served as sensors. The discussion about the optimal choice of Re-compound and sensor material leads deep into questions of solid state and low temperature physics beyond the scope of this review. In summary, the bolometer pulses from MIBETA showed shorter rise time (≈ 500 μs) and better energy resolution (≈ 25eV FWHM) when compared to the preceding experiment at Genoa. Also, the many control and calibration problems, connected to long term precision measurements have been tackled systematically in the MIBETA-Experiment. Figure 34 shows the Kurie-plot of the Re-β-spectrum obtained from about 5000hrs of running. The fit yields improved values for endpoint and half-life. Regarding the neutrino mass the result is [145]

$$m^2(\nu_e) = (-112 \pm 207_{stat} \pm 90_{syst})\,\text{eV}^2 \quad \Rightarrow m(\nu_e) \leq 15\,\text{eV at 90\% C.L.} \qquad (76)$$

The first successful steps taken by MANU and MIBETA into this extremely low energy domain of β-decay have encouraged the micro-calorimeter community to proceed with a competitive precision search for the neutrino mass. This ambitious project is planned in 2 steps, MARE I and MARE II [157,158]. MARE I is to meet the existing upper limit of 2 eV around 2010, MARE II is to challenge the KATRIN goal of 0.2 eV, starting in 2011. Each step requires an improvement in experimental sensitivity and accuracy by a factor of 100 over the present status. Simulations show that the first goal requires a total event number of order of $10^{10}$ taken at a resolution of 10 eV and a fraction of 'pile up' events not exceeding a few times $10^{-5}$. The latter is the product of the rise time of the pulse and of the decay rate from a single module; hence it limits this rate well below 1Bq at the presently achieved rise time. MARE I comprises two experiments, namely MANU II and MIBETA II, both based on their respective predecessors. MANU II will be an array of about 300 metallic Re crystals, connected to Ir-Au-transition edge sensors (TES) which exploit the very steep change of resistance at the onset of superconductivity at the critical temperature $T_c$. The MIBETA II array will comprise a similar number of $AgReO_4$ crystals; regarding sensors wide-ranging R&D on thermistors is being carried out at the present time with the expectation of reaching rise times of around 50μs.

The goal of MARE II requires first of all raising the event number by another factor of $10^4$. It is felt that this could hardly be done just by correspondingly multiplying the number of modules . Instead it is preferred to limit this number to below 100 000 ,and to use larger source crystals, measured by faster and better resolving cryogenic bolometers, which are yet to be developed. To that end, the MARE collaboration has started a widespread international R&D programme. Apart from thermistors and transition edge sensors, magnetic microcalorimeters (MMCs) look very promising. With these one measures by induction the change of magnetization with temperature of paramagnetic ions in a metal, as for example an erbium/gold alloy. MMCs, optimized for x-ray detection, have recently reached an energy resolution of 3.4 eV at 6 keV [159].

The decision to build MARE II will eventually depend on two prepositions: (i) the technological problems have been proven to be solvable; (ii) MARE I has yielded satisfactory scientific results. The latter concerns in particular systematic effects. So far they have not as yet reached that stage; the analysis of the present spectra has lead straight to $m^2(v_e)$-results fully compatible with zero (see (76)). But they may go up at higher sensitivity and have to be overcome then. One cannot yet exclude, for instance, the possibility that tiny fractions of the decay energy might be trapped in metastable states for times longer than those used for integrating the heat bolus, or might be radiated off through the surface. One should also give a second thought to the final state spectrum of the decay, although its influence is largely suppressed by summing up all of the released electromagnetic energy, as explained above. Still one should bear in mind that the heavier the atom is, radioactive decay usually shakes up the atomic shell the more. This is a consequence of the increasing difference in total atomic binding energy between neighbouring elements which amounts to 15292 eV for the Re/Os-pair [160]; it provokes a non-adiabatic rearrangement of atomic shells in the daughter (see sect. 3.4). In the unique case of rhenium, of course, the tiny total decay energy of 2470 eV allows only the excitation of outer shells.

*5.3 Final remarks*

The neutrino mass experiments at Mainz and Troitsk have improved the sensitivity on the neutrino mass $m(\nu_e)$ compared to previous tritium β-experiments by a factor 5 down to a value of 2eV. This was only possible by developing a new type of high-resolution and large acceptance spectrometer (MAC-E-Filter), as well as by detailed investigations of the systematics and their reduction in side experiments.

If research on radioactive decay has still got to play a role in fundamental physics, then it has to face tremendously increased demands in terms of the sensitivity and precision of experiments. Hence such experiments have to be transformed from the laboratory bench to really large scale undertakings requiring high investment in R&D, construction and running and the commitment of large collaborations for more than a decade. The search for the absolute neutrino mass from single and double β-decay is undergoing this transition just now. The present limits on the eV-scale have been obtained from medium sized experiments which could still be handled by single groups. Their potential has now reached its limits, but the groups have also paved the way towards the coming generation of large scale experiments.

KATRIN, the forthcoming electrostatic filter for β-spectroscopy of tritium, will break the 1 eV limit of neutrino mass sensitivity shortly after getting started and then will approach its sensitivity limit of 0.2 eV within a couple of years of running. At that point it will encounter the mass scales which are addressed by scenarios of degenerate neutrino mass eigenvalues and by the claim of evidence of neutrino-less double β-decay of $^{76}$Ge [67, 68]. Even if KATRIN does not find a finite mass signal, its new, refined mass limit will have exhausted virtually all the space open for degenerate neutrino masses. Any substantial progress in terrestrial neutrino mass determination will also be a valuable input and a means of crosschecking for our understanding of the role of neutrinos in astrophysics.

Regarding β-spectroscopy of tritium, KATRIN will probably have no competition, as the experiment requires enormous efforts and alternatives are, as yet, not apparent. This could lead to a somewhat uneasy situation, in particular, if a finite but small mass signal happens to appear. How can the requirement to independently check a new result be fulfilled? Here calorimetric detection of rhenium β-decay may well be able to play a role. Present results encourage further developing its technology and upgrading its scale. Envisaging a sensitivity which is competitive to KATRIN is not unrealistic, although there might still be a long way to go.

**Acknowledgements**

We would like to thank our colleagues and friends from the KATRIN, Mainz, Troitsk and MARE collaborations for fruitful discussions. Among them we would like especially to name Jochen Bonn. For the discussion on neutrino theory and cosmology, we would like to express our thanks to Fedor Simkovic and Steen Hannestad. We acknowledge the support of our research in this field by the German Ministry of Education and Research (BMBF), the Helmholtz Association, the Deutsche Forschungsgemeinschaft (DFG) and the universities of Mainz and Münster.

**References**

[1] Pauli W 1930 in a letter to a local meeting on radioactivity at Tübingen, Germany and *Rapports du Septième Conseil de Physique Solvay, Brussels,* 1933 (Paris: Gauthier-Villars (1934))
[2] Fermi E 1934 *Z. Physik* **88** 11
[3] Cowan C L *et al* 1956 Science **124** 103
[4] Goldhaber M *et al*. 1958 *Phys. Rev*. **109** 1015
[5] Yao W M *et al*. (Particle Data Group) 2006 *J. Phys.* G **33**, 1 (URL: http://pdg.lbl.gov)
[6] Garwiser E and Silk J 1998 Science 280 1405
[7] Komatsu E *et al*. (WMAP Collaboration) 2008 arXiv:0803.0547
[8] Davis R, Harmer D S and Hoffman K C 1968 *Phys. Rev. Lett*. **20** 1205
[9] Hirata K S *et al.* (Kamiokande Collaboration) 1988 *Phys. Lett.* **B 205**, 416
[10] Casper D *et al*. (IMB Collaboration) 1991 *Phys. Rev. Lett.* **66** 2561
[11] Fukuda Y *et al.* (Super-Kamiokande Collaboration) 1998 *Phys. Rev. Lett*. **81** 1562
[12] Hoska J *et al.* (Super-Kamiokande Collaboration) 2006 *Phys. Rev.* **D 73** 112001
[13] Hirata K S *et al.* (Kamiokande Collaboration) 1989 *Phys. Rev. Lett.* **63** 16
[14] Anselmann P *et al.* (GALLEX Collaboration) 1992 *Phys. Lett*. **B 285** 386
[15] Abdurashitov J N *et al*. (SAGE Collaboration) 1994 *Phys. Lett.* **B 328** 234
[16] Ahmad Q R *et al*. (SNO Collaboration) 2001 *Phys. Rev. Lett.* **87** 071301
[17] Arpesella C *et al*. (BOREXINO Collaboration) 2008 *Phys. Lett.* **B 658** 101
[18] Eguchi K *et al.* (KamLAND Collaboration) 2003 *Phys. Rev. Lett*. **90** 021802
[19] Alui E *et al*. (K2K Collaboration) 1994 *Phys. Rev. Lett.* **94** 081802
[20] Michael D G *et al*. (MINOS Collaboration) 2006 *Phys. Rev. Lett.* **97** 191801
[21] Curran S C, Angus J and Cockroft A 1949 Phys. Rev. **76** 853
[22] Kraus Ch, Bornschein B, Bornschein L, Bonn J, Flatt B, Kovalik A, Ostrick B, Otten E W, Schall J P, Thümmler Th, Weinheimer C 2005 *Eur. Phys. J.* C **40**, 447
[23] Lobashev V M *et al*. 1999 *Phys. Lett.* B **460** 227
[24] Bergkvist K E 1972 *Nucl. Phys.* B **39** 317
[25] Ljubimov V A, Novikov E G, Nozik V Z, Tretyakov E F and Kosik 1980 *Phys. Lett.* B **94 266**
[26] Tretyakov E F *et al*. 1976 *Izv. Akad. Nauk SSSR Ser. Fiz.* **40** 20
Tretyakov E F *et al*. 1976 *Proc. Int. Neutrino Conf.* (Aachen) ed H Faissner, H Reithler and P Zerwas Braunschweig: Viewg 1977) p 663
[27] Boris S *et al*. 1987 *Phys. Rev. Lett.* **58** 2019
[28] Lippmaa E, Pikver R, Suurmaa E, Past J, Puskar J, Koppel I and Tammik A 1985 *Phys. Rev. Lett*. **54** 285
[29] Van Dyck R S Jr., D.L. Farnham, P.B. Schwinberg, Phys. Rev. Lett **70**, 2888 (1993)
[30] Nagy Sz, Fritioff T, Björkhage M, Bergström I and Schuch R 2006 *Europhys. Lett.* **74** 404
[31] Fritschi M, Holzschuh E, Kündig, Petersen J W, Pixley RE and Stussi H 1986 *Phys. Lett.* **B 173** 485
[32] Claxton T A, Schafroth S and Meier P F 1992 *Phys. Rev.* **A 45** 6209
[33] Holzschuh E, Fritschi M and Kündig W 1992 Phys. Lett. **B 287** 381
[34] Wilkerson J F, Bowles T J, Browne J C, Maley M P, Robertson R G H, Cohen J S, Martin R L,

Knapp D A and Helffrich J A 1987 *Phys. Rev. Lett.* **58** 2023

35 Stoeffl W and Decman D J 1995 *Phys. Rev. Lett.* **75** 3237

36 Lobashev V M and Spivak P E 1985 *Nucl. Instr. and Meth. in Phys. Res*. **A 240** *305*

37 Belesev A I *et al.* 1995 *Physics Letters* **B 350** *263*

38 Osipowicz A *et al.* (KATRIN Collaboration) 2001 KATRIN: A next generation tritium beta decay experiment with sub-eV sensitivity for the electron neutrino mass (*Letter of Intent*) arXiv:hep-ex/0109033

39 Angrik J *et al.* (KATRIN Collaboration) 2004 KATRIN Design Report *Wissenschaftliche Berichte Forschungszentrum Karlsruhe* **7090** http://bibliothek.fzk.de/zb/berichte/FZKA7090.pdf

40 Robertson R G H, Bowles T J, Stephenson G J Jr., Wark D L, and Wilkerson J F 1991 *Phys. Rev. Lett*. **67**, 957

41 Kawakami H *et al.* 1991 *Phys. Lett.* **B 256** 105

42 Sun H C 1993 *et al. Chinese Journal of NuclearPhysics* **15** 261

43 Feldman G J and Cousins R D 1998 *Phys. Rev.* **D 57** 3873

44 Weinheimer C, 2003 *Neutrino Mass (Springer Tracts in Modern Physics* vol 190 *)* ed G Altarelli and K Winter (Berlin, Heidelberg: Springer) pp 25-54

45 Wilkerson J F and Robertson R G H 2001 *Current Aspects of Neutrino Physics* ed D O Caldwell, (Berlin, Heidelberg: Springer)

46 Holzschuh E 1992 *Rep. Prog. Phys*. **55**, 1035

47 Robertson R G H and Knapp D A 1988 *Ann. Rev. Nucl. Part. Sci.* **38** 185

48 This speculation traces back to Pontecorvo who conceived it for the two flavours $(\nu_e, \nu_\mu)$, known at that time: Pontecorvo B 1967 Zh. Eksp. Teor. Fiz. 53 1717

49 Apollonio M *et al*. (CHOOZ Collaboration) 1999 *Phys. Lett.* **B 466** 415

50 Boehm F *et al*. (Palo Verde Collaboration) 2001 *Phys.Rev* **D 64** 112001

51 Fogli G L *et al*. 2006 Prog. Part. Nucl. *Phys*. **57** 742

52 King S F 2004 Rep. Prog. Phys. **67** 107

55 Yanagida T 1979 Workshop on Unified Theories (KEK report 79-18, 1979), eds. Savada O and Sugamoto A, p95

56 Gell-Mann M, Ramond P, Slansky R 1979 Supergravity, eds. Freedman D, Niewenhuizen P van, North-Holland

57 Magg M and Wetterich C 1980 *Phys. Lett.* B**94** 61

58 Weinberg S 1979 *Phys. Rev. Lett*. **43** 1566

59 Mohapatra R N *et al.* 2007 *Rep. Prog. Phys*. **70** 1757

60 Lesgourgues J and Pastor S 2006 *Phys. Repts* **429** 307

61 Percival W J *et al*. 2007 *MNRAS* **381** 1053

62 Hannestad S, Mirizzi A, Raffelt G G and Wong Y Y Y 2007 JCAP 0708 015

63 Fogli G L, Lisi E, Marrone A, Melchiorri A, Palzzo A, Serra P, Silk J and Slozar A 2006 *arXiv:hep-ph*/0608060v1

64 Host O, Lahav O, Abdalla F B and Eitel K, 2007 *Phys. Rev*. **D 76** 113005

65 Hannestad S 1997 arXiv:0710.1952

66 Simkovic F, Faessler A, Rodin V, Vogel and J Engel P 2007 *preprint* arXiv. 0710.2055v2 [nucl-th]

67 Schechter J and Valle J 1981 *Phys. Rev*. **D 23** 1666

68 Klapdor-Kleingrothaus H V *et al.* (Heidelberg-Moscow Collaboration) 2001 *Eur. Phys. J.* **A 12** 147

69 Klapdor-Kleingrothaus H V, Dietz A, Harney H L and Krivosheina I V 2001 *Mod. Phys. Lett.* **A 16** 2409

70 Klapdor-Kleingrothaus H V, Krivosheina I V, Dietz A and Chkvorets O 2004 *Physics Letters* **B 586** 198

71 Aalseth C E *et al.* 2002 *Mod. Phys. Lett.* **A17** 1475

72 Feruglio F, Strumia A, Vissani F 2002 *Nucl. Phys*. **B637** 345

73 Klapdor-Kleingrothaus H V and Krivosheina I 2006 *Mod. Phys. Lett*. **A 21** 1547

74 Avignone F T, Elliott S R, Engel J 2008 *Rev. Mod. Phys*. **80** 481, arXiv: 0708.1033 al

75 Assamagan K *et al*,1996 *Phys. Rev*. **D 53** 6065

76 Barate R *et al.* 1998 *Eur. Phys. J.* **C 2**, 1 395

77 Armbruster B 2002 Phys. Rev. D65 112001

78 Aguilar-Arevalo A A *et al.* (MiniBooNE Collaboration) 2007 *Phys. Rev. Lett.* **98** 231801

79 Aguilar A 2001 Phys. Rev. D64 112007

80 Alexeyev E N, Alexeyeva L N, Krivosheina I V and Volchenko V I 1988 *Phys. Lett*. **B 205** 209

81 Bratton C B *et al*. (IMB Collaboration) 1988 *Phys. Rev*. **D 37** 3361

82 Hirata K *et al.* (Kamiokande Collboration) 1988 *Phys. Rev*. **D 38** 448

83 Loredo T J and Lamb D Q 2002 *Phys. Rev*. **D 65** 063002

84 Pascoli S and Petcov S T 2002 *Phys. Lett.* **B544** 239

85 Souers P C *Hydrogen Properties for Fusion Energy* 1986 (University of California Press, Berkeley, Cal., USA)

86 Jerziorski B *et al.,* 1985 Phys. Rev. **A 32** 2573

[87] Severijns N, Beck M, Naviliat-Cuncic O, 2006 Rev. Mod. Phys. **78** 991
[88] Simkovic F, Dvornicky R, Faessler A, 2007, arXiv:0712.3926 [hep-ph]
[89] Masood S S *et al.*, 2007 Phys. Rev. **C 76** 045501
[90] Weinheimer C *et al.*1993 *Phys. Lett* **B 300** 210
[91] Ciborowski J and Rembielinski J 1999 *Eur. Phys. J.* **C 8**, 157
[92] Repco W W and Wu C E 1983 *Phys. Rev.* **C 28** 2433
[93] Gardner S, Bernard V and Meißner U G 2004 *Phys. Lett.* **B 598** 188
[94] Stephenson G J and Goldman T 1998 *Phys. Lett*. **B 440** 89
[95] Ignatiev A Yu and McKellar B H J 2006 *Phys. Lett*. **B 633** 89
[96] Bonn J, Eitel K, Glück F, Sevilla-Sanchez D and Titov N 2006 *arXiv*:0704.3039v1 [hep-ph]
[97] Migdal A B 1977 Qualitative Methods in Quantum Theory W.A. Benjamin Inc. London
[98] Fackler O, Jeziorski B, Kolos W, Monkhorst H J and Szalewicz K 1985 *Phys. Rev. Lett.* **55** 1388
[99] Saenz A, Jonsell S and Froelich P 2000 *Phys. Rev. Lett.* **84** 242
[100] Fleischmann L, Bonn J, Degen B, Przyrembel M, Otten E W, Weinheimer C and Leiderer P 2000 *J. Low Temp. Phys.* **119** 615
[101] Fleischmann L, J Bonn, Bornschein B, Leiderer P, Otten E W, Przyrembel M and Weinheimer C 2000 *Euro. Phys. J.* **B 16** 521
[102] Doss N, Tennyson J, Saenz A, S. Jonsell, Phys. Rev. C **73**, 025502 (2006)
[104] Kolos W, Jeziorski B, Rychlewski J, Szalewicz, and O. Fackler 1988 *Phys. Rev.* **A 37** 2297
[105] Weinheimer C, Degen B, Bleile A, Bonn J, Bornschein L, Kazachenko O, Kovalik A and Otten E W, 1999 *Phys. Lett.* **B 460** 219
[106] Otten E W 1994 *Neutrinos in astro, particle and nuclear physics: Proc Int. School of Nuclear Physics (Erice, Italy, 1993) (Prog. Part. Nucl. Phys.* **32** 153) ed A Faessler
[107] see *e. g.*: Fritioff T, Carlberg C, Douysset G, Schuch R and Bergström I 2005 *Eur. Phys. J.* **D 15** 141 and ref.s therein
[108] Blaum K, Mainz, 2006 *priv. commun.*
[109] see also the discussion in: Otten E W, Bonn J and Weinheimer C 2006 *International Journal of Mass Spectrometry* **251** 371

[110] Hamilton D R and Gross L 1950 *Rev. Sci. Instr.* **21** 912

[112] Hamilton D R, Parker Alford W and Gross L 1953 *Phys. Rev.* **92** 1521
[113] Kruit P and Read F 1983 *J. Phys.* **E 16** 313
[114] Picard A *et al*. 1992 *Nucl. Instumr. Meth.* **B 63** 345
[115] Jackson J D 2004 Classical Electrodynamics, John Wiley and Sons Ltd.
[116] Picard A *et al*. 1992 *Z.Phys.* **A 342** 71
[117] Venos D, Dragoun O, Spalek A and Vobecky M 2006 *Nucl. Instr. Meth Phys. Res.* **A 560** 352
[118] see e.g.: Lehnert B 1964 *Dynamics of charged particles* (Amsterdam: North Holland,)
[119] Bonn J, Bornschein L, Degen B, Otten E W and Weinheimer C 1999 *Nucl. Instrum. Meth.* **A 421** 256
[120] Schall J P 2001 *Diploma Thesis* Mainz University
 Schwamm F *Dissertation* Karlsruhe University and to be published
[121] Lobashev V L, Troitsk, priv. commun.
[122] Sevilla-Sanchez D 2005 *Diploma Thesis* Mainz University and to be published
[123] Müller B 2002 *Diploma thesis, Mainz University,* Flatt B 2004 Dissertation*, Mainz University* and to be published
[124] Conradt R, Albrecht U, Herminghaus S and Leiderer P 1994 *Physica* **B 194-196** 679
[125] Classen J, Eschenröder K and Weiss G 1995 *Phys. Rev.* **B 52** 11475
 Conradt R, Dissertation, Konstanz University, 1996
[126] V.N. Aseev *et al.* 2000 *Euro. Phys. J.* **D 10** 39
[127] Lobashev V M 1998 *Prog. Part. Nucl. Phys.* **40** 337
[131] Barth H *et al.* 1998 *Neutrinos in astro, particle and nuclear physics: Proc Int. School of Nuclear Physics (Erice, Italy, 1997) (Prog. Part. Nucl. Phys.* **40** 353) ed A Faessler
[132] Bornschein B, Bonn J, Bornschein L, Otten E W and Weinheimer C 2003 J. Low Temperature Phys. **131** 69
[133] Bonn J et al 2001 Nucl. Phys. B (Proc. Suppl.) **91** 273
[134] Stephenson G J and Goldman T 1998 *Phys. Lett.* **B 440** 89
[135] Lobashev V M 2002 *Prog. Part. Nucl. Phys.* **48** 123
[136] Saenz A 2002 Humboldt University, Berlin, *priv. commun.*

[137] Lobashev V M 2003 *(Proc. 17. Int. Conf. on Nuclear Physics in Astrophysics (Debrecen, Hungary, 2002) Nucl. Phys.* **A 719** 153

[138] Lobashev V M 2005 *private commun.*

[139] Lobashev V M 1998 *Neutrinos in astro, particle and nuclear physics: Proc Int. School of Nuclear Physics (Erice, Italy, 1997) (Prog. Part. Nucl. Phys.* **40** 337) ed A Faessler

[140] Nastoyashchii A F, Titov N A, Morozov I N, Glück F and Otten E W 2005 *American Nuclear Society: Fusion Science and Technology* **48** 743

[141] Lobashev V M 2005 *Proc. 11. Int. Workshop on Neutrino Telescopes , (Venice, Italy)* ed Baldo Ceolin M (Edizioni Papergraf, 2005) pp. 507 -517

[142] Weinheimer C 2006 Prog. Part. Nucl. Phys. **57** 22

[143] Ellis C D and Wooster W A *Proc. Roy. Soc. Lond*. **A 117** 109
Meitner L and Orthmann W 1930 *Z. Physik* **60** 143

[144] Galeazzi M *et al.* 2001 *Phys. Rev*. **C 63** 014302

[145] Sisti M *et al.l.* 2004 *Nucl. Instr. Meth*. **A 520** 125

[146] Cosulich E *et al.* 1992 *Phys. Lett*. **B 295** 143

[151] Gatti F, Fontanelli F, Galeazzi M, Swift A M and Vitale S 1999 *Nature* **397** 137

[152] Simpson J J 1985 *Phys. Rev. Lett.* **54** 1891

[153] S. E. Koonin, Nature (London) **354**, 468 (1991)

[154] Gatti F 2001 *(Neutrino 2000, Sudbury Canada: Nucl. Phys. B (Proc. Suppl.)* **91** (2001) 293*)* eds J Law, R W Ollerhead and J J Simpson

[156] Nucciotti A, *et al.* 2002 *Proc. 9. Int. Workshop on Low Temperature Detectors (Madison, Wisconsin, USA 2001) (Melville, NY: American Institute of Physics,* Vol. 605) eds Scott Porter F, D McCammon D, Galeazzi M and Stahle C K p 453.

[157] Monfardini A *et al.* arXiv:hep-ex/0509038v1 A. Monfardini *et al*. in Proc. of 11th Intern. Workshop on Low Temperature Detectors (LTD-11), Tokyo, Japan, 31 Jul - 5 Aug 2005; *Prog. Part. Nucl. Phys.* **57** (2006) pp. 68-70, *Nucl. Instrum. Meth*. **A559** (2006) pp. 346-348; e-Print: hep-ex/0509038.

[158] A detailed proposal by Gatti F *et al*. may be downloaded from the webpage: http://mare.dfm.uninsubria.it/frontend/exec.php

[159] Fleischmann A, Daniyarov T, Rotzinger H, Enss C, Seidel G M 2003 *Physica* **B 329-33** 1594

[160] Rodrigues G C, Indelicato P Santos J P, Patt P and Parente F 2004 *Atom. Data Nucl. Data Tables* **86** 117